\documentclass[11pt,a4paper,preview]{article}
\usepackage[utf8]{inputenc}
\usepackage{amsmath}
\usepackage{amsfonts}
\usepackage{amssymb}
\usepackage{amsthm}

\usepackage{graphicx}

\usepackage{xcolor}
\usepackage[left=2.5cm,right=2.5cm,top=2cm,bottom=2cm]{geometry}
\usepackage{hyperref}
\usepackage{apacite}
\usepackage{bbm}
\usepackage{booktabs}
\usepackage{color}
\usepackage{multicol}
\usepackage[justification=centering]{caption}
\usepackage{subcaption}
\usepackage{multirow}
\usepackage{stackrel}
\usepackage{hhline}
\usepackage{makecell}
\usepackage{nicefrac}
\usepackage[hang]{footmisc}
\usepackage{authblk}
\usepackage{todonotes}
\usepackage{csquotes}
\usepackage{chngcntr} 
\usepackage[margin=true,inline=false,index=true]{fixme} 
\fxsetup{theme=color}
\definecolor{fxtarget}{rgb}{0.8000,0.0000,0.0000}
\allowdisplaybreaks 

\counterwithin{figure}{section}
\counterwithin{table}{section}

\numberwithin{equation}{section}
\theoremstyle{definition}
\newtheorem{assumption}{Assumption}[section]

\newtheorem{lemma}[assumption]{Lemma}

\newtheorem{corollary}[assumption]{Corollary}
\newtheorem{proposition}[assumption]{Proposition}

\newcommand{\EQ}[1]{\mathbb{E}_{\mathbb{Q}}\left[ #1\right]}
\newcommand{\EQtil}[1]{\mathbb{E}_{\tilde{\mathbb{Q}}}\left[ #1\right]}

\newcommand{\indF}[2]{\mathbbm{1}_{#2}\left( #1\right)} 
\newcommand{\tin}{t \in [0, T]}
\newcommand{\Q}{\mathbb{Q}}
\newcommand{\Qtilde}{\mathbb{\tilde{\Q}}}
\newcommand{\R}{\mathbb{R}}

\newcommand{\Ztildet}[1]{\widetilde{Z}({#1})}

\newcommand{\Ztillamt}[1]{\widetilde{Z}_{\lambda}({#1})}

\newcommand{\ZtillamastT}{\widetilde{Z}_{\lambda^{\ast}}(T)}

\newcommand{\phib}{\bar{\varphi}}
\newcommand{\pib}{\bar{\pi}}

\newcommand{\la}{\lambda^{\ast}}
\newcommand{\gamlam}{\gamma_{\lambda}}
\newcommand{\gamlamast}{\gamma_{\lambda^{\ast}}}

\DeclareMathOperator*{\argmin}{argmin}

\newcommand{\lrr}[1]{\left( #1 \right)}

\title{Decrease of capital guarantees in life insurance products: can reinsurance stop it?}

\author[a]{Marcos Escobar-Anel}
\author[b]{Yevhen Havrylenko\thanks{ Corresponding author.\\ E-Mail addresses: \href{mailto:marcos.escobar@uwo.ca}{marcos.escobar@uwo.ca} (M. Escobar-Anel), \href{mailto:yevhen.havrylenko@tum.de}{yevhen.havrylenko@tum.de} (Y. Havrylenko), 
\href{mailto:michel.kschonnek@tum.de}{michel.kschonnek@tum.de} (M. Kschonnek), \href{mailto:zagst@tum.de}{zagst@tum.de} (R. Zagst)} }
\author[b]{Michel Kschonnek}
\author[b]{Rudi Zagst}
\affil[a]{Department of Statistical and Actuarial Sciences, Western University, 1151 Richmond street, London, Canada}
\affil[b]{Chair of Mathematical Finance, Department of Mathematics, Technical University of Munich, Parkring 11, 85748 Garching bei M\"unchen, Germany}
\date{}                     
\begin{document}
\maketitle

\begin{abstract}
We analyze the potential of reinsurance for reversing the current trend of decreasing capital guarantees in life insurance products. Providing an insurer with an opportunity to shift part of the financial risk to a reinsurer, we solve the insurer's dynamic investment-reinsurance optimization problem under simultaneous Value-at-Risk and no-short-selling constraints.
We introduce the concept of guarantee-equivalent utility gain and use it to compare life insurance products with and without reinsurance. Our numerical studies indicate that the optimally managed reinsurance allows the insurer to offer significantly higher capital guarantees to clients without any loss in the insurer's expected utility. 
The longer the investment horizon and the less risk-averse the insurer, the more prominent the reinsurance benefit.
\end{abstract}




\section{Introduction}$\,$
\textbf{Motivation.} The capital guarantee is an important feature of various life insurance products. However, it is restrictive for shareholders of insurance companies, since higher capital guarantees for clients usually mean lower upside potential for investments, whence lower potential return on equity for shareholders. Since the financial crisis of 2007-2008, insurance companies have been facing substantial challenges in offering products with capital guarantees, for example, low interest rates and stricter regulations. As a consequence, global insurers began decreasing guarantee levels embedded in their products. For instance, starting in 2021, Allianz provides only $60\%$ to $90\%$ capital guarantee in their new life insurance products\footnote{See e.g., \url{https://www.sueddeutsche.de/wirtschaft/lebensversicherung-allianz-kuenftig-ohne-beitragsgarantie-1.5056917}}, although the company maintains old policies that have a non-negative guaranteed rate of return on all client's contributions.

In theory, the capital guarantee can be achieved via a Constant Proportion Portfolio Insurance (CPPI) strategy or an Option-Based Portfolio Insurance (OBPI) strategy. In a  CPPI strategy, the capital guarantee is ensured via investing a large proportion of the portfolio value in bonds. Also, a CPPI-fund can be combined with another riskier fund to improve portfolio performance while still reaching the guarantee. This approach is implemented e.g., in the so-called Drei-Topf-Hybrid\footnote{In English \enquote{Three-Pot-Hybrid}} (DTH) products \cite{Hambardzumyan2019}. In contrast to CPPI-investors, OBPI-investors secure the guarantee on the invested capital by holding a put option on their portfolio. OBPI strategies require the managed portfolio and the portfolio underlying the put option to be equal.
An example of an OBPI-based product is \enquote{ERGO Rente Garantie} launched in early 2010s. Purchasing this equity-linked product, clients can choose a guarantee of either $80\%$ or $100\%$ of their contributions. According to the product description\footnote{See Slide 13 in \url{https://www.yumpu.com/de/document/view/22247401/expertenwissen-zur-ergo-rente-garantie}}, clients' contributions are invested in fixed income assets as well as a target volatility fund (TVF) and are reinsured by Munich Re\footnote{See \url{https://www.focus.de/finanzen/steuern/ergo-ergo-rente-garantie_id_3550999.html}}.

Declining capital guarantees and the absence of papers analyzing the role of reinsurance in the design of life insurance products with capital guarantees motivate our research. In this paper, we answer the following questions in the context of equity-linked life insurance products with capital guarantees. When is reinsurance needed in the management of such products? What is the impact of reinsurance on capital guarantee levels? How does one optimize the asset allocation and the reinsurance strategy? 

To better model practical limitations, our framework allows the reinsurance of a subset of available financial assets and a subset of admissible investment strategies of insurers\footnote{This is different from an OBPI strategy where the put option's underlying is the actually managed portfolio}. This feature is motivated by the realistic situation when reinsurers sell protection only on specific portfolios or indices whose risks they understand and/or can control sufficiently well. One example would be when the reinsurance is on a well-known index and the actual investment opportunity for the insurer is based on an own strategic portfolio or on an exchange-traded fund that is not equal to the index. In another example insurers would invest part of their money beyond the reinsured portfolio in a riskier fund to have more upside potential.

\textbf{Implementation.} For simplicity, we consider one insurance company and one reinsurance company. The insurer can invest clients' money in a riskless asset as well as a  risky asset and can purchase a reinsurance contract.
The reinsurance contract is modelled as a put option on a constant-mix portfolio highly correlated with the investable risky asset.  The insurer maximizes its expected utility from terminal wealth given two constraints. First, the probability that the terminal portfolio value is below the capital guarantee to the client must be less than or equal to some threshold probability, i.e., a Value-at-Risk (VaR) constraint. Second, the insurer cannot have negative positions in risky assets and reinsurance, i.e., a no-short-selling constraint.

The above-mentioned portfolio optimization problem has three aspects different from standard settings: a traded put option (reinsurance), a no-short-selling constraint and a VaR-constraint. We solve our optimization problem in three steps. First, we link the problem in the original market with a reinsurance contract to a problem in the market containing only basic assets without optional features. This approach is inspired by \citeA{Korn1999}, where it is applied to the optimal control of a portfolio including stock options, and \citeA{Hambardzumyan2019}, where it is used in the optimal control of a DTH product. Second, we transform the optimization problem with both VaR- and allocation constraints in the financial market with basic assets to the one with only VaR-constraint but in an auxiliary financial market, following \citeA{Cvitanic1992}. Third, we transform that VaR-constrained problem to a VaR-unconstrained one and solve the latter using ideas from \citeA{Basak2001} and \citeA{Kraft2013}.

After solving the original investment-reinsurance problem in a semi-closed form, we show analytically when reinsurance is needed, i.e., when the investment in reinsurance of the optimal strategy is positive. Next, we calibrate our model to the German market and conduct a suboptimality analysis. There, we analyze the optimal investment-reinsurance strategy, the optimal investment strategy without reinsurance and a constant-mix strategy representative for an average life insurance company. First, we compare these strategies with respect to wealth-equivalent utility loss studied in \citeA{Larsen2012}. Second, we introduce a novel suboptimality measure --- the guarantee-equivalent utility gain --- and compare strategies with respect to it. After the suboptimality analysis, we investigate the impact of model parameters on the investment strategies and also give insights into measuring the level of reinsurance in the insurer's portfolio.



	\textbf{Contribution.} We contribute to the existing literature as follows,
	\begin{itemize}
	    \item actuarial literature:
	    \begin{itemize}
	        \item designing a realistic and workable framework for finding optimal investment-reinsurance strategies for equity-linked insurance products with capital guarantees. The framework combines put options, regulatory VaR and no-short-selling constraints, and a separation between insurable and reinsurable funds;
	        {
	        \item detecting market conditions and the asset manager's proficiency, for which (partial) reinsurance of the capital guarantee is advantageous. In particular, we find that reinsurance is optimal when the risky asset in the insurer's investment portfolio has a high correlation with and has a higher Sharpe ratio than the reinsurable asset;
	        \item establishing that optimal reinsurance significantly increases capital guarantees, while slightly decreasing product costs. In a typical example, a 10-year equity-linked insurance product that follows an optimal investment strategy uses no reinsurance and offers a terminal guarantee of $100\%$ (equivalent to $0\%$ annualized guaranteed return)
	        of the client's initial endowment can be replaced by a product that follows an optimal investment-reinsurance strategy with a terminal guarantee of $110\%$ (i.e., $0.96\%$ annualized guaranteed return)
	        of the initial capital. This is with  no loss in the insurer's expected utility. Moreover, our proposal allows the insurer to guarantee $128\%$ (equivalent to approximately $2.5\%$ annualized guaranteed return)
	        of the client's initial endowment without any decrease in the insurer's expected utility in comparison to the one obtained by a constant-mix $85\%$ bonds and $15\%$ stocks strategy, which is representative of investment strategies of life insurers;
	        }
	    \end{itemize}
	    \item portfolio optimization literature:
	    \begin{itemize}
	        \item solving explicitly the portfolio optimization problem with simultaneous VaR-constraint and no-short-selling constraint in a financial market with a traded put option on a well-known portfolio. The methodology presented here can be adapted to other types of terminal wealth constraints (e.g., an expected shortfall constraint), other types of allocation constraints (e.g., specific assets are not allowed to be traded), and other derivative-like traded assets, e.g., a call option, an OBPI fund, etc.
	    \end{itemize}
	\end{itemize}

	\textbf{Literature overview.} 
	
	\textit{Insurance focus.} \citeA{Mueller1985} derived the optimal investment-reinsurance strategy of a pension fund in a static portfolio optimization framework with exponential utility function and no constraints on terminal wealth or allocation. For dynamic portfolio optimization, the literature on reinsurance is mainly focused on the insurers' overall surplus processes and not on specific products or investment portfolios (e.g., \citeA{Luo2008}, \citeA{Bai2010}, \citeA{Liang2011}, \citeA{Li2014}). To the best of our knowledge, there are no publications on the optimal dynamic investment-reinsurance strategies for life insurance products with capital guarantees. Our research fills this gap.
    
    {Several papers study optimal investment strategies for life insurance products with capital guarantee in financial markets without reinsurance opportunities.} We mention only the most recent. \citeA{Chen2019} analyze optimal investment strategies when the capital guarantee is embedded in a piece-wise linear payoff of the decision maker such that the whole payoff is fairly priced. \citeA{Dong2019} and \citeA{Dong2020} consider loss-averse insurers and derive optimal investment strategies under both no-short-selling and terminal portfolio value constraints. In the former paper, the capital guarantee is modelled as a hard lower bound on the terminal portfolio value, whereas in the latter paper the guaranteed amount is part of the VaR-constraint. \citeA{Hambardzumyan2019} is centered around DTH products in the expected utility framework. The researchers consider the insurer who can invest in a risk-free bank account, a CPPI-fund as well as a free fund. Imposing  also a capital guarantee as a hard lower bound on the terminal portfolio value, the authors derive the optimal trading strategies. In our framework, the insurer can purchase reinsurance instead of a CPPI fund, has a no-short-selling constraint and a VaR-constraint, which is more general than the strict lower bound constraint.
	
	\textit{Portfolio optimization focus.} The classic continuous-time dynamic portfolio optimization problem of a utility-maximizing investor in a Black-Scholes market was first considered in an unconstrained setting in \citeA{merton1969} and \citeA{merton1971}, and has since been extended in a myriad of different ways. A fruitful branch is the addition of constraints restricting an investor's portfolio choice. One of the most natural constraints is a strict lower bound on the (terminal) portfolio wealth, which was considered in \citeA{Tepla2001} and \citeA{Korn2005} or a probabilistic lower bound, also-called VaR-constraint, which was considered in \citeA{Basak2001} and \citeA{Boyle2007}. Such constraints on the terminal {wealth}  can  be linked to a corresponding unconstrained portfolio optimization problem through the addition of a suitable Lagrange multiplier on the terminal utility, which is solvable via the Martingale approach. This link was investigated by \citeA{Kraft2013} using Hamilton–Jacobi–Bellman (HJB) equations and Black-Scholes partial differential equations (PDEs), which allow them to express the optimal terminal wealth as a function of the optimal unconstrained terminal wealth. The methodology developed in \citeA{Kraft2013} is not designed for constraints on portfolio allocation such as no-short-selling constraints, non-traded asset constraints, no-borrowing constraints or other general convex constraints on the allocation. However, by embedding an allocation-constrained portfolio optimization problem into a family of unconstrained portfolio optimization problems in different auxiliary markets, \citeA{Cvitanic1992} were able to derive closed-form solutions for {constant relative risk aversion} (CRRA) utility functions by solving the HJB PDE for an associated dual optimization problem. \\
	
	Thus far, combinations of constraints on terminal wealth and portfolio allocation have rarely been studied in the existing literature due to the increased complexity of the associated HJB PDEs.
	Exceptions to this  can be found in the work of \citeA{Bardhan1994}, \citeA{Dong2019}, \citeA{Wahl2019} as well as \citeA{Dong2020}. Specifically, in \citeA{Dong2020} the authors consider a portfolio optimization problem of a pension fund manager with S-shaped utility, convex cone allocation constraints as well as VaR-constraints motivated by a defined contribution pensions plan. They remove the VaR-constraint by adding a Lagrange multiplier to the utility function and concavify the resulting function. For every Lagrange multiplier, the resulting portfolio optimization problem is allocation-constrained and wealth-unconstrained and can thus be solved via the HJB PDE of the associated dual problem, as shown in \citeA{Zheng2011}. Finally, the authors of \citeA{Dong2020} show that an optimal Lagrange-multiplier exists such that an optimal solution for the original VaR-constrained problem can be obtained by this methodology. Although we consider a similar problem set-up as in \citeA{Dong2020}, our solution approach is different, providing a new path to solving these challenging problems. We build on the results in \citeA{Basak2001} to show that in a class of allocation-unconstrained, but VaR-constrained portfolio optimization problems the optimal solution can be expressed as a function of the VaR-unconstrained optimal terminal wealth, which is in the spirit of the ideas presented in  \citeA{Kraft2013}. Then, we show that the same auxiliary market as in \citeA{Cvitanic1992} admits the optimal solution to the original problem. This way we avoid determining the solution to the dual HJB PDE explicitly, which in some cases may not be viable. Further, we demonstrate that the optimal terminal portfolio value of the allocation- and VaR-constrained portfolio optimization problem can be expressed as a function of the optimal terminal wealth of the allocation-constrained and wealth-unconstrained portfolio optimization problem, i.e., the derivative structure proposed in \citeA{Kraft2013} is preserved under allocation constraints.
	
	\textbf{Paper structure.} In Section \ref{sec_problem_setting} we describe formally the financial market and state the optimization problem. The solution is presented in Section \ref{sec_solution_to_OP}. In  Section \ref{sec_numerical_studies} we calibrate our model parameters to the German market and conduct numerical studies. First, we conduct a suboptimality analysis of the strategies with respect to the initial wealth as well as the capital guarantee. Second, we calculate numerically the sensitivity of the optimal investment-reinsurance strategy with respect to the model parameters. In Section \ref{sec_conclusions} we draw conclusions and give an outlook for further research. In Appendix \ref{app:proofs_main}, we provide the proofs of the main theoretical results. Appendix \ref{app:proofs_auxiliary} contains the proofs of auxiliary theoretical results.

\section{Problem setting}\label{sec_problem_setting}$\,$
    
	Let $T>0$ be a finite time horizon and $(W(t))_{\tin} = ((W_1(t),W_2(t))')_{\tin}$ be a two-dimensional Wiener process on a filtered probability space $\left( \Omega, \mathcal{F}, \left(  \mathcal{F}_t\right)_{\tin}, \mathbb{Q}\right)$ that satisfies usual conditions of completeness and right-continuity. We denote by $\Q$ the real-world probability measure and by $\Qtilde$ the risk-neutral probability measure. The basic assets in the market are described by the following dynamics w.r.t. $\Q$:
	\begin{equation}
	\begin{aligned}
		dS_0(t) &= S_0(t) r dt\,\, \text{ bank-account}\\
		dS_1(t) &= S_1(t) (\mu_1 dt + \sigma_1dW_1(t))\,\, \text{fund the insurer can  invest in but cannot be reinsured}\\
		dS_2(t) &= S_2(t) (\mu_2 dt + \sigma_2(\rho dW_1(t) + \sqrt{1 - \rho^2}dW_2(t))\,\,\text{reinsurable fund}
	\end{aligned}
	\end{equation}
	with $S_0(0) = s_0,\,S_1(0) = s_1,\,S_2(0) = s_2$ and constants $r, \mu_1, \mu_2 \in \mathbbm{R}, \sigma_1,\sigma_2 > 0, \rho \in [-1,1]$.
	
	{We refer to this market as a linear market, since the value of any portfolio consisting of $S_0, S_1, S_2$ is linear with respect to prices of the basic assets.}
	Denoting $\bar{1} = (1, 1)'$, we use the following notation:
	\begin{equation*}
		\begin{aligned}
			\sigma &:=
			\left(
			\begin{matrix}
			\sigma_1 & 0 \\ 
			\sigma_2 \rho &  \sigma_2 \sqrt{1 - \rho^2}
			\end{matrix}
			\right)
			& 
			\mu &:=
			\left(
			\begin{matrix}
			\mu_1 \\ 
			\mu_2
			\end{matrix}
			\right)\\
			\gamma &:= \sigma^{-1} (\mu - r\bar{1})&
			\Ztildet{t} &:= e^{-\lrr{r + 0.5||\gamma||^2}t - \gamma' W(t)}
		\end{aligned}
	\end{equation*}
	and assume that $\mu - r\bar{1} \neq 0$.
	
	We assume that there are three parties: a client, a reinsurer and an insurer. The insurer receives at $t=0$ the client's initial endowment denoted by $v_0 > 0$, invests its on the client's behalf, and promises to pay back to the client at time $T$ at least the capital guarantee $G_T>0$.
	
	We assume that only constant-mix (CM) portfolios can be reinsured. This choice is motivated by the equivalence of constant-mix strategies and target-volatility strategies in a Black-Scholes market. Recall that the reinsurable risky portfolio in \enquote{ERGO Rente Garantie} is a target volatility fund. For CM strategies, the reinsurer can evaluate sufficiently well the potential loss in advance and can easily price it.
	So for $\pi_B^{CM} \in [0, 1]$, we denote by $\pi_B(t) = (0, \pi_B^{CM})',\,\tin,$ the relative portfolio process related to the risky assets in the CM strategy that the reinsurer can reinsure via a put option. Under this strategy, the proportion of wealth invested in $S_0$ equals  $1 -  \pi_B^{CM},\, \tin$, whereas the proportion of wealth invested in $S_2$ equals $\pi_B^{CM},\,\tin$. The dynamics of the corresponding CM portfolio value is given by:
	\begin{equation}\label{eq:CM_PF_dynamics}
		\begin{aligned}
		dV^{v_0, \pi_B}(t) &= (1 - \pi_B^{CM}) V^{v_0, \pi_B}(t) \frac{dS_0(t)}{S_0(t)} + \pi_B^{CM} V^{v_0, \pi_B}(t) \frac{dS_2(t)}{S_2(t)}\\
		& = V^{v_0, \pi_B}(t) ((r + \pi_B^{CM} (\mu_2 - r)) dt + \pi_B^{CM} \sigma_2 (\rho dW_1(t) + \sqrt{1 - \rho^2}dW_2(t)).
		\end{aligned}
	\end{equation}

	Let $P(t)$ be the fair price at time $t$ of a put option with the payoff $(G_T - V_T^{v_0, \pi_B})^+$.
	Then:
		
	\begin{equation}
		P(t) = S_0(t)\EQtil{S_0(T)^{-1}(G_T - V^{v_0, \pi_B}(T))^+|\mathcal{F}_t}
	\end{equation}
	
	The stochastic differential equation (SDE) of the put price is given by the dynamics of the corresponding replicating strategy:
	\begin{equation}\label{eq:put_dynamics}
	dP(t) = \underbrace{\frac{\partial P(t) }{\partial V^{v_0, \pi_B}(t)}}_{\text{delta hedge}} dV^{v_0, \pi_B}(t) + \underbrace{\left(\frac{P(t)}{S_0(t)} -  \frac{ V^{v_0, \pi_B}(t)\frac{\partial P(t) }{\partial V^{v_0, \pi_B}(t)}}{S_0(t)} \right)}_{\text{money left}}dS_0(t)
	\end{equation}
	
	The delta-hedge of the put option on the CM portfolio is given by:
	\begin{equation}\label{eq:delta_put}
		\begin{aligned}
			\frac{\partial P(t) }{\partial V^{v_0, \pi_B}(t)} & = & \Phi(d_+) - 1,
		\end{aligned}
	\end{equation}
	where $\Phi(\cdot)$ is the cumulative distribution function of a standard normal random variable and
	\begin{equation}\label{eq:d_plus}
		d_+ := d_+(V^{v_0, \pi_B}(t), T-t) = \frac{\ln\left(V^{v_0, \pi_B}(t) / G_T\right) + \left(r + 0.5 \left(\pi_B^{CM}\sigma_2\right)^2\right)(T - t)}{\pi_B^{CM} \sigma_2 \sqrt{T-t}}.
	\end{equation}
	
	Using \eqref{eq:put_dynamics}, \eqref{eq:CM_PF_dynamics} and \eqref{eq:delta_put}, we get  the dynamics of the value of the reinsurance (put option) in terms of its basic underlying assets:
	\begin{equation*}
		\begin{aligned}
		dP(t) & =  (\Phi(d_+) - 1) V^{v_0, \pi_B}(t)  ((r + \pi_B^{CM} (\mu_2 - r)) dt + \pi_B^{CM} \sigma_2 (\rho dW_1(t) + \sqrt{1 - \rho^2}dW_2(t))) \\
		& \quad  + \left(\frac{P(t)}{S_0(t)} -  \frac{ V^{v_0, \pi_B}(t)(\Phi(d_+) - 1)}{S_0(t)} \right) S_0(t) r dt \\
		& =  \lrr{(\Phi(d_+) - 1)V^{v_0, \pi_B}(t)  \pi_B^{CM} (\mu_2 - r)  + rP(t)}dt\\
		& \quad + (\Phi(d_+) - 1)V^{v_0, \pi_B}(t)  \pi_B^{CM} \sigma_2 \rho dW_1(t) \\
		& \quad + (\Phi(d_+) - 1)V^{v_0, \pi_B}(t)  \pi_B^{CM} \sigma_2 \sqrt{1 - \rho^2}dW_2(t).
		\end{aligned}
	\end{equation*}

	To sum up, the insurer can invest in a risk-free asset $S_0$, a risky asset $S_1$, and a reinsurance (put option) $P(t)$ with the underlying $V^{v_0, \pi_B}(T)$ and strike $G_T$. {We refer to the market consisting of $S_0, S_1, P$ as a non-linear market, since herein the portfolio value is in general a non-linear function w.r.t. the prices of the basic assets $S_0, S_1, S_2$.}
	
	Let $\bar \pi(t) = (\bar \pi_1(t), \bar \pi_2(t))',\,\tin,$ be the insurer's relative portfolio process with respect to assets $S_1(t)$, $P(t)$, with $\bar \pi_0(t) = 1 - \bar \pi_1(t) - \bar \pi_2(t),\,\tin$. Let $\bar \varphi(t)$ be the corresponding investment strategy at $\tin$, i.e., number of bonds, shares or reinsurance contracts. The portfolio value has the following dynamics:
	\begin{equation*}
	    d\bar{V}^{v_0, \bar \pi}(t) = (1 - \bar \pi_1(t) - \bar \pi_2(t))\frac{\bar{V}^{v_0, \bar \pi}(t)}{S_0(t)}dS_0(t) + \bar \pi_1(t)\frac{\bar{V}^{v_0, \bar \pi}(t)}{S_1(t)}dS_1(t) + \bar \pi_2(t) \frac{\bar{V}^{v_0, \bar \pi}(t)}{P(t)}dP(t)
	\end{equation*}
	with $\bar{V}^{v_0, \bar \pi}(0)=v_0$.
	
	Analogously, let $\pi(t) = (\pi_1(t),  \pi_2(t))',\,\tin,$ be the relative portfolio processes w.r.t. assets $S_1(t), S_2(t)$, with $\pi_0(t) = 1 - \pi_1(t) - \pi_2(t),\,\tin$, and $\varphi(t),\,\tin,$ be the corresponding investment strategy. Then the portfolio value in the financial market consisting of $S_0, S_1, S_2$ has the following dynamics:
	\begin{equation*}
	    dV^{v_0, \pi}(t) = (1 -  \pi_1(t) -  \pi_2(t))\frac{V^{v_0,  \pi}(t)}{S_0(t)}dS_0(t) +  \pi_1(t)\frac{V^{v_0,  \pi}(t)}{S_1(t)}dS_1(t) +  \pi_2(t) \frac{V^{v_0, \pi}(t)}{S_2(t)}dS_2(t)
	\end{equation*}
	with $V^{v_0, \pi}(0)=v_0$.
	
	The relative portfolio processes and the investment strategies are linked in the following way:
	\begin{equation}
	    \label{eq:link_pi_phi}
	    \varphi_i(t) = \frac{\pi_i(t)V^{v_0, \pi}(t)}{S_i(t)},\quad \phib_j(t) = \frac{\pib_j(t)\bar{V}^{v_0, \pib}(t)}{S_j(t)}, \quad \phib_2(t) = \frac{\pib_2(t)\bar{V}^{v_0, \pib}(t)}{P(t)}
	\end{equation}
	where $i \in \{0, 1, 2\}$, $j \in \{0,1\}$, $V^{v_0, \pi}(t)$ and $\bar{V}^{v_0, \pib}(t)$ are values of the corresponding portfolios at time $\tin$. 
	
	We assume that the insurer has a power utility function $U(x)=\frac{1}{b}x^b$ for $b < 1, \ b \neq 0$. The insurer has to fulfill a Value-at-Risk (VaR) constraint, which is widely used in the life insurance literature (see e.g., \citeA{Dong2020}, \citeA{Guan2016}, \citeA{Nguyen2020}) and is motivated by solvency regulations and the management rules of insurance companies. We denote by $\varepsilon \in [0,1]$ the probability of not achieving a guarantee by the insurer.
	
	We also add a no-short-selling constraint to the insurer's optimization problem. The motivation for it is twofold. First, shorting reinsurance is against the nature of the reinsurance business. The no-short-selling constraint on the reinsurance prevents the insurer from using reinsurance for speculation purposes. Second, shorting assets is quite uncommon for insurance companies due to regulations (see \citeA{Dong2020}).
	
	Define:
	\begin{equation*}
	    \begin{aligned}
	        \bar{\mathcal{U}}(v_0) &:= \left\{\left. \pib=\lrr{(\pib_1(t), \pib_2(t))'}_{\tin} \right\rvert \pib \text{ is prog. meas.},\, \int_0^T ||\bar \pi (t)\bar{V}^{v_0, \bar \pi}(t)||^2\,dt < \infty\,\mathbb{Q}\text{-a.s.} \right\};\\
	        \bar{K}_V(\varepsilon) &:= \{ \bar{V}^{v_0, \pib}(T)| \mathbb{Q}\lrr{\bar{V}^{v_0, \pib}(T) < G_T} \leq \varepsilon\}\quad\text{and}\quad
	        \bar{K}_{\pi} := [0, +\infty) \times [0, +\infty),
	    \end{aligned}
	\end{equation*}
	where \enquote{prog. meas.} stands for progressively measurable. Then the set of the insurer's admissible relative portfolio processes w.r.t. $S_1$ and $P$ is given by:
	\begin{equation*}
	    \bar{\mathcal{A}}\lrr{v_0, \bar{K}_V(\varepsilon), \bar{K}_{\pi}} := \left\{ \pib \in \bar{\mathcal{U}}(v_0)\,|\,\bar{V}^{v_0, \pib}(T) \in \bar{K}_V(\varepsilon),\,\pib(t) \in \bar{K}_{\pi} \, \forall \, \tin \, \mathbb{Q}\text{-a.s.}\right\}
	\end{equation*}
	Note: $\bar{\mathcal{A}}\lrr{v_0, \bar{K}_V(1), \mathbb{R}^2}=\bar{\mathcal{U}}(v_0)$. The optimization problem of the insurer under the no-short-selling constraint and VaR constraint is as follows: 
	\begin{equation}\label{OP_bar_P_epsilon_bar_K}\tag{$\bar P_{\varepsilon, \bar{K}_{\pi}}$}
		\begin{aligned}
			\max_{\pib}\, &\mathbb{E}\biggl[ U(\bar{V}^{v_0, \pib}(T)) \biggr] \quad \text{s.t.} \quad \pib \in  \bar{\mathcal{A}}\lrr{v_0, \bar{K}_V(\varepsilon), \bar{K}_{\pi}}
		\end{aligned}
	\end{equation}
	
	The notation $\lrr{\bar P_{\varepsilon, \bar{K}_{\pi}}}$ indicates that control variables in this optimization problem are relative portfolio processes w.r.t. assets $S_0, S_1, P$, there is a VaR-type terminal wealth constraint with probability $\varepsilon$ and there is an investment strategy constraint $\pib \in \bar{K}_{\pi}$. Special cases of this notation are $\lrr{\bar P_{1, \bar{K}_{\pi}}}$, i.e., the optimization problem does not have a terminal wealth constraint, and $\lrr{\bar P_{0, \bar{K}_{\pi}}}$, i.e., the optimization problem has a hard lower bound, also known as the portfolio insurance constraint.

\section{Solution to the optimization problem}\label{sec_solution_to_OP}$\,$

	In this section, we first provide an overview of our approach to solving \eqref{OP_bar_P_epsilon_bar_K}. In the solution procedure, we make several transformations of the problem eventually linking the solution to the original problem and the solution to a simpler problem that has neither terminal-wealth constraints nor allocation constraints. After the general overview, we describe the transformations in detail, each addressing a specific challenge of the optimization problem: reinsurance, no-short-selling constraint, VaR constraint. At the end of this section, we answer one of the main questions of our paper -- in which situations is it optimal for the insurer to buy reinsurance?
    We solve Problem \eqref{OP_bar_P_epsilon_bar_K} as follows:
    \begin{enumerate}
        \item Transform the original problem with traded reinsurance to an allocation-constrained VaR-constrained problem in the financial market with basic assets $S_0, S_1, S_2$ (\textit{reinsurance}).
        \item Transform the problem from Step 1 to an allocation-unconstrained VaR-constrained problem in an auxiliary financial market of basic assets (\textit{no-short-selling constraint}).
        \item Solve the allocation-unconstrained VaR-constrained problem from Step 2 and use it to recover the solution to the original problem (\textit{VaR-constraint}).
    \end{enumerate}
    
\begin{figure}[!htp]
         \centering
         \includegraphics[width=\linewidth]{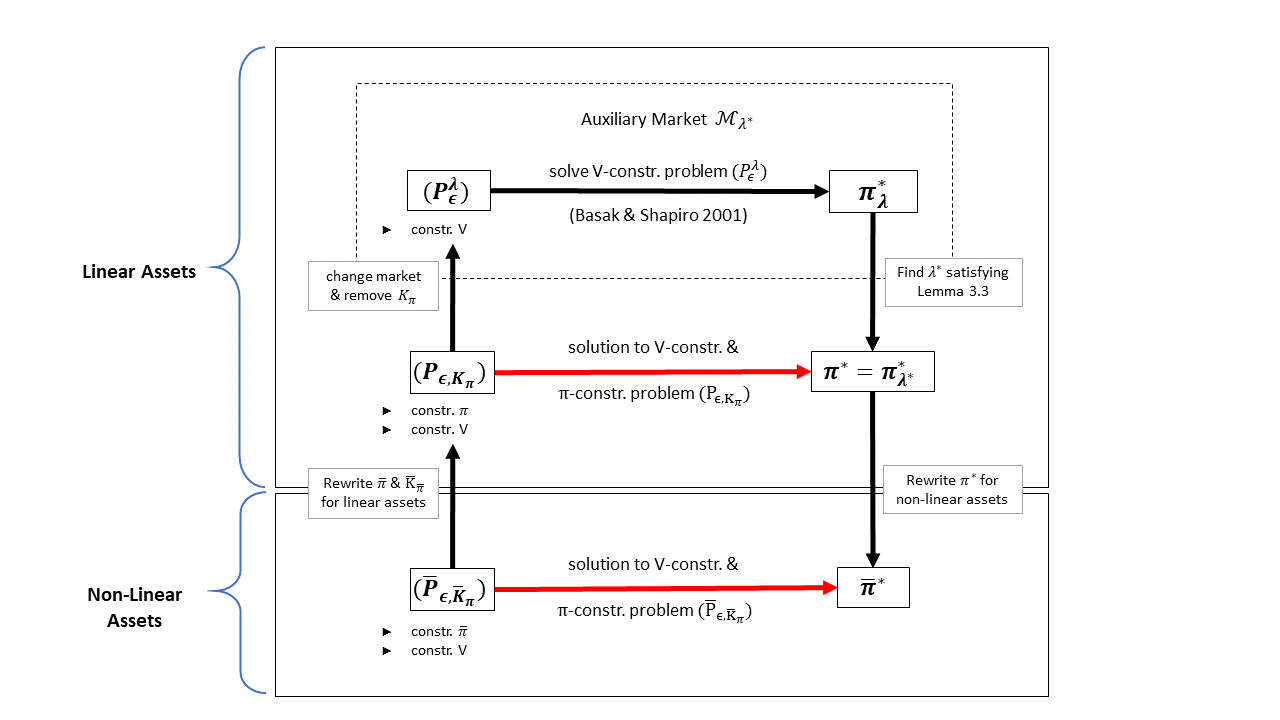}
         \caption{Visualization of the solution procedure for Problem (\ref{OP_bar_P_epsilon_bar_K})}\label{fig:solution_approach_visualization}
    \end{figure}

    Figure \ref{fig:solution_approach_visualization} visualizes our approach\footnote{New elements of notation are explained in the corresponding subsections of Section \ref{sec_solution_to_OP}}.

	\subsection{Relation between portfolios in non-linear and linear markets}$\,$
	
	The next proposition links the dynamics of a portfolio that consists of $S_0, S_1, P$ to the dynamics of a portfolio of the basic assets $S_0, S_1, S_2$.
	
	\begin{proposition}\label{prop:pi_relations}
	    If $\pi$ and $\pib$ satisfy the following relation:
	    \begin{equation}\label{eq:pi_relations}
    		\left\{
    		\begin{aligned}
    			\pib_1(t) &=  \pi_1(t)\\
    			\pib_2(t) &=  \pi_2(t)\frac{P(t)}{\pi_B^{CM} V^{v_0, \pi_B}(t)(\Phi(d_+) - 1)}
    		\end{aligned}
    		\right.
	\end{equation}
	    then:
	    \begin{equation}\label{eq:pi_pib_linked_V}
	        \bar{V}^{v_0, \bar \pi}(t) = V^{v_0, \pi}(t)\qquad \forall \tin\, \,\Q-\text{a.s.}
	    \end{equation}
	\end{proposition}
	\begin{proof}
	See Appendix \ref{app:proofs_main}.
	\end{proof}
	
	For convenience, we denote:
	\begin{equation}\label{eq:A_t}
		\begin{aligned}
			A(t) &:=
			\left(
			\begin{matrix}
			1 & 0 \\ 
			0 & \frac{P(t)}{\pi_B^{CM} V^{v_0, \pi_B}(t)(\Phi(d_+) - 1)}
			\end{matrix}
			\right)
		\end{aligned},\,\tin.
	\end{equation}
	Observe that:
	 \begin{equation}\label{relation_K_bar_K} 
	    \begin{aligned}
	        \forall& \tin,  \ \forall
    	    \lrr{
    	    \begin{matrix}
    		    \pib_1(t) \\ 
    		    \pib_2(t)
    		\end{matrix}
    		}
    		\in \bar{K}: \\
    		&\quad A^{-1}
    		(t)
    		\lrr{
    		\begin{matrix}
    		    \pib_1(t) \\ 
    		    \pib_2(t)
    		\end{matrix}
    		} =
    		\lrr{
    		\begin{matrix}
    		    \pib_1(t) \\ 
    		    \pib_2(t)\pi_B^{CM} V^{v_0, \pi_B}(t)(\Phi(d_+) - 1)/P(t)
    		\end{matrix}
    		}  \in [0,\infty) \times (-\infty, 0]=:K_{\pi}.
	    \end{aligned}
	 \end{equation}
	 
	Therefore, we define:
	\begin{equation*}
	    \begin{aligned}
	        &\mathcal{U}(v_0) := \left\{\left. \pi=\lrr{(\pi_1(t), \pi_2(t))'}_{\tin} \,\right\rvert\, \pib \text{ is prog. meas.},\, \int_0^T || \pi (t)V^{v_0,  \pi}(t)||^2\,dt < \infty\,\mathbb{Q}\text{-a.s.} \right\};\\
	        &K_V(\varepsilon) := \{ V^{v_0, \pi}(T)\,|\,\mathbb{Q}\lrr{V^{v_0, \pi}(T) < G_T} \leq \varepsilon\};\\
	        &\mathcal{A}\lrr{v_0, K_V(\varepsilon), K_{\pi}} := \left\{ \pi \in \mathcal{U}(v_0)\,|\,V^{v_0, \pi}(T) \in K_V(\varepsilon),\,\pi(t) \in K_{\pi} \, \forall \, \tin \, \mathbb{Q}\text{-a.s.}\right\}
	    \end{aligned}
	\end{equation*}
	 and consider the following optimization problem in the market with basic assets $S_0, S_1, S_2$:
	 \begin{equation}\label{OP_P_epsilon_K}\tag{$P_{\varepsilon, K_{\pi}}$}
		\begin{aligned}
			\max_{\pi}\, &\mathbb{E}\biggl[ U(V^{v_0, \pi}(T)) \biggr] \quad
			\text{s.t.} \quad   \pi \in  \mathcal{A}\lrr{v_0, K_V(\varepsilon), K_{\pi}} 
		\end{aligned}
	\end{equation}
	The next proposition links the solution to the original Problem \eqref{OP_bar_P_epsilon_bar_K} and the solution to the transformed Problem \eqref{OP_P_epsilon_K}.
	\begin{proposition}\label{prop:sol_VaR_pi_link_original_transformed}
	    Let $\pi^*$ be the optimal solution to Problem \eqref{OP_P_epsilon_K}. Then the portfolio process:
	    \begin{equation*}
	        \pib^*(t) := A(t)\pi^*(t)
	    \end{equation*}
	    is the solution to Problem \eqref{OP_bar_P_epsilon_bar_K}, where $A(t)$ is the transformation matrix from \eqref{eq:A_t}.
	\end{proposition}
	\begin{proof}
	    See Appendix \ref{app:proofs_main}.
	\end{proof}
	{Due to Proposition \ref{prop:sol_VaR_pi_link_original_transformed}, we focus on solving \eqref{OP_P_epsilon_K} in the sequel of this chapter.}
	
\subsection{Solving \eqref{OP_P_epsilon_K} in the market with basic assets}$\,$
 
    We deal with the additional allocation constraints  and VaR constraints on terminal wealth by borrowing and combining two popular approaches from the literature: The auxiliary market approach from \citeA{Cvitanic1992} as well as the idea of using option-like terminal payoffs on the unconstrained terminal wealth to eliminate terminal wealth constraints from \citeA{Kraft2013}. This is how we are going to proceed:
    
    \begin{enumerate}
        \item Extend the auxiliary market framework from \citeA{Cvitanic1992} to our setting and derive an optimality condition, which links the solutions of a family of wealth-constrained and allocation-unconstrained portfolio optimization problems \eqref{eq:OP_I_VaR_pi_original_post_transformation_auxiliary_market} to the solution of the wealth- and allocation-constrained portfolio optimization problem \eqref{OP_P_epsilon_K}.
        \item Use the results from \citeA{Basak2001} to derive the solutions to the wealth-constrained and allocation-unconstrained portfolio optimization problems \eqref{eq:OP_I_VaR_pi_original_post_transformation_auxiliary_market}.
        \item Find the optimal auxiliary market  $\mathcal{M}_{\lambda^{\ast}}$, compute the corresponding optimal portfolio and verify its optimality for the primal problem (\ref{OP_P_epsilon_K}) via the previously derived optimality condition.
    \end{enumerate}

    \subsubsection{{Auxiliary market with VaR- and allocation constraints}}$\,$

    In the literature, the classic approach to portfolio optimization under the presence of allocation constraints is the auxiliary market approach from \citeA{Cvitanic1992}. Despite the presence of the additional VaR-constraints, the concept of auxiliary market proves to be vital in solving  (\ref{OP_P_epsilon_K}). For setting up the auxiliary markets, we need to introduce the support function
    \begin{equation*}
        \delta(x) = - \underset{y \in K_{\pi}}{\inf}\big( x'y \big)
    \end{equation*}
    as well as the barrier cone
    \begin{equation*}
        X_{K_{\pi}}:= \{ x \in \mathbb{R}^2 \ | \ \delta(x)<\infty \}
    \end{equation*}
    In our setting, with $K_{\pi} = [0, \infty)\times (-\infty,0]$, the infimum in the definition of $\delta$ is attained by $y=0$, if $x \in K_{\pi}$ and is $-\infty$ otherwise. Hence, $X_{K_{\pi}}=K_{\pi}$ and $\delta(x)=0$ $\forall x \in K_{\pi}$. \\
    
    Further, we introduce the class of $\mathbb{R}^2$-valued dual processes $\mathcal{D}$, defined through
    \begin{equation*}
        \mathcal{D}:= \Big \{ \lambda=\big(\lambda(t)\big)_{\tin} \ \text{prog. measurable} \ \Big | \ E \Big[\int_0^T \Vert \lambda(t) \Vert^2dt\Big]< \infty, \ E\Big[\int_0^T \delta(\lambda(t))dt\Big]<\infty \Big \}
    \end{equation*}
    
    The second integrability condition implies that $\lambda(t) \in X_{K_{\pi}} = K_{\pi}$ $\Q \otimes \mathcal{L}[0,T]$-a.s., $\mathcal{L}[0,T]$ denoting the Lebesgue measure on $\tin$. In particular, every constant  process with value in $K_{\pi}$, from now on referred to as dual vector, is contained in $\mathcal{D}$. \\ 
    
    For each $\lambda \in \mathcal{D}$, we define the auxiliary market $\mathcal{M}_{\lambda}$,  where the assets $S^{\lambda}_0, S^{\lambda}_1$ and $S^{\lambda}_2$ follow the dynamics

	\begin{equation*}
	\begin{aligned}
		dS^{\lambda}_0(t) := &S^{\lambda}_0(t) \big(r+ \delta(\lambda(t))\big) dt =S^{\lambda}_0(t) r dt\\
		dS^{\lambda}_1(t) := &S^{\lambda}_1(t) \big(\big(\mu_1 + \lambda_1(t) + \delta(\lambda(t))\big)  dt + \sigma_1dW_1(t)\big)= S^{\lambda}_1(t) \big((\mu_1 + \lambda_1(t))  dt + \sigma_1dW_1(t)\big) \\
		dS^{\lambda}_2(t) := &S^{\lambda}_2(t) \big (\big(\mu_2+ \lambda_2(t)+ \delta(\lambda(t))\big) dt + \sigma_2(\rho dW_1(t) + \sqrt{1 - \rho^2})dW_2(t)\big) \\ 
		= &S^{\lambda}_2(t) \big ((\mu_2+ \lambda_2(t)) dt + \sigma_2(\rho dW_1(t) + \sqrt{1 - \rho^2})dW_2(t)\big)
	\end{aligned}
	\end{equation*}
	
	In $\mathcal{M}_{\lambda}$, the market price of risk and the pricing kernel are stochastic processes given by:
	\begin{equation*}
	    \gamlam(t) = \sigma^{-1}(\mu + \lambda(t)  - r \bar{1})\qquad \Ztillamt{t} = e^{-rt - 0.5\int_{0}^{t}||\gamlam(s)|^2\,ds - \int_{0}^{t} \gamlam'(s)\, dW(s)},\, \tin.
	\end{equation*}
	As we will see later, it is sufficient for our problem to consider dual vectors  { $\lambda\in K_{\pi}$} \footnote{{Note that for any dual vector $\lambda \in K_{\pi}$ the market $\mathcal{M}_{\lambda}$ is a two-dimensional Black-Scholes market with deterministic market coefficients. In particular, $\mathcal{M}_0$ is the standard market with assets $(S_0,S_1,S_2)$.}}. For such cases the market price of risk and the pricing kernel are simplified to
	\begin{equation}\label{eq:gamma_Z_for_auxiliary_market}
	    \gamlam = \sigma^{-1}(\mu + \lambda  - r \bar{1})\qquad \Ztillamt{t} = e^{-\lrr{r + 0.5||\gamlam||^2}t - \gamlam'\, W(t)},\, \tin
	\end{equation}
	and $\mathcal{M}_{\lambda}$ admits a unique risk-neutral probability measure $\tilde{\Q}_{\lambda}$\footnote{One obtains the risk-neutral measure for the original market $\mathcal{M}$ as $\tilde{\Q}_0 = \tilde{\Q}$.} with density
	$$\frac{d\tilde{\Q}_{\lambda}}{d \Q} \Bigg |_{\mathcal{F}_T} := Z_{\lambda}(T):= e^{- 0.5||\gamlam||^2T - \gamlam'\, W(T)}.$$
  The asset $S^{\lambda}_0$ represents the bank account, whereas the assets $S^{\lambda}_1$ and $S^{\lambda}_2$ represent the fund and market index from our original setting but with partially changed drift coefficients. Clearly, changing the drift coefficients of the basic assets in $\mathcal{M}_{\lambda}$ has an effect on the wealth process of an investor trading in $\mathcal{M}_{\lambda}$. Indeed, it is straightforward to show that the wealth process $V_{\lambda}^{v_0, \pi}(T)$, corresponding to trading in $\mathcal{M}_{\lambda}$ according to $\pi$ with initial wealth $v_0$, satisfies the SDE
	 \begin{equation}\label{eq:dynamics_V_lambda}
 	\begin{aligned}
 dV_{\lambda}^{v_0, \pi}(t) &= (1 - \pi_1(t) - \pi_2(t))\frac{V_{\lambda}^{v_0, \pi}(t)}{S^{\lambda}_0(t)}dS^{\lambda}_0(t) + \pi_1(t)\frac{V_{\lambda}^{v_0, \pi}(t)}{S^{\lambda}_1(t)}dS^{\lambda}_1(t) + \pi_2(t) \frac{V_{\lambda}^{v_0, \pi}(t)}{S^{\lambda}_2(t)}dS^{\lambda}_2(t) \\
 &= (1 - \pi_1(t) - \pi_2(t))\frac{V_{\lambda}^{v_0, \pi}(t)}{S_0(t)}dS_0(t) + \pi_1(t)\frac{V_{\lambda}^{v_0, \pi}(t)}{S_1(t)}dS_1(t) + \pi_2(t) \frac{V_{\lambda}^{v_0, \pi}(t)}{S_2(t)}dS_2(t)\\
 &\quad + V_{\lambda}^{v_0, \pi}(t)\underbrace{\big( \lambda(t)'\pi(t) \big)}_{\geq 0, \ \text{if }\pi(t) \in K_{\pi} }dt,
    \end{aligned}
 \end{equation}

    which is the same SDE as in the original market, but with an additional drift term. Due to $\lambda(t) \in [0,\infty)\times(-\infty, 0]=K_{\pi}$ $\Q \otimes \mathcal{L}[0,T]$-a.e., the additional drift to the wealth process $V_{\lambda}^{v_0, \pi}$ in $\mathcal{M}_{\lambda}$ is guaranteed to be non-negative if $\pi(t) \in K_{\pi}$ $\Q$-a.s. $\forall t$. Hence, the insurer always performs at least as good in $\mathcal{M}_{\lambda}$ as it would have in the original market, i.e., $V_{\lambda}^{v_0, \pi}(T) \geq V^{v_0, \pi}(T)$, provided that it abides by the allocation constraints $K_{\pi}$. The two wealth processes $V_{\lambda}^{v_0, \pi}$ and $V^{v_0, \pi}$ coincide if and only if $\lambda(t)'\pi(t) = 0$ $\Q \otimes \mathcal{L}[0,T]$-a.e.. \\
    
    Since we have changed the dynamics of the wealth process $V_{\lambda}^{v_0, \pi}$ of an investor trading in $\mathcal{M}_{\lambda}$, we need to adjust the class of admissible portfolio processes as well. For this purpose we define for every $\lambda \in \mathcal{D}$
    \begin{align*}
	        \mathcal{U}_{\lambda}(v_0) &:= \left\{\left. \pi=\lrr{(\pi_1(t), \pi_2(t))'}_{\tin} \right\rvert \pi \text{ is prog. meas.},\, \int_0^T || \pi (t)V_{\lambda}^{v_0, \pi}(t)||^2\,dt < \infty\,\mathbb{Q}\text{-a.s.} \right\};\\
	        K_{V_{\lambda}}(\varepsilon) &:= \{ V_{\lambda}^{v_0, \pi}(T)| \mathbb{Q}\lrr{V_{\lambda}^{v_0, \pi}(T) < G_T} \leq \varepsilon\}; \\
	         \mathcal{A}_{\lambda}\lrr{v_0, K_{V_{\lambda}}(\varepsilon)} &:= \left\{ \pi \in \mathcal{U}_{\lambda}(v_0)\,|\,V_{\lambda}^{v_0, \pi}(T) \in K_{V_{\lambda}}(\varepsilon)  \right\}.
    \end{align*}
    
 Consider now the following portfolio optimization problem, which is an allocation-unconstrained formulation in $\mathcal{M}_{\lambda}$: 
 
\begin{equation}\label{eq:OP_I_VaR_pi_original_post_transformation_auxiliary_market}\tag{$P^{\lambda}_{\varepsilon}$}
		\begin{aligned}
			\max_{\pi}\, &\mathbb{E}\biggl[ U(V_{\lambda}^{v_0, \pi}(T)) \biggr] \quad 
			\text{s.t.} \,   \pi \in  \mathcal{A}_{\lambda}\lrr{v_0, K_{V_{\lambda}}(\varepsilon)}
		\end{aligned}
\end{equation}
	
	For any fixed dual control process $\lambda \in \mathcal{D}$ we see that every portfolio process admissible for the original problem (\ref{OP_P_epsilon_K}) is also admissible for the optimization problem in the auxiliary market (\ref{eq:OP_I_VaR_pi_original_post_transformation_auxiliary_market}) and yields at least the same terminal wealth (or expected utility) as in the original problem. This leads to the following  condition, which can be used to verify that the optimal portfolio processes for $(P^{\la}_{\varepsilon})$ and (\ref{OP_P_epsilon_K}) coincide for a particular  dual process $\la \in \mathcal{D}$.
	
    \begin{lemma}\label{lem: condition b}
       Let $\lambda \in \mathcal{D}$, $\pi_{\lambda}$ be the optimal portfolio process for (\ref{eq:OP_I_VaR_pi_original_post_transformation_auxiliary_market}) in $\mathcal{M}_{\lambda}$ and $V_{\lambda}^{v_0, \pi_{\lambda}}$ be the corresponding wealth process. If
    	\begin{equation} \label{eq: Condition b}
    	\pi_{\lambda}(t)'\lambda(t) = 0 \ \text{and } \ \pi_{\lambda}(t) \in K_{\pi} \quad  \Q-a.s., \ \forall \tin
    	\end{equation}
    	then $\pi_{\lambda}$ is admissible and optimal for the original problem (\ref{OP_P_epsilon_K}) and $V_{\lambda}^{v_0, \pi_{\lambda}}(t)$ = $V^{v_0, \pi_{\lambda}}(t)$ a.s. $\forall \tin$.
    \end{lemma}
    \begin{proof}[Proof]
    The proof of this lemma is straightforward and uses the structure of Problems \eqref{eq:OP_I_VaR_pi_original_post_transformation_auxiliary_market} and \eqref{OP_P_epsilon_K}, dynamics of the wealth process \eqref{eq:dynamics_V_lambda} and Condition \eqref{eq: Condition b}. 
    \end{proof}
      
  In the following, a dual process $\lambda^{\ast}$ and the corresponding auxiliary market $\mathcal{M}_{\la}$, is referred to as optimal, if they satisfy \eqref{eq: Condition b}. Unfortunately, Lemma \ref{lem: condition b} only provides a convenient condition to verify optimality for a candidate dual process $\lambda^{\ast}\in \mathcal{D}$, but not a constructive way of finding such a $\lambda^{\ast}$. \\
 
  For a setting without additional terminal wealth constraints, i.e., for $\varepsilon = 1$,  \citeA{Cvitanic1992} were able to prove several equivalent optimality conditions that offer a way of computing the optimal $\lambda^{\ast}$ for the case of an investor following a power utility function. They also derived an explicit form for $\lambda^{\ast}$ using stochastic control methods. However, to solve (\ref{OP_P_epsilon_K}), we do not need to prove similar equivalencies, but we will have the opportunity to use a selection of their results from the wealth-unconstrained setting $(P_{1, K_{\pi}})$, when solving (\ref{OP_P_epsilon_K}). As a matter of fact, we show that the optimal $\la$ is the same for both settings. Interestingly, due to the market coefficients $\mu, \ r,$ and $\sigma$ being constant, it is sufficient to only consider constant dual vectors $\lambda \in K_{\pi}$ from the start, as the optimal $\lambda^{\ast}$ is constant.
  \\
 
 The required results from \citeA{Cvitanic1992} are summarized in the following corollary. 

\begin{corollary}\label{cor: summary karatzas} Consider the optimization problems $(P_{1, K_{\pi}})$ and  $(P^{\lambda}_{1})$. Furthermore, set
	
	\begin{equation}\label{eq: optimal lambda}
	    \lambda^{\ast} := \underset{x \in K_{\pi}}{\argmin}\Vert \gamma +\sigma^{-1}x\Vert^2.
	\end{equation}

Then, the optimal portfolio process $\hat{\pi}_{\lambda}$ for $(P^{\lambda}_{1})$, for any dual vector $\lambda \in K_{\pi}$, is given as 

\begin{equation}\label{def: pi hat lambda}
	\hat{\pi}_{\lambda}(t) := \hat{\pi}_{\lambda} :=  \frac{1}{1-b}C^{-1}(\mu+\lambda-r\cdot \mathbbm{1}) \quad \text{with } C := \sigma \cdot \sigma'.
\end{equation}

Furthermore, for the particular dual vector $\la \in K_{\pi}$, $(P_{1, K_{\pi}})$ and $(P^{\la}_{1})$ have the same solution (i.e., optimal portfolio process) $\hat{\pi}_{\la}$, which satisfies \eqref{eq: Condition b}.
\end{corollary}

\begin{proof}
	See Appendix \ref{app:proofs_main}.
\end{proof}

\subsubsection{VaR-constrained and allocation-unconstrained portfolio optimization}  $\,$

    {As noted in \citeA{Kraft2013},  the introduction of terminal wealth constraints on a portfolio optimization problem with initial wealth $v_0>0$ frequently results in an optimal terminal portfolio value that is a derivative on the unconstrained optimal portfolio value with a possibly lower initial capital $v_f$, i.e., $0 < v_f \leq  v_0$. { In other words, we can express the optimal terminal portfolio value in the constrained problem 
    as a function $f(\cdot)$ of the terminal optimal unconstrained portfolio value $V^{v_f, \pi^*_{uc}}(T)$, where $\pi^*_{uc}$ is the corresponding optimal unconstrained relative portfolio process}. Notable examples include
    \begin{itemize}
        \item Lower bound $G_T$ on terminal wealth (\citeA{Tepla2001}, \citeA{Korn2005}) 
         $$f(V^{v_f, \pi^*_{uc}}(T)) = G_T + (V^{v_f, \pi^*_{uc}}(T)-G_T)^{+}$$
        \item VaR constraint with boundary $G_T$ and level of confidence $\varepsilon$
        (\citeA{Basak2001})
        $$f(V^{v_f, \pi^*_{uc}}(T)) = V^{v_f, \pi^*_{uc}}(T) + (G_T-V^{v_f, \pi^*_{uc}}(T))\mathbbm{1}_{[k^{\varepsilon}, G_T]}(V^{v_f, \pi^*_{uc}}(T))$$
        for appropriate parameters $0 \leq k^{\varepsilon} \leq G_T$, determined by the budget constraint and the confidence level.
    \end{itemize}
    }
 
 As long as the derivative payoff $f$  satisfies sufficient regularity conditions, we can determine the optimal portfolio process via delta-hedging (see Lemma \ref{lem: delta-hedge} in Appendix \ref{app:proofs_auxiliary} for details).
Below we give a closed-form expression for the optimal portfolio process $\pi^{\ast}_{\lambda}$ corresponding to (\ref{eq:OP_I_VaR_pi_original_post_transformation_auxiliary_market}) for a dual vector $\lambda \in K_{\pi}$, as calculated by \citeA{Basak2001}, provided that the solution exists.

\begin{corollary}[VaR-constraints in $\mathcal{M}_{\lambda}$]\label{cor: pp VaR constraints, auxiliary market}  Consider the Black-Scholes market $\mathcal{M}_{\lambda}$ for a dual vector $\lambda \in K_{\pi}$ and the portfolio optimization problem under VaR constraints (\ref{eq:OP_I_VaR_pi_original_post_transformation_auxiliary_market}). Let $\hat{\pi}_{\lambda}$ be the optimal portfolio process for $(P^{\la}_{1})$ as defined in \eqref{def: pi hat lambda}. Let the parameters $0 \leq k^{\varepsilon}_{\lambda} < G_T$, $0\leq v_{f_{\lambda}}\leq v_0$ be determined so that
\begin{equation*}
    D^{f_{\lambda}}_{\lambda}:= f_{\lambda}(V^{v_{f_{\lambda}}, \hat{\pi}_{\lambda}}_{\lambda}(T)),
\end{equation*}
with $f_{\lambda}$ given as 
\begin{equation*}
     f_{\lambda}(V):= V + \big(G_T - V\big)\mathbbm{1}_{[k^{\varepsilon}_{\lambda},G_T]}(V),
\end{equation*}
satisfies the system of equations in $(k^{\varepsilon}_{\lambda}, v_{f_{\lambda}})$
\begin{equation*}
    \begin{aligned}
        e^{-rT}\mathbbm{E}_{\tilde{\Q}_{\lambda}}[f_{\lambda}(V^{v_{f_{\lambda}}, \hat{\pi}_{\lambda}}_{\lambda}(T))] \ &= \ v_0 \\
        \Q (f_{\lambda}(V_{\lambda}^{v_{f_{\lambda}}, \hat{\pi}_{\lambda}}(T)) < G_T)\ &= \ \varepsilon.
    \end{aligned}
\end{equation*}
Then, $D^{f_{\lambda}}_{\lambda}$ is the optimal terminal wealth for (\ref{eq:OP_I_VaR_pi_original_post_transformation_auxiliary_market}).

The value of the derivative $D^{f_{\lambda}}_{\lambda}$ at time $\tin$, given $V^{v_{f_{\lambda}},\hat{\pi}_{\lambda}}_{\lambda}(t)= V$, can be expressed as a function

\begin{equation}\label{eq: price of claim Df at time t}
\begin{aligned}
   D^{f_{\lambda}}_{\lambda}(t,V) =  V \
                -& \Bigg[ V  \Phi(-d^{\lambda}_1(G_T,V,t)) - G_Te^{-r(T-t)}\Phi(-d^{\lambda}_2(G_T,V,t)) \Bigg] \\
                +& \Bigg[ \ V  \Phi(-d^{\lambda}_1(k^{\varepsilon}_{\lambda},V,t)) - G_Te^{-r(T-t)}\Phi(-d^{\lambda}_2(k^{\varepsilon}_{\lambda},V,t)) \Bigg],
\end{aligned}
\end{equation}
where
\begin{equation}\label{eq:Gamma_d_1_d_2}
    \begin{aligned}
        \Gamma_{\lambda}(t) \ &= \ \frac{b}{1-b} \Bigg(r+ \frac{\Vert \gamma_{\lambda} \Vert^2}{2} \Bigg)(T-t) + \Big( \frac{b}{1-b}\Big)^2 \frac{\Vert \gamma_{\lambda} \Vert^2}{2}(T-t), \\
        d^{\lambda}_2(x,V,t) \ &= \ \frac{(b-1)\ln \big(\frac{x}{V}\big) + (b - 1)\Gamma_{\lambda}(t)+ \big( r- \frac{\Vert \gamma_{\lambda} \Vert^2}{2}\big)(T-t)}{\Vert \gamma_{\lambda} \Vert \sqrt{T-t}}, \\
        d^{\lambda}_1(x,V,t) \ &= \ d^{\lambda}_2(x,V,t) + \frac{1}{1-b}\Vert \gamma_{\lambda} \Vert\sqrt{T-t}.
    \end{aligned}
\end{equation}

Lastly, $D^{f_{\lambda}}_{\lambda}$ is attained by the portfolio process

\begin{equation*}
    \pi^{\ast}_{\lambda}(t):= \pi^{\ast}_{\lambda}(t,V^{v_{f_{\lambda}}, \hat{\pi}_{\lambda}}_{\lambda}(t)) = \alpha^{f_{\lambda}}_{\lambda}(t,V^{v_{f_{\lambda}}, \hat{\pi}_{\lambda}}(t)) \cdot \hat{\pi}_{\lambda},
\end{equation*}

with

\begin{equation}\label{eq:alphaVaR}
\begin{aligned}
        \alpha^{f_{\lambda}}_{\lambda}(t,V) \ &= \ 1-\frac{G_Te^{-r(T-t)}\big(\Phi(-d^{\lambda}_2(G_T,V,t))-\Phi(-d^{\lambda}_2(k^{\varepsilon}_{\lambda},V,t))\big)}{D^{f_{\lambda}}_{\lambda}(t,V)} \\ 
        &\qquad + \frac{(1-b)(G_T-k^{\varepsilon}_{\lambda})e^{-r(T-t)}\phi(d^{\lambda}_2(k^{\varepsilon}_{\lambda},V,t))}{D^{f_{\lambda}}_{\lambda}(t,V)  \ \Vert \gamma_{\lambda} \Vert \sqrt{T-t}} \geq 0 
\end{aligned}
\end{equation}

\end{corollary}
\begin{proof}
See Appendix \ref{app:proofs_main}.
\end{proof}

Note that  $\alpha^{f_{\lambda}}_{\lambda}(t, V)>0$ for $t \in [0, T)$ due to 

\begin{equation*}
\begin{aligned}
   D^{f_{\lambda}}_{\lambda}(t,V) \overset{(\ref{eq: price of claim Df at time t})}{=}& V \
      \Bigg[1-\underbrace{\Phi(-d^{\lambda}_1(G_T,V,t))}_{\ < 1}+ \underbrace{\Phi(-d^{\lambda}_1(k^{\varepsilon}_{\lambda},V,t))}_{> 0}\Bigg]\\
    & \quad + G_Te^{-r(T-t)} \Bigg[ \Phi(-d^{\lambda}_2(G_T,V,t)) -\Phi(-d^{\lambda}_2(k^{\varepsilon}_{\lambda},V,t)) \Bigg] \\
    > \hspace{2mm}& \ G_Te^{-r(T-t)}\big(\Phi(-d^{\lambda}_2(G_T,V,t))-\Phi(-d^{\lambda}_2(k^{\varepsilon}_{\lambda},V,t)) \geq 0,
\end{aligned}
\end{equation*}
as $k^{\varepsilon}_{\lambda} \in [0, G_T]$.

 \subsubsection{VaR-constrained and allocation-constrained portfolio optimization }$\,$

This section concludes the previous derivations by combining the results from \citeA{Cvitanic1992} and \citeA{Basak2001} to solve (\ref{OP_P_epsilon_K}). In short, we prove that the optimal payoff for (\ref{OP_P_epsilon_K}) is a derivative of the optimal payoff for $(P_{1, K_{\pi}})$ with some initial wealth $v_{f_{\la}} \leq v_0$, by verifying \eqref{eq: Condition b} for $\lambda = \la$, as in Lemma \ref{cor: summary karatzas}. \\

\begin{proposition} \label{prop: generalization no-shortsell}
	Set
	\begin{equation*}
	\lambda^{\ast} := \underset{x \in K_{\pi}}{\argmin} \Vert \gamma + \sigma^{-1}x \Vert. 
	\end{equation*}
	Denote by $\hat{\pi}_{\la}$ the optimal portfolio process for $(P_1^{\la})$, which is defined in \eqref{def: pi hat lambda}. 
	 Let the parameters $0 \leq k^{\varepsilon}_{\la} < G_T$, $0 < v_{f_{\la}}\leq v_0$ be determined so that the derivative $D_{f_{\la}}$ on the optimal terminal wealth for $(P_{1, K_{\pi}})$
	\begin{equation*}
	D^{f_{\la}} := {f_{\la}}(V^{v_{f_{\la}},\hat{\pi}_{\la}}(T))
	\end{equation*}
with ${f_{\la}}(V) := V + (G_T-V)\mathbbm{1}_{[k^{\varepsilon}_{\la}, G_T]}(V)$, satisfies the system of equations in $(k^{\varepsilon}_{\la}, v_{f_{\la}})$
\begin{equation}\label{eq: VaR constr. and budget}
        \begin{aligned}
          e^{-rT}{\mathbbm{E}_{\tilde{\Q}}[f(V^{v_{f_{\la}}, \hat{\pi}_{\la}}(T))]} = {\mathbbm{E}_{\Q}[f(V^{v_{f_{\la}}, \hat{\pi}_{\la}}(T))\tilde{Z}(T)]} \ &= \ v_0\\
        \Q(f(V^{v_{f_{\la}}, \hat{\pi}_{\la}}(T)) < G_T)\ &= \ \varepsilon.
        \end{aligned}
\end{equation}
Then, $D^f$ is the optimal terminal wealth for (\ref{OP_P_epsilon_K}). The corresponding  optimal portfolio process $\pi^{\ast}$ for  (\ref{OP_P_epsilon_K}) is given as 
	
	\begin{equation}\label{eq:piStar_pHat_link}
	\pi^{\ast}(t) := \pi^{\ast}(t, V^{v_{f_{\la}},\hat{\pi}_{\la}}(t)) = \alpha^{f_{\la}}_{\la}(t, V^{v_{f_{\la}},\hat{\pi}_{\la}}(t)) \cdot  \hat{\pi}_{\la},
	\end{equation}
	
	with $\alpha^{f_{\la}}_{\la} > 0$, as in Corollary \ref{cor: pp VaR constraints, auxiliary market}.
	\end{proposition}
	\begin{proof}
	See Appendix \ref{app:proofs_main}.
	\end{proof}

    The proof of Proposition \ref{prop: generalization no-shortsell} uses three facts:
    \begin{itemize}
        \item $\hat{\pi}_{\la} \in K_{\pi}$, according to \citeA{Cvitanic1992}
        \item $\pi^{\ast}_{\la} = \alpha^{f_{\la}}_{\la}\cdot \hat{\pi}_{\la}$ according to \citeA{Basak2001}
        \item $\alpha^{f_{\la}}_{\la}\geq 0$
    \end{itemize}
    As it can be seen in Proposition 5 in \citeA{Basak2001}, all of these three facts are also true for a portfolio optimization problem with an expected shortfall constraint\footnote{{Considering an expected shortfall constraint with a threshold $G_T$ and a tolerance level $\varepsilon$, we would have $f(V^{v_f, \pi^*_{uc}}(T)) = \frac{G_T}{k^{\varepsilon}}V^{v_f, \pi^*_{uc}}(T)\mathbbm{1}_{[0,k^{\varepsilon}]}(V) + (G_T-V^{v_f, \pi^*_{uc}}(T))\mathbbm{1}_{(k^{\varepsilon}, G_T]}(V^{v_f, \pi^*_{uc}}(T))+ V^{v_f, \pi^*_{uc}}(T)\mathbbm{1}_{(G_T, \infty)}(V^{v_f, \pi^*_{uc}}(T))$ and  $k^{\varepsilon}$ being determined through budget constraint and confidence level.}}. Hence, our methodology can also be used to calculate the optimal portfolio process for such an investor with a no-short-selling constraint.
   
    Having obtained $\pi^*$ that solves \eqref{OP_P_epsilon_K}, we apply Proposition \ref{prop:sol_VaR_pi_link_original_transformed} to get the solution to the original problem \eqref{OP_bar_P_epsilon_bar_K}: $\pib^*:=\lrr{ A(t)\pi^*(t)}_{\tin}$, where $A(t)$ defined in \eqref{eq:A_t}. 
    
    \subsection{Reinsurance optimality}$\,$
 
    We now want to answer the question when it is optimal for an insurer to buy reinsurance in {the product under consideration}.
    We denote by
    $$SR_i^{\lambda}= \frac{\mu_i + \lambda - r}{\sigma_i},\,i \in \{1, 2\}$$
    the Sharpe ratio of the corresponding asset in the market $\mathcal{M}_{\lambda}$.
	
	\begin{proposition}\label{prop:optimal_reinsurance_positive}
		 It is optimal for the insurer to buy partial reinsurance if and only if:
		 \begin{equation}\label{eq:pos_put_weight_log}
			SR_2^{\lambda^{\ast}} < \rho \cdot SR_1^{\lambda^{\ast}},
		\end{equation}
		where $\la$ is given by \eqref{eq: optimal lambda} in Corollary \ref{cor: summary karatzas}.
	\end{proposition}
	\begin{proof}
	See Appendix \ref{app:proofs_main}.
	\end{proof}
	Condition \eqref{eq:pos_put_weight_log} holds, e.g., when the correlation between the basic risky assets is sufficiently high and the asset that is not reinsurable is performance seeking, i.e., has a higher Sharpe ratio than the Sharpe ratio of the reinsurable risky asset.

	\section{Numerical studies}\label{sec_numerical_studies}$\,$
	
    First, we explain how we choose the model parameters. Second, we analyze the potential benefits of reinsurance. In particular, we calculate how much capital can be saved and how much higher a guarantee can be offered to the client when reinsurance is used in the design of insurance products with capital guarantees. Finally, we address the question of measuring how much of the insurer's loss is covered by reinsurance and perform a sensitivity analysis of this measure as well as the optimal investment-reinsurance strategy w.r.t. the model parameters.
	
	
	\subsection{Model parametrization and numerical algorithms} \label{subsec_parametrization}$\,$
 
	We estimate our model parameters in accordance with the European market. We choose the estimation period from January 1, 2003, till June 8, 2020 to include both bearish (financial 2008-2009 crisis, COVID-19 pandemic in 2020) and bullish markets.
	
	To estimate the risk-free rate we use Euro OverNight Index Average (EONIA) daily quotes. Parameters of $S_1$ are calibrated to the TecDAX daily data, whereas parameters of $S_2$ are estimated using DAX daily data. In this way we model the following situation:
	\begin{enumerate}
	    \item the asset manager of a German insurer overweights the technological sector, the corresponding portfolio is more performance-seeking than the overall market;
	    \item the reinsurer agrees to sell protection only on the overall market index, represented by the DAX in our study.
	\end{enumerate}
	In general, the asset manager of the insurer focuses on subindustries of its expertise. For the reinsurer these industries may be too risky or it does not have enough expertise in those areas. Therefore, it does not reinsure the specific portfolio of the insurer. Note that in the US market a comparable example would be the S\&P 500 Health Care Index or the S\&P 500 Consumer Discretionary Index as $S_1$ and the S\&P 500 Index as $S_2$. For estimating the risk-free rate in the US market, one could use the Effective Federal Funds Rate (EFFR).
	
	In Table \ref{tab:parametrization_summary} we summarize model parametrization.
	
    \begin{table}[!ht]
        \centering
        \begin{tabular}{@{}lll@{}}
            \toprule
            Parameter & Value & Explanation  \\ \midrule
            $r$ & $1.02\%$&  EONIA\\
            $\mu_1$ & $17.52\%$ & TecDAX rate of return\\
            $\mu_2$ & $12.37\%$ & DAX rate of return\\
            $\sigma_1$ & $23.66\%$ & TecDAX volatility\\
            $\sigma_2$ & $21.98\%$ & DAX volatility\\
            $\rho$ & $80.12\%$ & TecDAX and DAX correlation \\
            $S_0(0)$ & $1$ & For convenience\\
            $S_1(0)$ & $1$ & For convenience\\
            $S_2(0)$ & $1$ & For convenience\\
            $v_0$ & $100$ & For convenience \\
            $T$ & $10$ & Long-term investment \\
            $G_T$ & $100$ & Representative guarantee in the German market\\
            $\varepsilon$ & $0.5\%$ & High client's confidence in the guarantee \\
            $b$ & $-9$ & Corresponds to an RRA coefficient of $10$\\
            $\pi_B^{CM}$ & $29.47\%$ & Optimal initial proportion of money invested in the risky asset \\
            & & in the case of no reinsurance \\ \bottomrule
        \end{tabular}
        \caption{Model parametrization summary}
        \label{tab:parametrization_summary}
    \end{table}
    
    We choose the capital guarantee $G_T$ as $100\%$ of the initial investment to reflect the current situation in the \enquote{German Market}. In the past decades, insurers offered a positive guaranteed rate of return on clients' paid contributions. Due to a low interest-rate environment and other challenges, insurers have recently started offering products with full guarantee on paid capital but without any positive rate of return. In some products only a partial guarantee is embedded, e.g.,
    the product ERGO Rente Guarantee allows a customer to choose between $80\%$ and $100\%$ of the invested capital. Allianz offers policyholders a choice of guarantee levels between $60\%$ and $90\%$ of clients' contributions.
    
    Our choice of $b=-9$ leads to an insurer's relative risk aversion (RRA) coefficient of $10$, which is motivated by several aspects. In general, the RRA coefficient is a compromise between common RRA coefficients in theoretical research on long-term portfolio optimization and empirical evidence on RRA of mutual funds. On the one hand, \citeA{Broeders2011} and \citeA{Chen2018b}, investigating longer-term investment strategies in continuous time, set the RRA to $3$ in the base case. \citeA{Brandt2005}, \citeA{Garlappi2010} and \citeA{Cong2017}, analyzing optimal asset allocation in discrete time, consider higher RRA coefficients in their numerical studies, namely from $5$ to $15$. On the other hand, empirical research shows that the median and mean RRA coefficient of mutual fund managers are $5.75$ and $2.43$ respectively, see Table I in \citeA{Koijen2014}. Since mutual funds have less restrictive regulatory constraints than insurers do, it is reasonable to assume that RRA of the latter will be higher. Therefore, we set the RRA coefficient to $10$ in the base case and investigate a range of RRA coefficients from $5$ to $15$ in the sensitivity analysis section. Note that the insurer's optimal 1-year investment strategy without reinsurance  --- the solution to $(\bar{P}_{0.5\%, [0,+\infty)\times\{0\}})$ for $T=1$ --- has approximately $15\%$ of portfolio value invested in the risky asset. This value belongs to the range $10\%-15\%$, which is a representative range for the proportion of wealth insurance companies invest in the risky assets such as listed and private equity according to \citeA{Gruendl2016}.
    
   Appendix \ref{app:proofs_auxiliary} contains some propositions relevant for the numerical studies. In Proposition \ref{prop:v_f_kepsilon_SNLE} we provide the explicit formula of the left-hand side of the system of non-linear equations (SNLE) from Corollary \ref{cor: pp VaR constraints, auxiliary market}. In Proposition \ref{prop:value_function_formula} we calculate explicitly the insurer's value function, which is needed for Subsection \ref{subsec:reinsurance_benefits}. These two propositions use auxiliary Lemma \ref{lem:expectation_ztilde_interval} and Lemma \ref{lem:inequalities_optimal_unconstrained_wealth}, 
    which are also provided for completeness.
    
    {For solving the SNLE, we convert it to a minimization problem and apply the Sequential Quadratic Programming approach. For finding the roots of standalone non-linear equations appearing in the welfare loss and the guarantee gain analysis, we use the bisection method.}

	\subsection{Utility equivalent benefits of reinsurance}\label{subsec:reinsurance_benefits}$\,$
 
	The first natural question is whether the insurer needs reinsurance at all.  We assume that the insurer chooses the constant-mix strategy $\pi_B=(0, \pi_B^{CM})'=(0, \pi^\ast_{DN,1}(0))'$ that has the same proportion of wealth invested in $S_2$ as the proportion of wealth invested in $S_1$ in the insurer's optimal investment strategy under the no-reinsurance constraint. Here $\pi^\ast_{DN}$ solves $(\bar{P}_{0.5\%, [0,+\infty)\times\{0\}})$ and $DN$ stands for \textbf{d}ynamic (strategy with) \textbf{n}o reinsurance. In the base case, this leads to $\pi^{CM}_B = \pi_{DN,1}^\ast(0) = 29.47\%$. The insurer's optimal relative portfolio process at time $t=0$ is given by:
	\begin{equation*}
	    \pib_0(0) = 63.95\% \qquad \pib_1(0) = 33.48\% \qquad \pib_2(0) = 2.57\%
	\end{equation*}
	The optimal initial investment in terms of asset units is given by:
	\begin{equation*}
	    \phib_0(0) = 65.85 \qquad \phib_1(0) = 33.48 \qquad \phib_2(0) = 0.67,
	\end{equation*}
	and the price of one put is approximately equal to $3.85$. We see that it is optimal for the insurer to buy partial reinsurance, which costs about $2.5\%$ of the initial portfolio value. Interestingly, the optimal initial reinsured proportion of the benchmark portfolio is $67\%$. As we will see in Section \ref{subsec:reinsurance_proportion}, this partial reinsurance still leads to a high level of the insurer's expected loss covered by the reinsurer.
	
	
	\subsubsection{Welfare loss analysis}\label{subsec_WEUL}$\,$
 
	In this subsection, we calculate the monetary benefit to the insurer if the insurer follows the optimal investment-reinsurance strategy instead of implementing a suboptimal one. We determine the wealth-equivalent utility loss (WEUL), denoted by $l_{\pib^\ast, \pi_{SS}}$, that represents the proportion of the initial wealth \enquote{lost} when a suboptimal strategy $\pi_{SS}$ instead of the optimal strategy $\pib^\ast$ is followed. In particular, $l_{\pib^\ast, \pi_{SS}}$ is the solution to the following equation:
	    \begin{equation}\label{def:welfare_loss}
	       \EQ{U(\bar{V}^{v_0(1 - l_{\pib^*, \pi_{SS}}), \pib^*}(T))} = \EQ{U(\bar{V}^{v_0, \pi_{SS}}(T))}.
	    \end{equation}
	    So if the insurance company would follow an optimal investment-reinsurance strategy, the company would have needed $100 \cdot l_{\pib^\ast, \pi_{SS}}\%$ less initial capital to match the expected utility from the suboptimal strategy $\pi_{SS}$. If the expected utility from a suboptimal strategy is acceptable for both the insurer and the client, then switching to the optimal strategy may decrease product costs due to the saved $100 \cdot l_{\pib^\ast, \pi_{SS}}\%$ of the initial investment.
	    
	    We consider the following suboptimal strategies $\pi_{SS}$:
	    \begin{enumerate}
	        \item  the optimal dynamic strategy of the insurer under the no-reinsurance constraint, i.e., $\pi^\ast_{DN}$ that solves $(\bar{P}_{0.5\%, [0,+\infty)\times\{0\}})$. If the VaR-constraint is non-binding, then this is the Merton investment strategy.
	        \item the $(15\%, 0\%)'$ constant-mix strategy that approximates the long-term investment strategy of an average life insurer according to \citeA{Gruendl2016}, which we denote by $\pi_{CN}$ where $CN$ stands for \textbf{c}onstant-mix (strategy with) \textbf{n}o reinsurance.
	    \end{enumerate}
	    We obtain the following WEULs:
	    \begin{enumerate}
	        \item $l_{\pib^\ast,\pi^\ast_{DN}}=25$bp\footnote{bp stands for \enquote{basis point}, $1$bp$= 0.01\%=0.0001$}, i.e., replacing a product with the optimal no-reinsurance strategy with a product with the optimal investment-reinsurance strategy can make the product $25$bp cheaper to the customer without any loss in the insurer's expected utility;
	        \item $l_{\pib^\ast,\pi_{CN}}=588$bp, i.e., a product with optimal investment-reinsurance strategy requires $5.88\%$ less initial capital to reach the same expected utility as the suboptimal constant-mix $(15\%, 0\%)'$ strategy yields.
	    \end{enumerate}
        
\begin{figure}[!ht]
        \centering
        \begin{subfigure}{.5\textwidth}
          \centering
          \includegraphics[width=\linewidth]{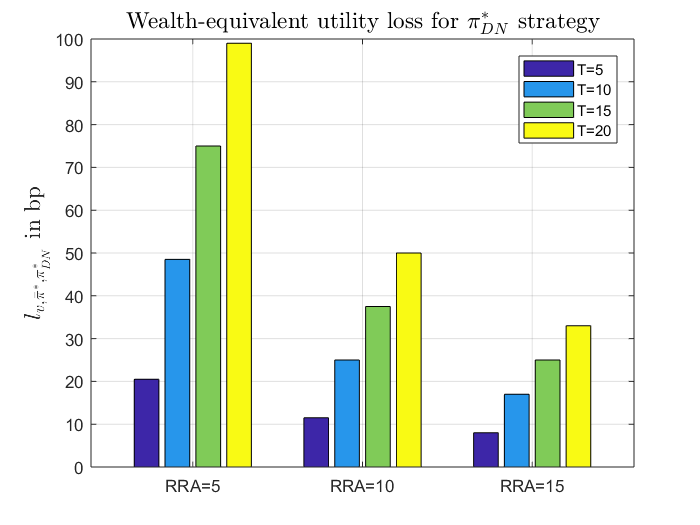}
          \caption{WEUL w.r.t. $\pi_{DN}^*$}
          \label{sfig:SA_WEUL_for_piDN}
        \end{subfigure}%
        \begin{subfigure}{.5\textwidth}
          \centering
          \includegraphics[width=\linewidth]{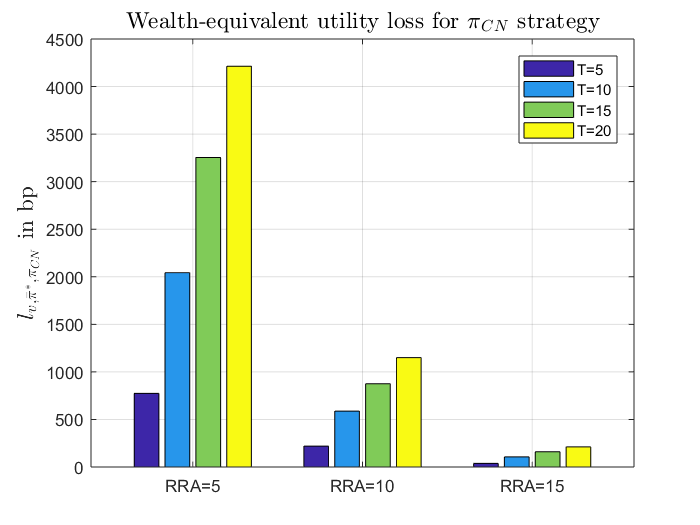}
          \caption{WEUL w.r.t. $\pi_{CN}$}
          \label{sfig:SA_WEUL_for_piCN}
        \end{subfigure}
        \caption{Impact of risk aversion and investment horizon on WEUL}
        \label{fig:SA_WEUL_depending_on_b_T}
        \end{figure}

	    
	    In Figure \ref{fig:SA_WEUL_depending_on_b_T}, we show the impact of the insurer's risk aversion and investment horizon on WEUL. The more risk averse the insurer, the less WEUL. For $\pi^\ast_{DN}$, this measure exhibits roughly linear dependence on $RRA$, whereas for $\pi_{CN}$ it shows a rather convex behaviour w.r.t $RRA$. The longer the investment period, the larger the WEUL. For both considered suboptimal strategies, WEUL shows approximately linear dependence on $T$. For very risk-averse insurers and short investment horizons, product costs that can be saved by using optimal reinsurance are relatively low. However, less risk-averse insurers offering mid to long-term equity-linked products with capital guarantees can decrease the corresponding product costs significantly, especially in comparison to products with underlying strategy $\pi_{CN}$. For example, for an insurer with $RRA=5$ and $T=15$ the cost reduction is about $32\%$.
	    
	    Overall, the results in this section indicate that the inclusion of dynamic reinsurance in the design of life insurance products with a capital guarantee decreases the product costs for the clients.
	    The actual \enquote{loss} of capital from investing suboptimally in practice may be different, as there are transaction costs, safety loadings in pricing reinsurance contracts, discrete trading times, jumps in asset prices, etc.
	    
	     For a broader view, we also provide for each of the these three strategies the corresponding risk-return profile and the probability that the terminal portfolio value falls below the guarantee $G_T$ in Table \ref{tab:strategies_performance}.
	    \begin{table}[!ht]
        \centering
        \begin{tabular}{@{}llll@{}}
        \toprule
         & $\pib^*$ & $\pi^\ast_{DN}$ & $\pi_{CN}$ \\ \midrule
        Annualized return & $6.11\%$  & $6.06\%$ & $3.56\%$ \\
        Annualized standard deviation of return & $12.85\%$ & $12.71\%$ & $5.05\%$ \\
        Probability of not reaching $G_T$ & $0.5\%$ & $0.5\%$ & $0.0011\%$ \\
        \bottomrule
        \end{tabular}
        \caption{Strategies' risk-return profiles and probabilities of not reaching $G_T$ }
        \label{tab:strategies_performance}
        \end{table}
        
        We see that the optimal dynamic investment strategy with reinsurance and the one without reinsurance have very similar risk-return profiles and fully use the available risk budget in the optimization problem, i.e., the corresponding underperformance probabilities are equal to the VaR probability $0.5\%$. The CN strategy, on the other side, does not fully use the available risk budget and thus loses more than $2.5\%$ in performance (annualized return).


	    \subsubsection{Guarantee gain analysis}\label{subsec_GEUG}$\,$
 
	    Here we measure the benefit of the optimal investment-reinsurance strategy in terms of a potential increase in the capital guarantee. We calculate the guarantee-equivalent utility gain (GEUG), denoted by $g_{\pib^\ast,\pi_{SS}}$, that indicates the proportion by which the terminal guarantee $G_T$ to the client can be increased such that the expected utility of the insurer following $\pib^\ast$ and the correspondingly higher guarantee is equal to the expected utility of the insurer following the suboptimal strategy $\pi_{SS}$ with the original guarantee $G_T$. Denote by $\bar{V}^{v_0, \pib^\ast}(T|G_T)$ the portfolio value at time $T$ with the initial capital $v_0$, relative portfolio process $\pib^\ast$ and guarantee $G_T$. Then, $g_{\pib^\ast,\pi_{SS}}$ is the solution to the following equation:
	    \begin{equation}\label{def:guarantee_loss}
	       \EQ{U(\bar{V}^{v_0, \pib^*}(T|(1 + g_{\pib^\ast,\pi_{SS}})\cdot G_T))} = \EQ{U(\bar{V}^{v_0, \pi_{SS}}(T|G_T))},
	    \end{equation}
	   
	   We obtain the following GEUGs:
	   {
	    \begin{enumerate}
	        \item $g_{\pib^\ast,\pi^\ast_{DN}}=10.08\%$, i.e., a product with optimal investment strategy without reinsurance and with a guarantee of $100\%$ of the client's initial endowment at product maturity ($0\%$ annualized guaranteed return) can be replaced --- without any loss in the insurer's expected utility ---  by a product with optimal investment-reinsurance strategy and a guarantee of $110\%$  of the client's initial endowment ($0.96\%$ annualized guaranteed return);
	        \item $g_{\pib^\ast,\pi_{CN}}=28.09\%$, i.e., a product with a constant-mix $(15\%, 0\%)'$ investment strategy without reinsurance and a guarantee of $100\%$ of the initial endowment at product maturity can be replaced --- without any loss in the insurer's expected utility --- by a product with optimal investment-reinsurance strategy and a guarantee of $128\%$  of the client's initial contribution ($2.5\%$ annualized guaranteed return).
	    \end{enumerate}
	    }

\begin{figure}[!ht]
        \centering
        \begin{subfigure}{.5\textwidth}
          \centering
          \includegraphics[width=\linewidth]{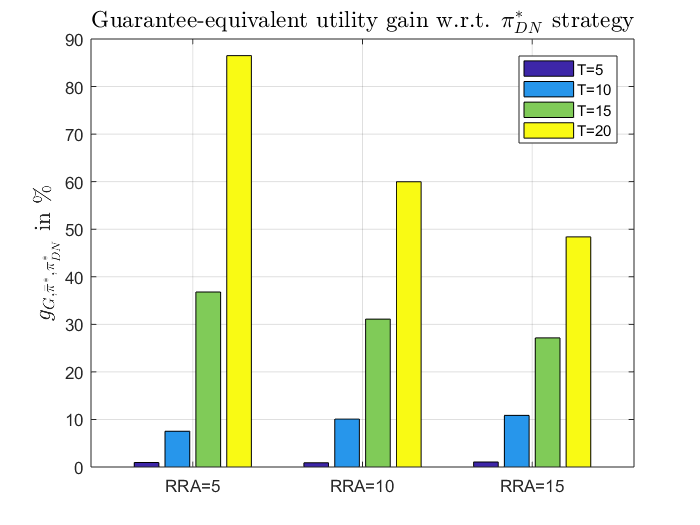}
          \caption{GEUG w.r.t. $\pi_{DN}^*$}
          \label{sfig:SA_GEUG_for_piDN}
        \end{subfigure}%
        \begin{subfigure}{.5\textwidth}
          \centering
          \includegraphics[width=\linewidth]{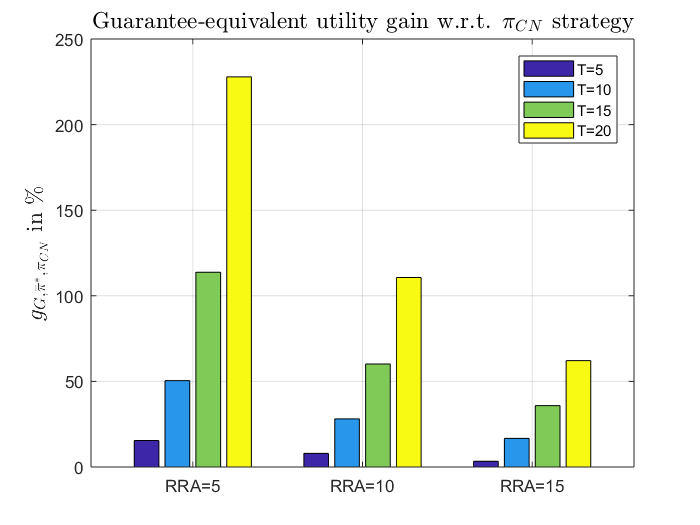}
          \caption{GEUG w.r.t. $\pi_{CN}$}
          \label{sfig:SA_GEUG_for_piCN}
        \end{subfigure}
        \caption{Impact of risk aversion and investment horizon on GEUG}
        \label{fig:SA_GEUG_depending_on_b_T}
        \end{figure}

	    Figure \ref{fig:SA_GEUG_depending_on_b_T} illustrates how the insurer's risk aversion and investment horizon influence GEUG. The more risk averse the insurer, the less GEUG. With increasing risk aversion, the optimally behaving insurer invests more in bonds and less in stocks and reinsurance, as it will be shown in Section \ref{subsec:SA_for_piBarStar}. Since a risk-free investment has a comparably low rate of return, GEUG decreases. We also observe that GEUG is convex w.r.t. $RRA$ in both cases, $\pi^\ast_{DN}$ and $\pi_{CN}$. The longer the investment period, the larger the GEUG. This dependence also illustrates convexity w.r.t. $T$ for both considered suboptimal strategies. We see that even very risk-averse insurers with short to mid-term equity-linked products can significantly increase their guarantee levels without any loss in expected utility. {For $RRA=15$ and $T=5$, the  guarantee in the product following the optimal investment-reinsurance strategy is increased from $100\%$ to $108\%$ ($1.55\%$ guaranteed annualized return) in comparison to a product with the strategy $\pi_{CN}$\footnote{For $T=5$, $\mathbb{Q}\lrr{\bar{V}^{v_0, \pi_{CN}}(T) < G_T} \approx 1.5\%$}. For less risk-averse insurers and products with longer investment horizons, GEUG is even more prominent. For example, for $RRA=5$ and $T=15$, the insurer with the optimally managed reinsurance can guarantee that the client's terminal payoff equals at least $135\%$ of the initial contribution (about $2\%$ annualized guaranteed return) and achieve the same expected utility as for a product with the underlying strategy $\pi^\ast_{DN}$ and a guarantee of only $100\%$ of the initial contribution.}
	    
	    Overall, the inclusion of dynamic reinsurance in the design of life insurance products with capital guarantees can lead to substantial increase in the guarantee levels that insurance companies can offer to their clients. The actual guarantee gain in practice may be different due to reasons mentioned at the end of the previous subsection.
	
	\subsection{Reinsurance proportion}\label{subsec:reinsurance_proportion}
	$\,$
 
	In this section we briefly address a natural question about reinsurance: how much protection against the insurer's loss does the reinsurance provide? The precise measurement of the reinsurance level/proportion is challenging as:
 \begin{enumerate}
     \item  the underlying portfolio in the reinsurable portfolio is not the same as the insurer's actual portfolio due to different assets ($S_1 \neq S_2$);
     \item the corresponding relative portfolio processes are different ($\pi^{CM}_B \neq \pib^*_1(0) / (1 - \pib^*_2(0))$), where the former term is the proportion of money invested in $S_1$ in the benchmark portfolio and the latter term is the proportion of investment in $S_1$ in the insurer's portfolio after subtracting money spent on reinsurance.
     \item the initial capital of the reinsured benchmark portfolio is slightly higher than the capital invested in $S_0$ and $S_1$ by the insurer due to the purchase of reinsurance.
 \end{enumerate}
 
The first insight into the reinsurance level can be captured by the number of put options in the insurer's portfolio $\phib^*_2(0)$, i.e., 1 put approximately hedges the portfolio of the insurer. Approximately because of the above mentioned points 1 to 3. We could also look at the number of puts adjusted by the correlation between the insurer's portfolio and the reinsured portfolio $\rho \phib^*_2(0)$.
 
 
 
 

 In the literature on reinsurance, two types of reinsurance are differentiated: proportional reinsurance and excess-of-loss reinsurance. In the former reinsurance type, the insurer's total loss is shared proportionally between the insurer and the reinsurer. However, the reinsurance is written on the exact portfolio the insurer has. Motivated by it, we consider the proportion of expected loss coverage (PELC), which we define as follows:
 \begin{equation*}
     \begin{aligned}
         PELC_t & = \frac{\textit{Amount of reinsurance at $t$} \times \textit{Expected coverage from one reinsurance contract}}{\textit{Expected total loss of insurer}} \\
         & {=} \ \frac{\EQ{\phib^*_2(t)\lrr{G_T - V_T^{v_0, \pi_B}}^+|\mathcal{F}_t}}{\EQ{\lrr{G_T - V_T^{v_0, \pib^*}}^+|\mathcal{F}_t}} =  \frac{\phib^*_2(t)\EQ{\lrr{G_T - V_T^{v_0, \pi_B}}^+|\mathcal{F}_t}}{\EQ{\lrr{G_T - V_T^{v_0, \pib^*}}^+|\mathcal{F}_t}}\\
     \end{aligned}
 \end{equation*}
 
 
 The calculation of PELC requires Monte-Carlo simulations due to the sophisticated optimal investment-reinsurance strategy of the insurer, which is needed to estimate the denominator of PELC.


For the base case we have:
\begin{equation*}
    \phib^*_2(0) = 0.67,\quad \rho \phib^*_2(0) = 0.54,\quad PELC_0 = 138.52\%, \quad \rho PELC_0 = 110.82\%.
\end{equation*}

We see that the initial optimal reinsurance strategy implies buying $67\%$  reinsurance, which leads to a correlation-corrected $PELC_0$ slightly higher than $100\%$. Taking into account the above mentioned challenges 1-3, we find these numbers reasonable.

	\subsection{Sensitivity analysis of optimal investment-reinsurance strategies}\label{subsec:SA_for_piBarStar}$\,$
	
	{In this subsection, we summarize the impact of changes in model parameters on the optimal  investment-reinsurance strategy. Not to make the paper unnecessarily longer, we provide figures depicting the sensitivity analysis results only for the risk-aversion parameter and the VaR probability threshold.}
	
\begin{figure}[!ht]
    \centering
    \begin{subfigure}{.5\textwidth}
      \centering
      \includegraphics[width=\linewidth]{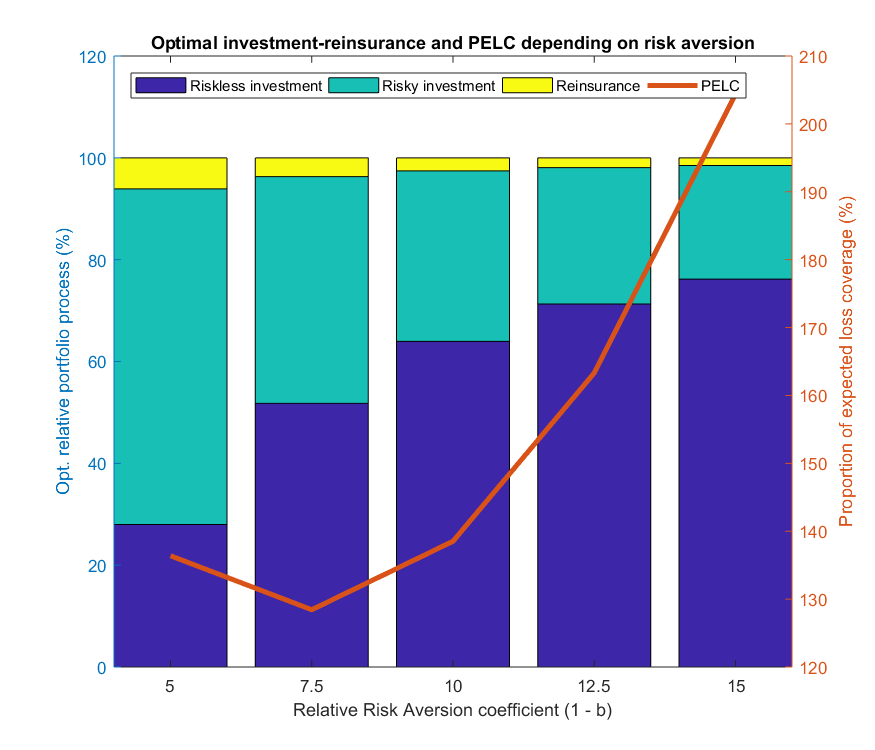}
      \caption{Sensitivity of $\pib^*$ and $PELC_0$ w.r.t. $b$}
      \label{sfig:SA_b}
    \end{subfigure}%
    \begin{subfigure}{.5\textwidth}
      \centering
      \includegraphics[width=\linewidth]{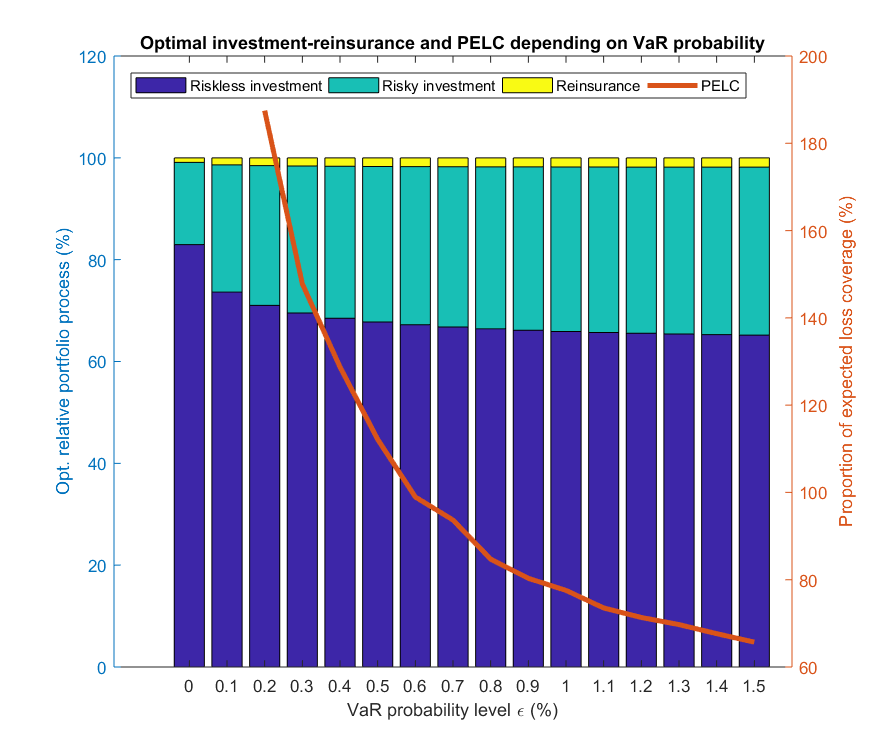}
      \caption{Sensitivity of $\pib^*$ and $PELC_0$ w.r.t. $\varepsilon$}
      \label{sfig:SA_epsilon}
    \end{subfigure}
    \caption{Sensitivity of the optimal investment-reinsurance strategy w.r.t. risk aversion and VaR probability}
    \label{fig:SA_piBarStar_PELC_for_b_and_epsilon}
    \end{figure}

	As mentioned in Subsection \ref{subsec_parametrization}, we explore the risk-aversion parameter values $1 - b = RRA \in \{5, 7.5, 10, 12.5, 15\}$. The higher the RRA coefficient, the less the optimally behaving insurer invests in the risky assets and the less money is spent on reinsurance. However, $PELC_0$ increases as the put option becomes cheaper due to the decreasing riskiness of the benchmark portfolio. This is illustrated in Subfigure \ref{sfig:SA_b}.
	
	For the VaR-probability $\varepsilon \in \{0\%, 0.1\%, 0.2\%, ..., 1.5\%\}$, we observe that the higher $\varepsilon$, the more money is invested in both the risky asset and the reinsurance. However, the $PELC_0$ gradually decreases due to the increasing riskiness of the insurer's optimal investment strategy and inability to hedge out all the residual risk arising due to a less risky reinsurable benchmark portfolio. Since for shorter investment horizons the influence of the VaR-constraint is more prominent, we illustrate for $T=5$ the sensitivity of $\pib^\ast$ and $PELC_0$ w.r.t. $\varepsilon$ in Subfigure \ref{sfig:SA_epsilon}. 
	
	Varying the weight of the risky asset in the benchmark portfolio $\pi^{CM}_B \in \{ \pi^\ast_{DN,1}(0) - 15\%, \pi^\ast_{DN,1}(0) - 10\%, ..., \pi^\ast_{DN,1}(0) + 15\%\}$, we find no change in the optimal investment strategy with respect to the risky asset. However, more money is invested in reinsurance and the $PELC_0$ increases.
	
    The higher the interest rate $r \in \{-2\%, -1\%, 0\%, 1\%, 2\% \}$, the less money the optimally behaving insurer invests in the risky asset. However, more money is invested in reinsurance, which, in conjunction with a decreasing price for the reinsurance, leads to an increase of $PELC_0$. 
	
	For an increasing investment horizon $T \in \{1, 5, 10, 15, 20\}$ we observe that both the optimal initial investment in the risky assets as well as the proportion of initial wealth invested in reinsurance increase. The $PELC_0$ gradually increases too.
	
	When the terminal capital guarantee $G_T \in \{0.7 \cdot v_0\,,0.8 \cdot v_0,\dots, 1.1 \cdot v_0\}$ increases, the insurer's optimal investment in stocks slightly decreases. Simultaneously, the insurer invests more money in reinsurance, even though the price of the reinsurance contract surges. The $PELC_0$ gradually decreases.
	
	In all numerical studies we also observe that the inclusion of reinsurance in the product design increases the insurer's optional initial risky asset exposure by up to $10\%$ in comparison to the optimal investment strategy in the no-reinsurance case. 

	\section{Conclusion and further research}\label{sec_conclusions}$\,$
	
	This paper solves the optimal investment-reinsurance problem under both no-short-selling and VaR-constraints, which is relevant for insurers offering equity-linked products with capital guarantees. This portfolio optimization advancement allows for an analysis of the potential of reinsurance to reverse the trend of decreasing capital guarantees embedded in the above mentioned products, which is currently observed in the insurance sector. We find that fairly priced dynamic reinsurance can allow insurers -- without loss in their expected utility --  decrease product costs and offer significantly higher capital guarantees to their clients. We reach the former conclusion by analyzing the wealth-equivalent utility loss. We infer the latter benefit of reinsurance using the concept of a guarantee-equivalent utility gain. In our main example, we show that the inclusion of the optimally managed reinsurance to an equity-linked insurance product running for 10 years can allow the insurer to guarantee (at product maturity with $99.5\%$ probability) that the client will receive at least $110\%$ of the client's initial contribution without any loss of the insurer's expected utility. Moreover, such a product with optimal reinsurance can allow the insurer to guarantee (at product maturity with $99.5\%$ probability) that the client will receive at least $128\%$  of the client's initial endowment without any loss of the insurer's expected utility in comparison to the one obtained by a constant-mix strategy $85\%$ bonds and $15\%$ stocks.

    Our inference relies on the model assumptions. Therefore, it might be interesting to see how it changes when more complex models are considered, e.g., stochastic interest rates. The extension of the class of reinsurable strategies may be another fascinating line of research, e.g., reinsurance could be modelled as a passport option on the insurer's actual portfolio. A passport option would give the insurer the right to follow any admissible trading strategy, while the reinsurer would be obliged to cover any net losses on the strategy.  Usually, such options are quite expensive due to the flexibility of the underlying portfolio.

    \section*{Acknowledgement}
    Yevhen Havrylenko and Rudi Zagst acknowledge the financial support of the ERGO Center of Excellence in Insurance at the Technical University of Munich promoted by ERGO Group.

\bibliographystyle{apacite} 
\bibliography{Dynamic_reinsurance}

\begin{thebibliography}{}

\bibitem [\protect \citeauthoryear {%
Bai%
\ \BBA {} Guo%
}{%
Bai%
\ \BBA {} Guo%
}{%
{\protect \APACyear {2010}}%
}]{%
Bai2010}
\APACinsertmetastar {%
Bai2010}%
\begin{APACrefauthors}%
Bai, L.%
\BCBT {}\ \BBA {} Guo, J.%
\end{APACrefauthors}%
\unskip\
\newblock
\APACrefYearMonthDay{2010}{Jul}{01}.
\newblock
{\BBOQ}\APACrefatitle {Optimal dynamic excess-of-loss reinsurance and
  multidimensional portfolio selection} {Optimal dynamic excess-of-loss
  reinsurance and multidimensional portfolio selection}.{\BBCQ}
\newblock
\APACjournalVolNumPages{Science China Mathematics}{53}{7}{1787--1804}.
\newblock
\begin{APACrefURL} \url{https://doi.org/10.1007/s11425-010-4033-4}
  \end{APACrefURL}
\PrintBackRefs{\CurrentBib}

\bibitem [\protect \citeauthoryear {%
Bardhan%
}{%
Bardhan%
}{%
{\protect \APACyear {1994}}%
}]{%
Bardhan1994}
\APACinsertmetastar {%
Bardhan1994}%
\begin{APACrefauthors}%
Bardhan, I.%
\end{APACrefauthors}%
\unskip\
\newblock
\APACrefYearMonthDay{1994}{}{}.
\newblock
{\BBOQ}\APACrefatitle {{Consumption and investment under constraints}}
  {{Consumption and investment under constraints}}.{\BBCQ}
\newblock
\APACjournalVolNumPages{Journal of Economic Dynamics and
  Control}{18}{5}{909-929}.
\newblock
\begin{APACrefURL} \url{https://doi.org/10.1016/0165-1889(94)90038-8}
  \end{APACrefURL}
\PrintBackRefs{\CurrentBib}

\bibitem [\protect \citeauthoryear {%
Basak%
\ \BBA {} Shapiro%
}{%
Basak%
\ \BBA {} Shapiro%
}{%
{\protect \APACyear {2001}}%
}]{%
Basak2001}
\APACinsertmetastar {%
Basak2001}%
\begin{APACrefauthors}%
Basak, S.%
\BCBT {}\ \BBA {} Shapiro, A.%
\end{APACrefauthors}%
\unskip\
\newblock
\APACrefYearMonthDay{2001}{}{}.
\newblock
{\BBOQ}\APACrefatitle {Value-at-Risk-Based Risk Management: Optimal Policies
  and Asset Prices} {Value-at-risk-based risk management: Optimal policies and
  asset prices}.{\BBCQ}
\newblock
\APACjournalVolNumPages{The Review of Financial Studies}{14}{2}{371-405}.
\newblock
\begin{APACrefURL} \url{http://dx.doi.org/10.1093/rfs/14.2.371}
  \end{APACrefURL}
\PrintBackRefs{\CurrentBib}

\bibitem [\protect \citeauthoryear {%
Bian%
, Miao%
\BCBL {}\ \BBA {} Zheng%
}{%
Bian%
\ \protect \BOthers {.}}{%
{\protect \APACyear {2011}}%
}]{%
Zheng2011}
\APACinsertmetastar {%
Zheng2011}%
\begin{APACrefauthors}%
Bian, B.%
, Miao, S.%
\BCBL {}\ \BBA {} Zheng, H.%
\end{APACrefauthors}%
\unskip\
\newblock
\APACrefYearMonthDay{2011}{}{}.
\newblock
{\BBOQ}\APACrefatitle {Smooth Value Functions for a Class of Nonsmooth Utility
  Maximization Problems} {Smooth value functions for a class of nonsmooth
  utility maximization problems}.{\BBCQ}
\newblock
\APACjournalVolNumPages{SIAM Journal on Financial Mathematics}{2}{1}{727-747}.
\PrintBackRefs{\CurrentBib}

\bibitem [\protect \citeauthoryear {%
Boyle%
\ \BBA {} Tian%
}{%
Boyle%
\ \BBA {} Tian%
}{%
{\protect \APACyear {2007}}%
}]{%
Boyle2007}
\APACinsertmetastar {%
Boyle2007}%
\begin{APACrefauthors}%
Boyle, P.%
\BCBT {}\ \BBA {} Tian, W.%
\end{APACrefauthors}%
\unskip\
\newblock
\APACrefYearMonthDay{2007}{}{}.
\newblock
{\BBOQ}\APACrefatitle {Portfolio management with constraints} {Portfolio
  management with constraints}.{\BBCQ}
\newblock
\APACjournalVolNumPages{Mathematical Finance}{17}{3}{319-343}.
\newblock
\begin{APACrefURL}
  \url{https://onlinelibrary.wiley.com/doi/abs/10.1111/j.1467-9965.2007.00306.x}
  \end{APACrefURL}
\PrintBackRefs{\CurrentBib}

\bibitem [\protect \citeauthoryear {%
Brandt%
, Goyal%
, Santa-Clara%
\BCBL {}\ \BBA {} Stroud%
}{%
Brandt%
\ \protect \BOthers {.}}{%
{\protect \APACyear {2005}}%
}]{%
Brandt2005}
\APACinsertmetastar {%
Brandt2005}%
\begin{APACrefauthors}%
Brandt, M\BPBI W.%
, Goyal, A.%
, Santa-Clara, P.%
\BCBL {}\ \BBA {} Stroud, J\BPBI R.%
\end{APACrefauthors}%
\unskip\
\newblock
\APACrefYearMonthDay{2005}{05}{}.
\newblock
{\BBOQ}\APACrefatitle {{A Simulation Approach to Dynamic Portfolio Choice with
  an Application to Learning About Return Predictability}} {{A Simulation
  Approach to Dynamic Portfolio Choice with an Application to Learning About
  Return Predictability}}.{\BBCQ}
\newblock
\APACjournalVolNumPages{The Review of Financial Studies}{18}{3}{831-873}.
\newblock
\begin{APACrefURL} \url{https://doi.org/10.1093/rfs/hhi019} \end{APACrefURL}
\newblock
\begin{APACrefDOI} \doi{10.1093/rfs/hhi019} \end{APACrefDOI}
\PrintBackRefs{\CurrentBib}

\bibitem [\protect \citeauthoryear {%
Broeders%
, Chen%
\BCBL {}\ \BBA {} Koos%
}{%
Broeders%
\ \protect \BOthers {.}}{%
{\protect \APACyear {2011}}%
}]{%
Broeders2011}
\APACinsertmetastar {%
Broeders2011}%
\begin{APACrefauthors}%
Broeders, D.%
, Chen, A.%
\BCBL {}\ \BBA {} Koos, B.%
\end{APACrefauthors}%
\unskip\
\newblock
\APACrefYearMonthDay{2011}{July}{}.
\newblock
{\BBOQ}\APACrefatitle {{A utility-based comparison of pension funds and life
  insurance companies under regulatory constraints}} {{A utility-based
  comparison of pension funds and life insurance companies under regulatory
  constraints}}.{\BBCQ}
\newblock
\APACjournalVolNumPages{Insurance: Mathematics and Economics}{49}{1}{1-10}.
\newblock
\begin{APACrefURL} \url{https://doi.org/10.1016/j.insmatheco.2011.01.011}
  \end{APACrefURL}
\PrintBackRefs{\CurrentBib}

\bibitem [\protect \citeauthoryear {%
Chen%
, Hieber%
\BCBL {}\ \BBA {} Nguyen%
}{%
Chen%
\ \protect \BOthers {.}}{%
{\protect \APACyear {2019}}%
}]{%
Chen2019}
\APACinsertmetastar {%
Chen2019}%
\begin{APACrefauthors}%
Chen, A.%
, Hieber, P.%
\BCBL {}\ \BBA {} Nguyen, T.%
\end{APACrefauthors}%
\unskip\
\newblock
\APACrefYearMonthDay{2019}{}{}.
\newblock
{\BBOQ}\APACrefatitle {Constrained non-concave utility maximization: An
  application to life insurance contracts with guarantees} {Constrained
  non-concave utility maximization: An application to life insurance contracts
  with guarantees}.{\BBCQ}
\newblock
\APACjournalVolNumPages{European Journal of Operational
  Research}{273}{3}{1119-1135}.
\newblock
\begin{APACrefURL} \url{https://doi.org/10.1016/j.ejor.2018.09.002}
  \end{APACrefURL}
\PrintBackRefs{\CurrentBib}

\bibitem [\protect \citeauthoryear {%
Chen%
, Nguyen%
\BCBL {}\ \BBA {} Stadje%
}{%
Chen%
\ \protect \BOthers {.}}{%
{\protect \APACyear {2018}}%
}]{%
Chen2018b}
\APACinsertmetastar {%
Chen2018b}%
\begin{APACrefauthors}%
Chen, A.%
, Nguyen, T.%
\BCBL {}\ \BBA {} Stadje, M.%
\end{APACrefauthors}%
\unskip\
\newblock
\APACrefYearMonthDay{2018}{}{}.
\newblock
{\BBOQ}\APACrefatitle {{Optimal investment under VaR-Regulation and Minimum
  Insurance}} {{Optimal investment under VaR-Regulation and Minimum
  Insurance}}.{\BBCQ}
\newblock
\APACjournalVolNumPages{Insurance: Mathematics and Economics}{79}{C}{194-209}.
\newblock
\begin{APACrefURL} \url{https://doi.org/10.1016/j.insmatheco.2018.01.008}
  \end{APACrefURL}
\PrintBackRefs{\CurrentBib}

\bibitem [\protect \citeauthoryear {%
Cong%
\ \BBA {} Oosterlee%
}{%
Cong%
\ \BBA {} Oosterlee%
}{%
{\protect \APACyear {2017}}%
}]{%
Cong2017}
\APACinsertmetastar {%
Cong2017}%
\begin{APACrefauthors}%
Cong, F.%
\BCBT {}\ \BBA {} Oosterlee, C\BPBI W.%
\end{APACrefauthors}%
\unskip\
\newblock
\APACrefYearMonthDay{2017}{{\APACmonth{03}}}{}.
\newblock
{\BBOQ}\APACrefatitle {Accurate and Robust Numerical Methods for the Dynamic
  Portfolio Management Problem} {Accurate and robust numerical methods for the
  dynamic portfolio management problem}.{\BBCQ}
\newblock
\APACjournalVolNumPages{Computational Economics}{49}{3}{433--458}.
\newblock
\begin{APACrefURL} \url{https://doi.org/10.1007/s10614-016-9569-0}
  \end{APACrefURL}
\PrintBackRefs{\CurrentBib}

\bibitem [\protect \citeauthoryear {%
Cvitanic%
\ \BBA {} Karatzas%
}{%
Cvitanic%
\ \BBA {} Karatzas%
}{%
{\protect \APACyear {1992}}%
}]{%
Cvitanic1992}
\APACinsertmetastar {%
Cvitanic1992}%
\begin{APACrefauthors}%
Cvitanic, J.%
\BCBT {}\ \BBA {} Karatzas, I.%
\end{APACrefauthors}%
\unskip\
\newblock
\APACrefYearMonthDay{1992}{11}{}.
\newblock
{\BBOQ}\APACrefatitle {Convex Duality in Constrained Portfolio Optimization}
  {Convex duality in constrained portfolio optimization}.{\BBCQ}
\newblock
\APACjournalVolNumPages{Ann. Appl. Probab.}{2}{4}{767--818}.
\newblock
\begin{APACrefURL} \url{https://www.jstor.org/stable/2959666} \end{APACrefURL}
\PrintBackRefs{\CurrentBib}

\bibitem [\protect \citeauthoryear {%
Dong%
\ \BBA {} Zheng%
}{%
Dong%
\ \BBA {} Zheng%
}{%
{\protect \APACyear {2019}}%
}]{%
Dong2019}
\APACinsertmetastar {%
Dong2019}%
\begin{APACrefauthors}%
Dong, Y.%
\BCBT {}\ \BBA {} Zheng, H.%
\end{APACrefauthors}%
\unskip\
\newblock
\APACrefYearMonthDay{2019}{}{}.
\newblock
{\BBOQ}\APACrefatitle {{Optimal investment of DC pension plan under
  short-selling constraints and portfolio insurance}} {{Optimal investment of
  DC pension plan under short-selling constraints and portfolio
  insurance}}.{\BBCQ}
\newblock
\APACjournalVolNumPages{Insurance: Mathematics and Economics}{85}{}{47 - 59}.
\newblock
\begin{APACrefURL} \url{https://doi.org/10.1016/j.insmatheco.2018.12.005}
  \end{APACrefURL}
\PrintBackRefs{\CurrentBib}

\bibitem [\protect \citeauthoryear {%
Dong%
\ \BBA {} Zheng%
}{%
Dong%
\ \BBA {} Zheng%
}{%
{\protect \APACyear {2020}}%
}]{%
Dong2020}
\APACinsertmetastar {%
Dong2020}%
\begin{APACrefauthors}%
Dong, Y.%
\BCBT {}\ \BBA {} Zheng, H.%
\end{APACrefauthors}%
\unskip\
\newblock
\APACrefYearMonthDay{2020}{}{}.
\newblock
{\BBOQ}\APACrefatitle {Optimal investment with S-shaped utility and trading and
  Value at Risk constraints: An application to defined contribution pension
  plan} {Optimal investment with s-shaped utility and trading and value at risk
  constraints: An application to defined contribution pension plan}.{\BBCQ}
\newblock
\APACjournalVolNumPages{European Journal of Operational Research}{281}{2}{341 -
  356}.
\newblock
\begin{APACrefURL} \url{https://doi.org/10.1016/j.ejor.2019.08.034}
  \end{APACrefURL}
\PrintBackRefs{\CurrentBib}

\bibitem [\protect \citeauthoryear {%
Escobar%
, Kriebel%
, Wahl%
\BCBL {}\ \BBA {} Zagst%
}{%
Escobar%
\ \protect \BOthers {.}}{%
{\protect \APACyear {2019}}%
}]{%
Wahl2019}
\APACinsertmetastar {%
Wahl2019}%
\begin{APACrefauthors}%
Escobar, M.%
, Kriebel, P.%
, Wahl, M.%
\BCBL {}\ \BBA {} Zagst, R.%
\end{APACrefauthors}%
\unskip\
\newblock
\APACrefYearMonthDay{2019}{}{}.
\newblock
{\BBOQ}\APACrefatitle {Portfolio optimization under {S}olvency {II}} {Portfolio
  optimization under {S}olvency {II}}.{\BBCQ}
\newblock
\APACjournalVolNumPages{Annals of Operations Research}{281}{1}{193-227}.
\newblock
\begin{APACrefURL} \url{https://doi.org/10.1007/s10479-018-2835-x}
  \end{APACrefURL}
\PrintBackRefs{\CurrentBib}

\bibitem [\protect \citeauthoryear {%
Garlappi%
\ \BBA {} Skoulakis%
}{%
Garlappi%
\ \BBA {} Skoulakis%
}{%
{\protect \APACyear {2010}}%
}]{%
Garlappi2010}
\APACinsertmetastar {%
Garlappi2010}%
\begin{APACrefauthors}%
Garlappi, L.%
\BCBT {}\ \BBA {} Skoulakis, G.%
\end{APACrefauthors}%
\unskip\
\newblock
\APACrefYearMonthDay{2010}{08}{}.
\newblock
{\BBOQ}\APACrefatitle {{Solving Consumption and Portfolio Choice Problems: The
  State Variable Decomposition Method}} {{Solving Consumption and Portfolio
  Choice Problems: The State Variable Decomposition Method}}.{\BBCQ}
\newblock
\APACjournalVolNumPages{The Review of Financial Studies}{23}{9}{3346-3400}.
\newblock
\begin{APACrefURL} \url{https://doi.org/10.1093/rfs/hhq045} \end{APACrefURL}
\PrintBackRefs{\CurrentBib}

\bibitem [\protect \citeauthoryear {%
Gr\"undl%
, Dong%
\BCBL {}\ \BBA {} Gal%
}{%
Gr\"undl%
\ \protect \BOthers {.}}{%
{\protect \APACyear {2016}}%
}]{%
Gruendl2016}
\APACinsertmetastar {%
Gruendl2016}%
\begin{APACrefauthors}%
Gr\"undl, H.%
, Dong, M.%
\BCBL {}\ \BBA {} Gal, J.%
\end{APACrefauthors}%
\unskip\
\newblock
\APACrefYearMonthDay{2016}{}{}.
\newblock
{\BBOQ}\APACrefatitle {The evolution of insurer portfolio investment strategies
  for long-term investing} {The evolution of insurer portfolio investment
  strategies for long-term investing}.{\BBCQ}
\newblock
\APACjournalVolNumPages{OECD Journal: Financial Market Trends}{2016}{1}{}.
\newblock
\begin{APACrefURL} \url{https://doi.org/10.1787/fmt-2016-5jln3rh7qf46}
  \end{APACrefURL}
\PrintBackRefs{\CurrentBib}

\bibitem [\protect \citeauthoryear {%
Guan%
\ \BBA {} Liang%
}{%
Guan%
\ \BBA {} Liang%
}{%
{\protect \APACyear {2016}}%
}]{%
Guan2016}
\APACinsertmetastar {%
Guan2016}%
\begin{APACrefauthors}%
Guan, G.%
\BCBT {}\ \BBA {} Liang, Z.%
\end{APACrefauthors}%
\unskip\
\newblock
\APACrefYearMonthDay{2016}{}{}.
\newblock
{\BBOQ}\APACrefatitle {Optimal management of DC pension plan under loss
  aversion and Value-at-Risk constraints} {Optimal management of dc pension
  plan under loss aversion and value-at-risk constraints}.{\BBCQ}
\newblock
\APACjournalVolNumPages{Insurance: Mathematics and Economics}{69}{}{224 - 237}.
\newblock
\begin{APACrefURL} \url{https://doi.org/10.1016/j.insmatheco.2016.05.014}
  \end{APACrefURL}
\PrintBackRefs{\CurrentBib}

\bibitem [\protect \citeauthoryear {%
Hambardzumyan%
\ \BBA {} Korn%
}{%
Hambardzumyan%
\ \BBA {} Korn%
}{%
{\protect \APACyear {2019}}%
}]{%
Hambardzumyan2019}
\APACinsertmetastar {%
Hambardzumyan2019}%
\begin{APACrefauthors}%
Hambardzumyan, H.%
\BCBT {}\ \BBA {} Korn, R.%
\end{APACrefauthors}%
\unskip\
\newblock
\APACrefYearMonthDay{2019}{}{}.
\newblock
{\BBOQ}\APACrefatitle {Dynamic hybrid products with guarantees—An optimal
  portfolio framework} {Dynamic hybrid products with guarantees—an optimal
  portfolio framework}.{\BBCQ}
\newblock
\APACjournalVolNumPages{Insurance: Mathematics and Economics}{84}{}{54 - 66}.
\newblock
\begin{APACrefURL} \url{https://doi.org/10.1016/j.insmatheco.2018.11.005}
  \end{APACrefURL}
\PrintBackRefs{\CurrentBib}

\bibitem [\protect \citeauthoryear {%
Koijen%
}{%
Koijen%
}{%
{\protect \APACyear {2014}}%
}]{%
Koijen2014}
\APACinsertmetastar {%
Koijen2014}%
\begin{APACrefauthors}%
Koijen, R\BPBI S.%
\end{APACrefauthors}%
\unskip\
\newblock
\APACrefYearMonthDay{2014}{}{}.
\newblock
{\BBOQ}\APACrefatitle {The Cross-Section of Managerial Ability, Incentives, and
  Risk Preferences} {The cross-section of managerial ability, incentives, and
  risk preferences}.{\BBCQ}
\newblock
\APACjournalVolNumPages{The Journal of Finance}{69}{3}{1051-1098}.
\newblock
\begin{APACrefURL} \url{https://doi.org/10.1111/jofi.12140} \end{APACrefURL}
\PrintBackRefs{\CurrentBib}

\bibitem [\protect \citeauthoryear {%
Korn%
}{%
Korn%
}{%
{\protect \APACyear {2005}}%
}]{%
Korn2005}
\APACinsertmetastar {%
Korn2005}%
\begin{APACrefauthors}%
Korn, R.%
\end{APACrefauthors}%
\unskip\
\newblock
\APACrefYearMonthDay{2005}{}{}.
\newblock
{\BBOQ}\APACrefatitle {Optimal portfolios with a positive lower bound on final
  wealth} {Optimal portfolios with a positive lower bound on final
  wealth}.{\BBCQ}
\newblock
\APACjournalVolNumPages{Quantitative Finance}{5}{3}{315-321}.
\newblock
\begin{APACrefURL} \url{https://doi.org/10.1080/14697680500167927}
  \end{APACrefURL}
\newblock
\begin{APACrefDOI} \doi{10.1080/14697680500167927} \end{APACrefDOI}
\PrintBackRefs{\CurrentBib}

\bibitem [\protect \citeauthoryear {%
Korn%
\ \BBA {} Trautmann%
}{%
Korn%
\ \BBA {} Trautmann%
}{%
{\protect \APACyear {1999}}%
}]{%
Korn1999}
\APACinsertmetastar {%
Korn1999}%
\begin{APACrefauthors}%
Korn, R.%
\BCBT {}\ \BBA {} Trautmann, S.%
\end{APACrefauthors}%
\unskip\
\newblock
\APACrefYearMonthDay{1999}{{\APACmonth{02}}}{}.
\newblock
{\BBOQ}\APACrefatitle {Optimal control of option portfolios and applications}
  {Optimal control of option portfolios and applications}.{\BBCQ}
\newblock
\APACjournalVolNumPages{OR-Spektrum}{21}{1}{123--146}.
\newblock
\begin{APACrefURL} \url{https://doi.org/10.1007/s002910050084} \end{APACrefURL}
\PrintBackRefs{\CurrentBib}

\bibitem [\protect \citeauthoryear {%
Kraft%
\ \BBA {} Steffensen%
}{%
Kraft%
\ \BBA {} Steffensen%
}{%
{\protect \APACyear {2013}}%
}]{%
Kraft2013}
\APACinsertmetastar {%
Kraft2013}%
\begin{APACrefauthors}%
Kraft, H.%
\BCBT {}\ \BBA {} Steffensen, M.%
\end{APACrefauthors}%
\unskip\
\newblock
\APACrefYearMonthDay{2013}{}{}.
\newblock
{\BBOQ}\APACrefatitle {{A Dynamic Programming Approach to Constrained
  Portfolios}} {{A Dynamic Programming Approach to Constrained
  Portfolios}}.{\BBCQ}
\newblock
\APACjournalVolNumPages{European Journal of Operational
  Research}{229}{2}{453--461}.
\newblock
\begin{APACrefURL} \url{https://doi.org/10.1016/j.ejor.2013.02.039}
  \end{APACrefURL}
\PrintBackRefs{\CurrentBib}

\bibitem [\protect \citeauthoryear {%
Larsen%
\ \BBA {} Munk%
}{%
Larsen%
\ \BBA {} Munk%
}{%
{\protect \APACyear {2012}}%
}]{%
Larsen2012}
\APACinsertmetastar {%
Larsen2012}%
\begin{APACrefauthors}%
Larsen, L\BPBI S.%
\BCBT {}\ \BBA {} Munk, C.%
\end{APACrefauthors}%
\unskip\
\newblock
\APACrefYearMonthDay{2012}{}{}.
\newblock
{\BBOQ}\APACrefatitle {The costs of suboptimal dynamic asset allocation:
  General results and applications to interest rate risk, stock volatility
  risk, and growth/value tilts} {The costs of suboptimal dynamic asset
  allocation: General results and applications to interest rate risk, stock
  volatility risk, and growth/value tilts}.{\BBCQ}
\newblock
\APACjournalVolNumPages{Journal of Economic Dynamics and
  Control}{36}{2}{266-293}.
\newblock
\begin{APACrefURL} \url{https://doi.org/10.1016/j.jedc.2011.09.009}
  \end{APACrefURL}
\PrintBackRefs{\CurrentBib}

\bibitem [\protect \citeauthoryear {%
Li%
, Rong%
\BCBL {}\ \BBA {} Zhao%
}{%
Li%
\ \protect \BOthers {.}}{%
{\protect \APACyear {2014}}%
}]{%
Li2014}
\APACinsertmetastar {%
Li2014}%
\begin{APACrefauthors}%
Li, D.%
, Rong, X.%
\BCBL {}\ \BBA {} Zhao, H.%
\end{APACrefauthors}%
\unskip\
\newblock
\APACrefYearMonthDay{2014}{}{}.
\newblock
{\BBOQ}\APACrefatitle {Optimal reinsurance–investment problem for maximizing
  the product of the insurer’s and the reinsurer’s utilities under a CEV
  model} {Optimal reinsurance–investment problem for maximizing the product
  of the insurer’s and the reinsurer’s utilities under a cev model}.{\BBCQ}
\newblock
\APACjournalVolNumPages{Journal of Computational and Applied
  Mathematics}{255}{}{671 - 683}.
\newblock
\begin{APACrefURL} \url{https://doi.org/10.1016/j.cam.2013.06.033}
  \end{APACrefURL}
\PrintBackRefs{\CurrentBib}

\bibitem [\protect \citeauthoryear {%
Liang%
, Yuen%
\BCBL {}\ \BBA {} Guo%
}{%
Liang%
\ \protect \BOthers {.}}{%
{\protect \APACyear {2011}}%
}]{%
Liang2011}
\APACinsertmetastar {%
Liang2011}%
\begin{APACrefauthors}%
Liang, Z.%
, Yuen, K\BPBI C.%
\BCBL {}\ \BBA {} Guo, J.%
\end{APACrefauthors}%
\unskip\
\newblock
\APACrefYearMonthDay{2011}{}{}.
\newblock
{\BBOQ}\APACrefatitle {Optimal proportional reinsurance and investment in a
  stock market with Ornstein–Uhlenbeck process} {Optimal proportional
  reinsurance and investment in a stock market with ornstein–uhlenbeck
  process}.{\BBCQ}
\newblock
\APACjournalVolNumPages{Insurance: Mathematics and Economics}{49}{2}{207 -
  215}.
\newblock
\begin{APACrefURL} \url{https://doi.org/10.1016/j.insmatheco.2011.04.005}
  \end{APACrefURL}
\PrintBackRefs{\CurrentBib}

\bibitem [\protect \citeauthoryear {%
Luo%
, Taksar%
\BCBL {}\ \BBA {} Tsoi%
}{%
Luo%
\ \protect \BOthers {.}}{%
{\protect \APACyear {2008}}%
}]{%
Luo2008}
\APACinsertmetastar {%
Luo2008}%
\begin{APACrefauthors}%
Luo, S.%
, Taksar, M.%
\BCBL {}\ \BBA {} Tsoi, A.%
\end{APACrefauthors}%
\unskip\
\newblock
\APACrefYearMonthDay{2008}{}{}.
\newblock
{\BBOQ}\APACrefatitle {On reinsurance and investment for large insurance
  portfolios} {On reinsurance and investment for large insurance
  portfolios}.{\BBCQ}
\newblock
\APACjournalVolNumPages{Insurance: Mathematics and Economics}{42}{1}{434 -
  444}.
\newblock
\begin{APACrefURL} \url{https://doi.org/10.1016/j.insmatheco.2007.04.002}
  \end{APACrefURL}
\PrintBackRefs{\CurrentBib}

\bibitem [\protect \citeauthoryear {%
Merton%
}{%
Merton%
}{%
{\protect \APACyear {1969}}%
}]{%
merton1969}
\APACinsertmetastar {%
merton1969}%
\begin{APACrefauthors}%
Merton, R.%
\end{APACrefauthors}%
\unskip\
\newblock
\APACrefYearMonthDay{1969}{02}{}.
\newblock
{\BBOQ}\APACrefatitle {{Lifetime Portfolio Selection under Uncertainty: The
  Continuous-Time Case}} {{Lifetime Portfolio Selection under Uncertainty: The
  Continuous-Time Case}}.{\BBCQ}
\newblock
\APACjournalVolNumPages{The Review of Economics and Statistics}{51}{}{247-57}.
\newblock
\begin{APACrefURL} \url{https://doi.org/10.2307/1926560} \end{APACrefURL}
\PrintBackRefs{\CurrentBib}

\bibitem [\protect \citeauthoryear {%
Merton%
}{%
Merton%
}{%
{\protect \APACyear {1971}}%
}]{%
merton1971}
\APACinsertmetastar {%
merton1971}%
\begin{APACrefauthors}%
Merton, R.%
\end{APACrefauthors}%
\unskip\
\newblock
\APACrefYearMonthDay{1971}{}{}.
\newblock
{\BBOQ}\APACrefatitle {{Optimum consumption and portfolio rules in a
  continuous-time model}} {{Optimum consumption and portfolio rules in a
  continuous-time model}}.{\BBCQ}
\newblock
\APACjournalVolNumPages{Journal of Economic Theory}{4}{3}{373-413}.
\newblock
\begin{APACrefURL} \url{https://doi.org/10.1016/0022-0531(71)90038-X}
  \end{APACrefURL}
\PrintBackRefs{\CurrentBib}

\bibitem [\protect \citeauthoryear {%
Müller%
}{%
Müller%
}{%
{\protect \APACyear {1985}}%
}]{%
Mueller1985}
\APACinsertmetastar {%
Mueller1985}%
\begin{APACrefauthors}%
Müller, H\BPBI H.%
\end{APACrefauthors}%
\unskip\
\newblock
\APACrefYearMonthDay{1985}{}{}.
\newblock
{\BBOQ}\APACrefatitle {Investment policies and reinsurance for pension funds}
  {Investment policies and reinsurance for pension funds}.{\BBCQ}
\newblock
\APACjournalVolNumPages{Insurance: Mathematics and Economics}{4}{2}{123 - 127}.
\newblock
\begin{APACrefURL} \url{https://doi.org/10.1016/0167-6687(85)90006-X}
  \end{APACrefURL}
\PrintBackRefs{\CurrentBib}

\bibitem [\protect \citeauthoryear {%
Nguyen%
\ \BBA {} Stadje%
}{%
Nguyen%
\ \BBA {} Stadje%
}{%
{\protect \APACyear {2020}}%
}]{%
Nguyen2020}
\APACinsertmetastar {%
Nguyen2020}%
\begin{APACrefauthors}%
Nguyen, T.%
\BCBT {}\ \BBA {} Stadje, M.%
\end{APACrefauthors}%
\unskip\
\newblock
\APACrefYearMonthDay{2020}{04}{}.
\newblock
{\BBOQ}\APACrefatitle {{Nonconcave Optimal Investment with Value-at-Risk
  Constraint: An Application to Life Insurance Contracts}} {{Nonconcave Optimal
  Investment with Value-at-Risk Constraint: An Application to Life Insurance
  Contracts}}.{\BBCQ}
\newblock
\APACjournalVolNumPages{SIAM Journal on Control and Optimization}{}{}{}.
\newblock
\begin{APACrefURL} \url{https://doi.org/10.1137/18M1217322} \end{APACrefURL}
\PrintBackRefs{\CurrentBib}

\bibitem [\protect \citeauthoryear {%
Teplá%
}{%
Teplá%
}{%
{\protect \APACyear {2001}}%
}]{%
Tepla2001}
\APACinsertmetastar {%
Tepla2001}%
\begin{APACrefauthors}%
Teplá, L.%
\end{APACrefauthors}%
\unskip\
\newblock
\APACrefYearMonthDay{2001}{}{}.
\newblock
{\BBOQ}\APACrefatitle {Optimal investment with minimum performance constraints}
  {Optimal investment with minimum performance constraints}.{\BBCQ}
\newblock
\APACjournalVolNumPages{Journal of Economic Dynamics and Control}{25}{10}{1629
  - 1645}.
\newblock
\begin{APACrefURL} \url{https://doi.org/10.1016/S0165-1889(99)00066-4}
  \end{APACrefURL}
\PrintBackRefs{\CurrentBib}

\end{thebibliography}


\begin{appendix}
\section{Proofs of main results}\label{app:proofs_main}
\begin{proof}[Proof of Proposition \ref{prop:pi_relations}]
The dynamics of the insurer's portfolio with respect to $S_0, S_1, P$ is given by:
	\begin{equation}\label{eq:PF_original}
	\begin{aligned}
		d\bar{V}^{v_0, \bar \pi}(t) &= \phib_0(t)dS_0(t) + \phib_1(t)dS_1(t) + \phib_2(t)dP(t)\\
		& =  \phib_0(t)S_0(t)r dt + \phib_1(t)S_1(t)(\mu_1 dt + \sigma_1 dW_1(t))\\
		& \quad + \phib_2(t) \left(((\Phi(d_+) - 1)V^{v_0, \pi_B}(t) \pi_B^{CM} (\mu_2 - r) + rP(t))dt\right.\\
		& \quad + \left. (\Phi(d_+) - 1)V^{v_0, \pi_B}(t) \pi_B^{CM} \sigma_2 \left(\rho dW_1(t) + \sqrt{1 - \rho^2}dW_2(t)\right)\right)\\
		& = \left( \phib_0(t)S_0(t) r + \phib_1(t) S_1(t)\mu_1 + \phib_2(t)\left(\left(\Phi(d_+) - 1\right)V^{v_0, \pi_B}(t) \pi_B^{CM}(\mu_2 - r) + rP(t)\right) \right)dt\\
		& \quad + \lrr{\phib_1(t)S_1(t)\sigma_1 + \phib_2(t)(\Phi(d_+) - 1)V^{v_0, \pi_B}(t) \pi_B^{CM}\sigma_2 \rho}dW_1(t) \\
		& \quad + \lrr{\phib_2(t)(\Phi(d_+) - 1)V^{v_0, \pi_B}(t) \pi_B^{CM}\sigma_2 \sqrt{1 - \rho^2}} dW_2(t)
	\end{aligned}
	\end{equation}
	
	The dynamics of a portfolio with respect to $S_0, S_1, S_2$ is given by:
	\begin{equation}\label{eq:PF_transformed}
		\begin{aligned}
			dV^{v_0, \pi}(t) &= \varphi_0(t)dS_0(t) + \varphi_1(t)dS_1(t) + \varphi_2(t)dS_2(t)\\
			& =  \varphi_0(t)S_0(t)r dt + \varphi_1(t)S_1(t) (\mu_1 dt + \sigma_1 dW_1(t)) \\
			& \quad + \varphi_2(t)  S_2(t) (\mu_2 dt + \sigma_2(\rho dW_1(t) + \sqrt{1 - \rho^2}dW_2(t))) \\
			& = \left(\varphi_0(t)S_0(t)r + \varphi_1(t)S_1(t)\mu_1 + \varphi_2(t) S_2(t)\mu_2\right)dt \\
			& \quad + \left(\varphi_1(t)S_1(t)\sigma_1 + \varphi_2(t)S_2(t)\sigma_2\rho\right)dW_1(t)\\
			& \quad + \lrr{\varphi_2(t)S_2(t)\sigma_2 \sqrt{1 - \rho^2}}dW_2(t)
		\end{aligned}
	\end{equation}
	
	Equating the coefficients next to the terms $dt$, $dW_1(t)$ and $dW_2(t)$ in \eqref{eq:PF_original} and \eqref{eq:PF_transformed} we get the following link:
	
    	
	\begin{equation}\label{eq:phi_relations}
	\left\{
	\begin{aligned}
		& \phib_0(t) = \varphi_ 0(t) + \varphi_2(t) \frac{S_2(t)}{S_0(t)} \lrr{1 - \frac{P(t)}{\pi_B^{CM} V^{v_0, \pi_B}(t)(\Phi(d_+) - 1)}}\\
		& \phib_1(t) = \varphi_1(t)\\
		& \phib_2(t) = \frac{\varphi_2(t)S_2(t)}{\pi_B^{CM} V^{v_0, \pi_B}(t) (\Phi(d_+) - 1)}
	\end{aligned}
	\right.
	\end{equation} 
	
	Using \eqref{eq:phi_relations} and the relation \eqref{eq:link_pi_phi} between
	the investment strategies $\varphi (\phib)$ and the relative portfolio processes $\pi (\pib)$, the claim follows.
\end{proof}

\begin{proof}[Proof of Proposition \ref{prop:sol_VaR_pi_link_original_transformed}]
	We prove the claim by contradiction.
	    
	Let $\pi^*$ be the solution to \eqref{OP_P_epsilon_K}. Then $\pib^* =\lrr{A(t)\pi^*(t)}_{\tin} \in \bar{\mathcal{A}}\lrr{v_0, \bar{K}_V(\varepsilon), \bar{K}_{\pi}}$ according to \eqref{relation_K_bar_K}. 
	    
	    Assume that $\exists \,\, \pib^{**} \in \bar{\mathcal{A}}\lrr{v_0, \bar{K}_V(\varepsilon), \bar{K}_{\pi}}$
	    such that:
	    \begin{equation}\label{ineq:false_assumption}
	        \EQ{U\lrr{\bar{V}^{v_0, \pib^{**}}(T)}} > \EQ{U\lrr{\bar{V}^{v_0, \pib^{*}}(T)}}.
	    \end{equation}
	    
	    Then $\pi^{**} := \lrr{A^{-1}(t)\pib^{**}(t)}_{\tin} \in \mathcal{A}\lrr{v_0, K_V(\varepsilon), K_{\pi}}$
	    and:
	    \begin{equation*}
	    \begin{aligned}
	        \EQ{U\lrr{V^{v_0, \pi^{**}}(T)}} \stackrel{\text{Prop.\ref{prop:pi_relations}}}{=}  \EQ{U\lrr{\bar{V}^{v_0, \pib^{**}}(T)}} & \stackrel{\eqref{ineq:false_assumption}}{>}  \EQ{U\lrr{\bar{V}^{v_0, \pib^{*}}(T)}}\\
	        & \stackrel{\text{Prop.\ref{prop:pi_relations}}}{=} \EQ{U\lrr{V^{v_0, \pi^{*}}(T)}},
	    \end{aligned}
	    \end{equation*}
	    which contradicts the optimality of $\pi^{*}$ for \eqref{OP_P_epsilon_K}. The claims follows.
	
	\end{proof}

\begin{proof}[Proof of Corollary \ref{cor: summary karatzas}]
Recall that we are considering a power utility function $U(x) = \frac{1}{b}x^b$ with $b<1, \ b \neq 0$ and the set of allocation constraints $K_{\pi}$ is a convex cone. 

For the case $0<b<1$, the statement of Corollary \ref{cor: summary karatzas} is an immediate consequence of Theorem 10.1 and Theorem 15.3 from \citeA{Cvitanic1992}. 

For the case $b<0$, we momentarily emphasize the explicit dependence of $\hat{\pi}_{\lambda}$ as in (\ref{def: pi hat lambda})  on $b$ as \enquote{$\hat{\pi}_{\lambda}(b)$}. Let $b = b^{-}<0$ and $b^{+}\in (0,1)$ be arbitrary but fixed. 

Then, $\hat{\pi}_{\lambda}(b^{+})$ solves $(P_{1, K_{\pi}})$ with risk aversion parameter $b^{+}$, and $\hat{\pi}_{\lambda}(b^{+})$ satisfies equations (\ref{eq: Condition b}). Clearly, this implies 

\begin{equation*}
    \hat{\pi}_{\la}(b^{-}) = \underbrace{\frac{1-b^{+}}{1-b^{-}}}_{\geq 0} \underbrace{\hat{\pi}_{\la}(b^{+})}_{\ \in K_{\pi}} \in K_{\pi},
\end{equation*}
because $K_{\pi}$ is a convex cone, as well as
\begin{equation*}
    \hat{\pi}_{\la}(b^{-})'\la = \frac{1-b^{+}}{1-b^{-}} \hat{\pi}_{\la}(b^{+})'\la \overset{(\ref{eq: Condition b})}{=}0.
\end{equation*}

Moreover, $\hat{\pi}_{\la}(b^{-})$ is the optimal portfolio process for $(P^{\la}_{1})$ with power utility with $b = b^{-}$. Hence, $\hat{\pi}_{\la}(b^{-})$ satisfies \eqref{eq: Condition b} and is optimal for the primal problem $(P_{1, K_{\pi}})$.
\end{proof}

\begin{proof}[Proof of Corollary \ref{cor: pp VaR constraints, auxiliary market}]
The problem considered in Corollary \ref{cor: pp VaR constraints, auxiliary market} is precisely as in Section 2 of \citeA{Basak2001}. We relate $(y,\tilde{Z}_{\lambda}(t))$ and $(v_{f_{\lambda}},V^{v_{f_{\lambda}}, \hat{\pi}_{\lambda}}_{\lambda}(t))$ for arbitrary $v_{f_{\lambda}}, y \geq 0$ in such a way that the statements of Corollary \ref{cor: pp VaR constraints, auxiliary market} follow directly from \citeA{Basak2001}. \\
 For any $v_{f_{\lambda}} \geq 0$, the optimal unconstrained wealth process $V^{v_{f_{\lambda}}, \hat{\pi}_{\lambda}}_{\lambda}(t)$, $\tin$ is given as 
\begin{align*}
    V^{v_{f_{\lambda}}, \hat{\pi}_{\lambda}}_{\lambda}(t) &= v_{f_{\lambda}}\exp \Big \{ \big(r+\underbrace{(\mu +\lambda -r\mathbbm{1})'\hat{\pi}_{\lambda}}_{= \frac{1}{1-b}\Vert \gamma_{\lambda} \Vert^2} - \underbrace{\frac{1}{2}\Vert \sigma'\hat{\pi}_{\lambda}\Vert^2}_{= \frac{1}{2(1-b)^2} \Vert \gamma_{\lambda}\Vert^2} \big)t + \underbrace{\hat{\pi}_{\lambda}'\sigma}_{= \frac{1}{1-b}\gamma_{\lambda}}W(t)\Big \} \\
    &= v_{f_{\lambda}}\exp \Big \{ \big(r-\frac{1}{b-1}\Vert \gamma_{\lambda} \Vert^2- \frac{1}{2(b-1)^2} \Vert \gamma_{\lambda}\Vert^2 \big)t -\frac{1}{b-1} \gamma_{\lambda}'W(t) \Big \} \\
    &= v_{f_{\lambda}} \big(\tilde{Z}_{\lambda}(t)\big)^{\frac{1}{b-1}}\exp \Big \{ \big(r-\frac{1}{b-1}\Vert \gamma_{\lambda} \Vert^2- \frac{1}{2(b-1)^2} \Vert \gamma_{\lambda}\Vert^2 + \frac{1}{b-1}(r+ \frac{1}{2}\Vert \gamma_{\lambda} \Vert^2) \big)t \Big \} \\
    &= v_{f_{\lambda}} \big(\tilde{Z}_{\lambda}(t)\big)^{\frac{1}{b-1}} \exp \Big \{\frac{b}{b-1}\big(r + \frac{1}{2}\Vert \gamma_{\lambda} \Vert^2 \big)t - \underbrace{\big(\frac{1}{2} + \frac{1}{b-1} + \frac{1}{2(b-1)^2}\big)}_{=\frac{b^2}{(b-1)^2}}\Vert \gamma_{\lambda}\Vert^2t \Big \} \\
    &= v_{f_{\lambda}} \big(\tilde{Z}_{\lambda}(t)\big)^{\frac{1}{b-1}} \exp \Big \{\underbrace{\frac{b}{b-1}\big(r + \frac{1}{2}\Vert \gamma_{\lambda} \Vert^2 \big)t - \big(\frac{b}{b-1}\big)^2\frac{1}{2}\Vert \gamma_{\lambda}\Vert^2t}_{\stackrel{\eqref{eq:Gamma_d_1_d_2}}{=} \Gamma_{\lambda}(t)- \Gamma_{\lambda}(0)} \Big \} \\
    &= v_{f_{\lambda}} \big(\tilde{Z}_{\lambda}(t)\big)^{\frac{1}{b-1}} \exp \Big \{ \Gamma_{\lambda}(t)- \Gamma_{\lambda}(0) \Big \}.
\end{align*}
We set $y:= \big(v_{f_{\lambda}}e^{\Gamma_{\lambda}(0)}\big)^{b-1}$ and can thus obtain any $y>0$ by choosing a suitable $v_{f_{\lambda}} >0$. Moreover, this gives us the relation
\begin{align}\label{eq:optimal_uncostrained_wealth}
    V^{v_{f_{\lambda}}, \hat{\pi}_{\lambda}}_{\lambda}(t)  = v_{f_{\lambda}} \big(\tilde{Z}_{\lambda}(t)\big)^{\frac{1}{b-1}} \exp \Big \{ \Gamma_{\lambda}(t)- \Gamma_{\lambda}(0) \Big \} = \big(y\tilde{Z}_{\lambda}(t)\big)^{\frac{1}{b-1}} e^{ \Gamma_{\lambda}(t)}.
\end{align}
By plugging \eqref{eq:optimal_uncostrained_wealth} into Proposition 1 and Proposition 3 from \citeA{Basak2001} and rewriting their definition of $d_1$ and $d_2$ in terms of $k^{\varepsilon}_{\lambda}$ and $G_T$, the statements of Corollary \ref{cor: pp VaR constraints, auxiliary market} follow immediately.
\end{proof}

\begin{proof}[Proof of Proposition \ref{prop: generalization no-shortsell}]\ \\
    As per auxuliary Lemma \ref{lem: delta-hedge} in Appendix \ref{app:proofs_auxiliary}, the portfolio $\pi^{\ast}$ that attains $D^{f}$ can be determined by delta-hedging and $\pi^{\ast}$ is of the form 
    
    \begin{equation*}
        \pi^{\ast}(t) = \alpha_f(t,V^{v_f,\hat{\pi}_{\la}}(t)) \cdot \hat{\pi}_{\la},\,\tin.
    \end{equation*}
    
    We continue by verifying that this candidate portfolio $\pi^{\ast}$ is optimal for (\ref{eq:OP_I_VaR_pi_original_post_transformation_auxiliary_market}) with $\lambda = \la$,  satisfies \eqref{eq: Condition b} and is therefore optimal for \eqref{OP_P_epsilon_K} by Lemma \ref{lem: condition b}: \\
    
    For this purpose, Corollary \ref{cor: summary karatzas} provides us with useful information about $\hat{\pi}_{\la}$ and $\la$:
    
	
	\begin{equation} \label{eq: generalization slackness condition}
	\hat{\pi}_{\la} \in K_{\pi} \quad \text{and}\quad (\la)'\hat{\pi}_{\la} = 0 \quad \Q-a.s..
	\end{equation}

    The multiplicative structure of $\pi^{\ast}$ and (\ref{eq: generalization slackness condition}) has the convenient implication that
    
    \begin{equation*}
(\la)'\pi^{\ast}(t) = \alpha^f(t,V^{v_f,\hat{\pi}_{\la}}(t)) \cdot (\la)'\hat{\pi}_{\la}= 0 \quad \forall \tin, \ \Q-a.s.,
    \end{equation*}
    
    which means that the wealth processes of an investor trading according to $\pi^{\ast}$ in $\mathcal{M}_{0}$ and $\mathcal{M}_{\la}$ coincide. In other words,
    \begin{equation*}
    V^{v_0, \pi^{\ast}}(t) = V_{\la}^{v_0, \pi^{\ast}}(t) \quad \forall \tin\, \Q-a.s..
    \end{equation*}
    
    Hence, the respective terminal values at time T for both processes satisfy the same budget equation
    and VaR-constraint
    \begin{equation*}
        \Q(f(V^{v_f, \hat{\pi}_{\la}}_{\la}(T)) < G_T) = \Q(f(V^{v_f, \hat{\pi}_{\la}}(T)) < G_T)  = \ \varepsilon.
    \end{equation*}
	
	By comparing $D^f$ with $D^f_{\la}$ from Corollary \ref{cor: pp VaR constraints, auxiliary market}, we realize that $f = f_{\la}$, $v_f = v_{f_{\la}}$, $k^{\epsilon} = k^{\varepsilon}_{\la}$ and thus
	
	\begin{equation*}
	    D^f = f(V^{v_f, \hat{\pi}_{\la}}(T))= f_{\la}(V^{v_{f_{\la}}, \hat{\pi}_{\la}}_{\la}(T)) = D^{f_{\la}}_{\la}
	\end{equation*}
    
     is the optimal terminal wealth for (\ref{eq:OP_I_VaR_pi_original_post_transformation_auxiliary_market}). Moreover, $\pi^{\ast}= \pi^{\ast}_{\la}$ is the corresponding optimal portfolio process and Corollary \ref{cor: pp VaR constraints, auxiliary market} provides us with an explicit expression for $\pi^{\ast}$ (and thus for $\alpha^f$):
    \begin{equation*}
        \begin{aligned}
             \alpha^f(t,V^{v_f,\hat{\pi}_{\la}}(t))  \cdot &\hat{\pi}_{\la} = \pi^{\ast}(t) = \pi^{\ast}_{\la}(t) = \alpha^{f_{\la}}_{\la}(t, V^{v_{f_{\la}}, \hat{\pi}_{\la}}_{\la}(t)) \cdot \hat{\pi}_{\la} = \alpha^{f_{\la}}_{\la}(t, V^{v_f, \hat{\pi}_{\la}}(t)) \cdot \hat{\pi}_{\la} \\
             & \Rightarrow \alpha^f(t,V) = \alpha^{f_{\la}}_{\la}(t,V) \geq  0 \quad \forall (t,V) \in [0,T]\times[0,\infty) \quad \text{or} \quad \hat{\pi}_{\la} = 0
        \end{aligned}
    \end{equation*}
    
   Since $\hat{\pi}_{\la} \in K_{\pi}$, $K_{\pi} = [0, \infty)\times (-\infty,0]$ is a convex cone and $\alpha^{f_{\la}}_{\la}(t,V) \geq 0$, we obtain that $\pi^{\ast}(t) \in K_{\pi}$ $\Q$-a.s. $\forall \tin$. \\
    
    In summary, we have shown that $\pi^{\ast}=(\pi^{\ast}(t))_{\tin}$ is optimal for (\ref{eq:OP_I_VaR_pi_original_post_transformation_auxiliary_market}), $\pi^{\ast}(t) \in K_{\pi}$  $\forall \tin$ $\Q$-a.s., and $(\la)'\pi^{\ast}(t) =  0$ $\forall \tin$ $\Q$-a.s.. Hence, $\pi^{\ast}$ and $\la$ satisfy \eqref{eq: Condition b} and  $\pi^{\ast}$ is optimal for the primal problem (\ref{OP_P_epsilon_K}).
\end{proof}
	
	\begin{proof}[Proof of Proposition \ref{prop:optimal_reinsurance_positive}]\ \\
	Observe that:
	\begin{equation*}
		\begin{aligned}
			\pib_2^*(t) > 0 &\stackrel[\eqref{eq:A_t}]{\text{Prop.}\ref{prop:sol_VaR_pi_link_original_transformed}}{\iff} \underbrace{\frac{P(t)}{\pi_B^{CM} V^{v_0, \pi_B}(t)(\Phi(d_+) - 1)}}_{< 0}\pi_2
			^*(t) > 0 \iff \pi_2^*(t) < 0\\ & \stackrel{\eqref{eq:piStar_pHat_link}}{\iff} \alpha^{f_{\la}}_{\la}(t, V^{\nu_f,\hat{\pi}_{\la}}(t)) \cdot  \hat{\pi}_{\la, 2} < 0 
			\stackrel[\alpha > 0]{\text{Prop.}\ref{prop: generalization no-shortsell}}{\iff} \hat{\pi}_{\la,2}(t) < 0 \\
			& \stackrel{\eqref{def: pi hat lambda}}{\iff} \left(\frac{1}{1-b}C^{-1}(\mu + \la -r\cdot \mathbbm{1})\right)'
			\left(
    		\begin{matrix}
    		 0 \\ 
    		 1
    		 \end{matrix}
    		 \right) < 0 .
		\end{aligned}
	\end{equation*}
			The inverse of the volatility matrix is given by:
	\begin{equation*}
		\sigma^{-1} = \frac{1}{\sigma_1 \sigma_2 \sqrt{1 - \rho^2}}
		\left(
		\begin{matrix}
		\sigma_2 \sqrt{1 - \rho^2} & 0 \\ 
		-\sigma_2 \rho & \sigma_1
		\end{matrix} 
		\right)
	\end{equation*}
    and thus:
	\begin{equation*}
		\begin{aligned}
			 C^{-1}(\mu + \la - r \bar{1}) &= \lrr{\sigma^{-1}}' \sigma^{-1}(\mu + \la - r \bar{1})\\
			 & =  \frac{1}{\sigma_1 \sigma_2 \sqrt{1 - \rho^2}}
		\left(
		\begin{matrix}
		\sigma_2 \sqrt{1 - \rho^2} & -\sigma_2 \rho \\ 
		0 & \sigma_1
		\end{matrix} 
		\right)  \frac{1}{\sigma_1 \sigma_2 \sqrt{1 - \rho^2}}
		\left(
		\begin{matrix}
		\sigma_2 \sqrt{1 - \rho^2} & 0 \\ 
		-\sigma_2 \rho & \sigma_1
		\end{matrix} 
		\right)\\
		& \quad \cdot (\mu + \la - r \bar{1}) \\
		&=   \frac{1}{\sigma_1^2 \sigma_2^2 (1 - \rho^2)}
		\left(
		\begin{matrix}
		\sigma_2^2 & -\sigma_1\sigma_2 \rho \\ 
		-\sigma_1\sigma_2 \rho & \sigma_1^2
		\end{matrix}
		\right)
		\left(
		\begin{matrix}
		 \mu_1 + \la_1 - r \\ 
		 \mu_2 + \la_2  - r
		 \end{matrix}
		 \right)  \\
		&=   \frac{1}{\sigma_1^2 \sigma_2^2 (1 - \rho^2)}
		\left(
		\begin{matrix}
		\sigma_2^2 (\mu_1 + \la_1- r) - \sigma_1\sigma_2\rho(\mu_2 + \la_2- r)  \\ 
		\sigma_1^2 (\mu_2 + \la_2 - r) - \sigma_1\sigma_2\rho(\mu_1 + \la_1 - r)
		\end{matrix}
		\right)
		\end{aligned}
	\end{equation*}
	
	Hence, we obtain:
	\begin{equation*}
	    \begin{aligned}
	    \left(\frac{1}{1-b}C^{-1}(\mu + \la -r\cdot \mathbbm{1})\right)'&
	        \left(
    		\begin{matrix}
    		 0 \\ 
    		 1
    		 \end{matrix}
    		 \right) < 0\\
	        &\stackrel[\sigma_1^2 \sigma_2^2 (1 - \rho^2) > 0]{1 - b > 0}{\iff} \sigma_1^2 (\mu_2 + \la_2 - r) - \sigma_1\sigma_2\rho(\mu_1 + \la_1 - r) < 0\\
			&\iff \frac{\mu_2 + \la_2- r}{\sigma_2} < \rho \cdot \frac{\mu_1 + \la_1 - r}{\sigma_1} \iff SR_2^{\lambda^{\ast}} < \rho \cdot SR_1^{\lambda^{\ast}}.
	    \end{aligned}
	\end{equation*}
		\end{proof}

\section{Proofs of auxiliary results}\label{app:proofs_auxiliary}
This appendix contains proofs of auxiliary lemmas and propositions needed for Sections \ref{sec_solution_to_OP} and \ref{sec_numerical_studies}.

\begin{lemma}[Delta-hedging in $\mathcal{M}_{\lambda}$]\label{lem: delta-hedge}
Fix a dual vector $\lambda \in K_{\pi}$ and consider the following derivative in $\mathcal{M}_{\lambda}$
\begin{equation*}
    D^f_{\lambda} := f(V_{\lambda}^{v_f, \pi}(T))
\end{equation*}

for a given function $f:[0,\infty)\rightarrow [0,\infty)$, initial wealth $v_f > 0$ and constant-mix strategy $\pi\in \R^2$. Let

\begin{align*}
    D^f_{\lambda}(t,V) := e^{-r(T-t)} \mathbb{E}_{\tilde{\Q}_{\lambda}}[D^f_{\lambda} | \ V^{v_f,\pi}_{\lambda}(t) = V ] 
\end{align*} denote the time-t value of $D^f_{\lambda}$, provided that $V_{\lambda}^{v_f, \pi}(t) = V$. Furthermore, assume that $D^f_{\lambda}(t,V) \in C^{(1 \times 2)}\big([0,T)\times (0,\infty)\big)$. Then, $D_{\lambda}^f$ can be attained by trading in $\mathcal{M}_{\lambda}$  according to the portfolio process 
\begin{equation*}
    \pi^{\ast}_{\lambda}(t):= \pi^{\ast}_{\lambda}(t,V_{\lambda}^{v_f, \\
    \pi}(t)) = \underbrace{\frac{\frac{d}{dV}D^f_{\lambda}(t,V_{\lambda}^{v_f,\pi}(t))\cdot V_{\lambda}^{v_f, \pi}(t)}{D^f_{\lambda}(t,V_{\lambda}^{v_f, \pi}(t))}}_{=: \alpha^f_{\lambda}(t,V_{\lambda}^{v_f, \pi}(t))}\cdot \pi = \alpha^f_{\lambda}(t,V_{\lambda}^{v_f, \pi}(t)) \cdot \pi
\end{equation*}
with initial wealth 
\begin{equation*}
v_0 = D^f_{\lambda}(0,v_f).    
\end{equation*}
\end{lemma}

\begin{proof} \ \\
	Since $\pi$ is a constant-mix strategy, the corresponding  wealth process $V^{v_f,\pi}_{\lambda}$ in $\mathcal{M}_{\lambda}$ has the dynamics of a geometric Brownian motion:
	
	\begin{equation*}
	\begin{aligned}
	V^{v_f,\pi}_{\lambda}(0) =& v_f \\
	\frac{dV^{v_f,\pi}_{\lambda}(t)}{V^{v_f,\pi}_{\lambda}(t)} =& [r + (\mu + \lambda-r\cdot \mathbbm{1})'\pi]dt + \pi'\sigma dW(t) \\
	=& r dt + \pi'\sigma \underbrace{(dW(t) + [\gamma + \sigma^{-1}\lambda] dt)}_{=: d\tilde{W}_{\lambda}(t)}
	\end{aligned}
	\end{equation*}
    Since $D^f_{\lambda} \in C^{(1,2)}([0,T)\times[0,\infty))$, we can use It\^{o}'s formula to determine the hedging portfolio for $D^f_{\lambda}$. Furthermore, from an application of Feynman-Kac Theorem we know that $D^f_{\lambda}(t, V)$ satisfies the following PDE:
	
	\begin{equation}\label{eq: Black-Scholes PDE}
	\begin{aligned}
	0 &= \frac{d}{dt} D^f_{\lambda} + \frac{1}{2} \Vert \sigma' \pi \Vert^2\cdot V^2 \cdot \frac{d^2}{d^2V} D^f_{\lambda} + r \cdot V \cdot \frac{d}{dV} D^f_{\lambda} - r \cdot D^f_{\lambda} \\
	f(V) &= D^f_{\lambda}(T, V)
	\end{aligned}
	\end{equation}
	
	Let $\pi^{\ast}_{\lambda}$ be the portfolio process that attains $D^f_{\lambda}$ in $\mathcal{M}_{\lambda}$. The existence of $\pi^{\ast}_{\lambda}$ is guaranteed by the market completeness of $\mathcal{M}_{\lambda}$. \\
	Then, for no-arbitrage reasons,  $V^{ v_0,\pi^{\ast}_{\lambda}}_{\lambda}(t) \overset{!}{=} D^f_{\lambda}(t,V^{v_f,\pi}_{\lambda}(t)) \ \mathcal{L}[0,T)\otimes\Q-a.e.$. In particular, $v_0$ is determined through
	
	\begin{equation*}
	v_0 = D^f_{\lambda}(0,V^{v_f,\pi}_{\lambda}(0)) = D^f_{\lambda}(0,v_f)
	\end{equation*}

	Further, $V_{\lambda}^{v_0, \pi^{\ast}_{\lambda}}(t)$ and $D^f_{\lambda}(t, V^{v_f,\pi}_{\lambda}(t))$ follow the dynamics
	
	\begin{equation*}
	dV_{\lambda}^{v_0, \pi^{\ast}_{\lambda}}(t) = V_{\lambda}^{v_0,\pi^{\ast}_{\lambda}}(t)\cdot [rdt + \pi^{\ast}_{\lambda}(t)'\sigma d\tilde{W}_{\lambda}(t)] = D^f_{\lambda}(t, V^{v_f,\pi}_{\lambda}(t)) \cdot [rdt + \pi^{\ast}_{\lambda}(t)'\sigma d\tilde{W}_{\lambda}(t)]
	\end{equation*}
	and 
	\begin{equation*}
	\begin{aligned}
	dD^f_{\lambda}(t, V^{v_f,\pi}_{\lambda}(t)) \overset{\text{It\^{o}}}{=}& \ \frac{d}{dt}D^f_{\lambda}(t, V^{v_f,\pi}_{\lambda}(t))dt + \frac{d}{dV}D^f_{\lambda}(t, V^{v_f,\pi}_{\lambda}(t))dV^{v_f,\pi}_{\lambda}(t) \\
		 & \qquad + \frac{1}{2}\frac{d^2}{d^2V}D^f_{\lambda}(t, V^{v_f,\pi}_{\lambda}(t))d \langle V^{v_f,\pi}_{\lambda} \rangle_t \\
	=& \frac{d}{dt}D^f_{\lambda}(t, V^{v_f,\pi}_{\lambda}(t))dt + r \cdot V^{v_f,\pi}_{\lambda}(t) \cdot \frac{d}{dV}D^f_{\lambda}(t, V^{v_f,\pi}_{\lambda}(t))dt\\
	 & \qquad +  \frac{d}{dV}D^f_{\lambda}(t, V^{v_f,\pi}_{\lambda}(t))\cdot V^{v_f,\pi}_{\lambda}(t) \cdot (\pi)'\sigma d\tilde{W}_{\lambda}(t)   \\
	& \qquad + \frac{1}{2}\Vert \sigma' \pi\Vert^2\cdot V^{v_f,\pi}_{\lambda}(t)^2 \cdot \frac{d^2}{d^2V}D^f_{\lambda}(t, V^{v_f,\pi}_{\lambda}(t))dt \\
	\overset{(\ref{eq: Black-Scholes PDE})}{=}&  r\cdot  D^f_{\lambda}(t, V^{v_f,\pi}_{\lambda}(t))dt + \frac{d}{dV}D^f_{\lambda}(t, V^{v_f,\pi}_{\lambda}(t))\cdot V^{v_f,\pi}_{\lambda} \cdot(\pi)'\sigma d\tilde{W}_{\lambda}(t).
	\end{aligned}
	\end{equation*}
	
	Matching the diffusion coefficients provides the condition 
	
	\begin{equation*}
	\begin{aligned}
	&D^f_{\lambda}(t,V^{v_f,\pi}_{\lambda}(t))\cdot \sigma' \pi^{\ast}_{\lambda}(t) \overset{!}{=} \frac{d}{dV}D^f_{\lambda}(t, V^{v_f,\pi}_{\lambda}(t)) \cdot V^{v_f,\pi}_{\lambda}(t) \cdot \sigma' \pi \quad \mathcal{L}[0,T]\otimes\Q-a.e. \\
	\Leftrightarrow \ & \pi^{\ast}_{\lambda}(t) \overset{!}{=} \underbrace{\frac{\frac{d}{dV}D^f_{\lambda}(t, V^{v_f,\pi}_{\lambda}(t))\cdot V^{v_f,\pi}_{\lambda}(t)}{D^f_{\lambda}(t,V^{v_f,\pi}_{\lambda}(t))}}_{ =: \ \alpha^f_{\lambda}(t,V^{v_f,\pi}_{\lambda}(t))} \cdot \pi = \alpha^f_{\lambda}(t,V^{v_f,\pi}_{\lambda}(t))\cdot \pi  \quad \mathcal{L}[0,T]\otimes\Q-a.e. .
	\end{aligned}
	\end{equation*}
\end{proof}

\begin{lemma}\label{lem:expectation_ztilde_interval}
    Let $p \in \mathbb{R}$, $-\infty \leq l \leq u \leq +\infty$, $X \stackrel{d}{=}N(0,1)$. Then:
    \begin{equation*}
      \EQ{e^{pX}\indF{X}{(l,u]}} = e^{\frac{p^2}{2}}(\Phi(u - p) - \Phi(l - p)).
    \end{equation*}
    In particular, the moment generating function of X is given by:
        \begin{equation}
            f_{X}(p) := \EQ{e^{pX}} = e^{\frac{1}{2}p^2}.
        \end{equation}
    \end{lemma}
\begin{proof}
    \begin{flalign*}
        \EQ{e^{pX}\indF{X}{(l,u]}} &= \int_{l}^{u}e^{px} e^{-\frac{x^2}{2}} \frac{1}{\sqrt{2\pi}} \, dx =  \int_{l}^{u}e^{px} e^{-\frac{1}{2}(x^2 - 2px + p^2 - p^2)}\frac{1}{\sqrt{2\pi}}\,dx\\
        & = e^{\frac{p^2}{2}} \int_{l}^{u} e^{-\frac{1}{2}(x - p)^2}\frac{1}{\sqrt{2\pi}}\,dx = e^{\frac{p^2}{2}} \int_{l - p}^{u - p} e^{-\frac{y^2}{2}}\frac{1}{\sqrt{2\pi}}\,dy \\
        & = e^{p^2/2}(\Phi(u - p) - \Phi(l - p)).
    \end{flalign*}
    The claim about the moment generating function follows, since:
    $$\lim_{l \downarrow -\infty}\Phi(l - p) = 0,\,\lim_{u \uparrow +\infty}\Phi(u - p) = 1.$$
\end{proof}


\begin{lemma}\label{lem:inequalities_optimal_unconstrained_wealth}
    Let $0 < l \leq u < +\infty$. Then:
    \begin{equation*}
        \begin{aligned}
           l & < & V^{v_{f_{\la}}, \hat{\pi}_{\la}}(T) & \leq & u \qquad \qquad \qquad \qquad \qquad \quad &\\
           \Longleftrightarrow -d^{\la}_2(l, v_{f_{\la}}, 0) - || \gamlamast || \sqrt{T} & < &  \frac{\gamlamast' W(T)}{|| \gamlamast || \sqrt{T}}
            & \leq & -d_2^{\la}(u, v_{f_{\la}}, 0) - || \gamlamast || \sqrt{T}.&
        \end{aligned}
    \end{equation*}
\end{lemma}
\begin{proof}
Using \eqref{eq:optimal_uncostrained_wealth}, we get:
    \begin{flalign*}
        l & < v_{f_{\la}} e^{-\Gamma_{\la}(0)}\left(\ZtillamastT\right)^{\frac{1}{b-1}} \leq u\\
        &\stackrel{\eqref{eq:gamma_Z_for_auxiliary_market}}{\Longleftrightarrow} l < v_{f_{\la}} e^{-\Gamma_{\la}(0)}\left(e^{-\lrr{r + 0.5||\gamlamast||^2}T - \gamlamast' W(T)}\right)^{\frac{1}{b-1}} \leq u\\
        & \stackrel{\frac{v_{f_{\la}}}{e^{\Gamma_{\la}(0)}} > 0}{\Longleftrightarrow} \frac{l e^{\Gamma_{\la}(0)}}{v_{f_{\la}}}<  e^{\frac{1}{1 - b}\lrr{r + 0.5||\gamlamast||^2}T + \frac{1}{1 - b} \gamlamast' W(T)} \leq \frac{u e^{\Gamma_{\la}(0)}}{v_{f_{\la}}} \\
        & \stackrel{\ln(\cdot)\uparrow}{\Longleftrightarrow} \ln \lrr{\frac{l e^{\Gamma_{\la}(0)}}{v_{f_{\la}}}}  - \frac{\lrr{r + 0.5||\gamlamast||^2}T}{1 - b}
        < \frac{\gamlamast' W(T)}{1 - b} \leq \ln \lrr{\frac{u e^{\Gamma_{\la}(0)}}{v_{f_{\la}}}}  - \frac{\lrr{r + 0.5||\gamlamast||^2}T}{1 - b} \\
        & \stackrel{\frac{|| \gamlamast || \sqrt{T}}{1 - b} > 0}{\Longleftrightarrow}
        \frac{(1 - b) \ln \lrr{\frac{l e^{\Gamma_{\la}(0)}}{v_{f_{\la}}}}  - \lrr{r + 0.5||\gamlamast||^2}T}{|| \gamlamast || \sqrt{T}}
        < \frac{\gamlamast' W(T)}{|| \gamlamast || \sqrt{T}}\\
        & \qquad \qquad \qquad \leq \frac{(1 - b) \ln \lrr{\frac{u e^{\Gamma_{\la}(0)}}{v_{f_{\la}}}}  - \lrr{r + 0.5||\gamlamast||^2}T}{|| \gamlamast || \sqrt{T}}\\
        & \stackrel{\eqref{eq:Gamma_d_1_d_2}}{\Longleftrightarrow} -d^{\la}_2(l, v_{f_{\la}}, 0) - || \gamlamast || \sqrt{T} <  \frac{\gamlamast' W(T)}{|| \gamlamast || \sqrt{T}}
        \leq -d_2^{\la}(u, v_{f_{\la}}, 0) - || \gamlamast || \sqrt{T} 
    \end{flalign*}
\end{proof}

\begin{proposition}[Explicit form of equations for calculation of $v_{f_{\la}}, k^{\varepsilon}_{\la}$]\label{prop:v_f_kepsilon_SNLE}$ $\\
    The explicit form of the budget constraint in \eqref{eq: VaR constr. and budget} is given by:
    \begin{equation*}
        \begin{aligned}
           & v_{f_{\la}}\cdot \left( 1 + \Phi(d_1^{\la}(G_T, v_{f_{\la}}, 0)) - \Phi(d_1^{\la}(k^{\varepsilon}_{\la}, v_{f_{\la}}, 0))\right) \\
           & + e^{-rT}G_T \left(\Phi(d_2^{\la}(k^{\varepsilon}_{\la}, v_{f_{\la}}, 0)) - \Phi(d_2^{\la}(G_T, v_{f_{\la}}, 0))  \right) - v_0 = 0;
        \end{aligned}
    \end{equation*}
    The explicit form of the probability constraint in \eqref{eq: VaR constr. and budget} is given by:
    \begin{equation*}
        \Phi(d_2^{\la}(k^{\varepsilon}_{\la}, v_{f_{\la}}, 0)) + \varepsilon - 1 = 0;
    \end{equation*}
\end{proposition}


\begin{proof}
    First, we simplify the budget constraint:
    \begin{flalign*}
        e^{-rT}&\mathbbm{E}_{\tilde{\Q}_{\la}}[f(V^{v_{f_{\la}}, \hat{\pi}_{\la}}(T))] \stackrel{\partial \Q/\partial \tilde{\Q}_{\la}}{=} \mathbbm{E}_{\Q}[\ZtillamastT f(V^{v_{f_{\la}}, \hat{\pi}_{\la}}(T))] \\
        & \stackrel{f \text{ def.}}{=} \EQ{\ZtillamastT\left( V^{v_{f_{\la}}, \hat{\pi}_{\la}}(T)  + \big(G_T - V^{v_{f_{\la}}, \hat{\pi}_{\la}}(T)\big)\mathbbm{1}_{[k^{\varepsilon}_{\la},G_T]}(V^{v_{f_{\la}}, \hat{\pi}_{\la}}(T))\right)}\\
        & \stackrel{\eqref{eq:optimal_uncostrained_wealth}}{=} \EQ{ \ZtillamastT v_{f_{\la}}  e^{-\Gamma_{\la}(0)} \left(\ZtillamastT\right)^{\frac{1}{b-1}} \indF{v_{f_{\la}} e^{-\Gamma_{\la}(0)}\left(\ZtillamastT\right)^{\frac{1}{b-1}}}{(0, k^{\varepsilon}_{\la})}}\\
        &\qquad + \EQ{ \ZtillamastT G_T \indF{v_{f_{\la}} e^{-\Gamma_{\la}(0)}\left(\ZtillamastT\right)^{\frac{1}{b-1}}}{[k^{\varepsilon}_{\la}, G_T]}} \\
        & \qquad + \EQ{ \ZtillamastT v_{f_{\la}}  e^{-\Gamma_{\la}(0)} \left(\ZtillamastT\right)^{\frac{1}{b-1}} \indF{v_{f_{\la}} e^{-\Gamma_{\la}(0)}\left(\ZtillamastT\right)^{\frac{1}{b-1}}}{(G_T, +\infty)}}\\
        & = E_1 + E_2 + E_3.
    \end{flalign*}

    Take $0 < l < k^{\varepsilon}_{\la}$ and calculate:
    \begin{flalign*}
        E_1(l) &= \EQ{ \ZtillamastT v_{f_{\la}}  e^{-\Gamma_{\la}(0)} \left(\ZtillamastT\right)^{\frac{1}{b-1}} \indF{v_{f_{\la}} e^{-\Gamma_{\la}(0)}\left(\ZtillamastT\right)^{\frac{1}{b-1}}}{(l, k^{\varepsilon}_{\la})}}\\
        & \stackrel{\eqref{eq:gamma_Z_for_auxiliary_market}}{=} v_{f_{\la}} e^{-\Gamma_{\la}(0)} \EQ{\left(e^{-\lrr{r + 0.5||\gamlamast||^2}T - \gamlamast' W(T)}\right)^{\frac{b}{b - 1}}\indF{v_{f_{\la}} e^{-\Gamma_{\la}(0)}\left(\ZtillamastT\right)^{\frac{1}{b-1}}}{(l, k^{\varepsilon}_{\la})}}\\
        & \stackrel{\text{Lem.}\,\ref{lem:inequalities_optimal_unconstrained_wealth}}{=} v_{f_{\la}} e^{-\Gamma_{\la}(0)}e^{\frac{b}{1 - b}\lrr{r + 0.5||\gamlamast||^2}T} \mathbb{E}_{\Q} \biggl[ e^{\frac{b}{1 - b}||\gamlamast||\sqrt{T} \frac{\gamlamast'W(T)}{|| \gamlamast ||\sqrt{T}}}  \\
        & \qquad \indF{\frac{\gamlamast' W(T)}{|| \gamlamast || \sqrt{T}}}{(-d^{\la}_2(l, v_{f_{\la}}, 0) - || \gamlamast || \sqrt{T}, -d^{\la}_2(k^{\varepsilon}_{\la}, v_{f_{\la}}, 0) - || \gamlamast || \sqrt{T}]}  \biggr]\\
        & = v_{f_{\la}} e^{-\Gamma_{\la}(0)}e^{\frac{b}{1 - b}\lrr{r + 0.5||\gamlamast||^2}T}
        \mathbb{E}_{\Q} \biggl[ e^{\frac{b}{1 - b}||\gamlamast||\sqrt{T} X}  \\
        & \qquad \indF{X}{(-d^{\la}_2(l, v_{f_{\la}}, 0) - || \gamlamast || \sqrt{T}, -d^{\la}_2(k^{\varepsilon}_{\la}, v_{f_{\la}}, 0) - || \gamlamast || \sqrt{T}]}  \biggr]\\
        & \stackrel{\text{Lem.}\,\ref{lem:expectation_ztilde_interval}}{=} v_{f_{\la}} e^{-\Gamma_{\la}(0)}e^{\frac{b}{1 - b}\lrr{r + 0.5||\gamlamast||^2}T}e^{\frac{1}{2}\left(\frac{b}{1 - b} ||\gamlamast|| \sqrt{T}\right)^2}\\
        &\qquad  \qquad \cdot \left( \Phi\biggl( -d^{\la}_2(k^{\varepsilon}_{\la}, v_{f_{\la}}, 0) - || \gamlamast || \sqrt{T} -  \frac{b}{b - 1} ||\gamlamast|| \sqrt{T}\right)  \\
        & \qquad \qquad  \qquad - \Phi\left( -d^{\la}_2(l, v_{f_{\la}}, 0) - || \gamlamast || \sqrt{T} -  \frac{b}{1 - b} ||\gamlamast|| \sqrt{T}\right) \biggr)\\
        & \stackrel[\eqref{eq:Gamma_d_1_d_2}]{\Phi(-x)=1 - \Phi(x)}{=} v_{f_{\la}} \cdot \biggl(1 -  \Phi\left( d^{\la}_2(k^{\varepsilon}_{\la}, v_{f_{\la}}, 0) +  \frac{1}{1 - b} ||\gamlamast|| \sqrt{T}\right) \\
        & \qquad \qquad - \left(1 -  \Phi\left( d^{\la}_2(l, v_{f_{\la}}, 0) +  \frac{1}{1 - b} ||\gamlamast|| \sqrt{T}\right)\right)\biggr)\\
        & \stackrel{\eqref{eq:Gamma_d_1_d_2}}{=} v_{f_{\la}}\cdot \left( \Phi\left( d^{\la}_1(l, v_{f_{\la}}, 0)\right) - \Phi\left( d^{\la}_1(k^{\varepsilon}_{\la}, v_{f_{\la}}, 0)\right) \right),
    \end{flalign*}
    where we used in the fourth equality $X \stackrel{d}{=} N(0, 1) \stackrel{d}{=} \frac{\gamlamast'W(T)}{|| \gamlamast ||\sqrt{T}}$. Hence, we get:
    \begin{equation*}
        \begin{aligned}
            E_1 &= \lim_{l\downarrow 0} E_1(l) = \lim_{l\downarrow 0} \lrr{v_{f_{\la}} \left( \Phi\left( d^{\la}_1(l, v_{f_{\la}}, 0)\right) - \Phi\left( d^{\la}_1(k^{\varepsilon}_{\la}, v_{f_{\la}}, 0)\right) \right)}\\
            & \stackrel{\Phi\,\text{is cts.}}{=}v_{f_{\la}} \left( \Phi\left(\lim_{l\downarrow 0} d^{\la}_1(l, v_{f_{\la}}, 0)\right) - \Phi\left( d^{\la}_1(k^{\varepsilon}_{\la}, v_{f_{\la}}, 0)\right) \right) \\
            & \stackrel[ b < 1]{\eqref{eq:Gamma_d_1_d_2}}{=}
           v_{f_{\la}} \left( 1 - \Phi\left( d^{\la}_1(k^{\varepsilon}_{\la}, v_{f_{\la}}, 0)\right) \right)
        \end{aligned}
    \end{equation*}
    where we also used in the last equality $\lim_{l\downarrow 0} \ln(l) = -\infty$ and $\lim_{u\uparrow +\infty}\Phi(u) = 1$.
    
    Replacing $0$ by $G_T$ and $k^{\varepsilon}_{\la}$ by $u$ and considering $\lim_{u\uparrow +\infty}$ we obtain:
    \begin{equation*}
        \begin{aligned}
            E_3 &= \lim_{u\uparrow +\infty} E_3(u) = v_{f_{\la}}\cdot \left( \Phi\left( d^{\la}_1(G_T, v_{f_{\la}}, 0)\right) - \Phi\left( \lim_{u\uparrow +\infty} d^{\la}_1(u, v_{f_{\la}}, 0)\right) \right)\\
            & \stackrel[ b < 1]{\eqref{eq:Gamma_d_1_d_2}}{=} v_{f_{\la}}\cdot \Phi\left( d^{\la}_1(G_T, v_{f_{\la}}, 0)\right),
        \end{aligned}
    \end{equation*}
    where we also used in the last equality $\lim_{u\uparrow +\infty} \ln(u) = +\infty$ and $\lim_{l\downarrow 0}\Phi(l) = 0$.
    
    Next we calculate $E_2$:
    \begin{flalign*}
        E_2 & = \EQ{\ZtillamastT G_T \indF{v_{f_{\la}} e^{-\Gamma_{\la}(0)}\left(\ZtillamastT\right)^{\frac{1}{b-1}}}{[k^{\varepsilon}_{\la}, G_T]} }\\
        & \stackrel{\eqref{eq:gamma_Z_for_auxiliary_market}}{=} G_T \EQ{e^{-\lrr{r + 0.5||\gamlamast||^2}T - \gamlamast' W(T)}\indF{v_{f_{\la}} e^{-\Gamma_{\la}(0)}\left(\ZtillamastT\right)^{\frac{1}{b-1}}}{[k^{\varepsilon}_{\la}, G_T]}}\\
        & \stackrel{\text{Lem.}\,\ref{lem:inequalities_optimal_unconstrained_wealth}}{=} G_T e^{-\lrr{r + 0.5||\gamlamast||^2}T} \mathbb{E}_{\Q} \biggl[e^{-||\gamlamast||\sqrt{T} \frac{\gamlamast'W(T)}{|| \gamlamast ||\sqrt{T}}} \\
        & \qquad \cdot \indF{\frac{\gamlamast' W(T)}{|| \gamlamast || \sqrt{T}}}{(-d^{\la}_2(k^{\varepsilon}_{\la}, v_{f_{\la}}, 0) - || \gamlamast || \sqrt{T}, -d^{\la}_2(G_T, v_{f_{\la}}, 0) - || \gamlamast || \sqrt{T}]}\biggr]\\
        & \stackrel{\text{Lem.}\,\ref{lem:expectation_ztilde_interval}}{=} G_T e^{-\lrr{r + 0.5||\gamlamast||^2}T}e^{0.5||\gamlamast||^2 T} \biggl( \Phi \left(-d^{\la}_2(G_T, v_{f_{\la}}, 0) - || \gamlamast || \sqrt{T} -   \lrr{-||\gamlamast|| \sqrt{T}}\right)  \\
        & \qquad \qquad  \qquad - \Phi\left( -d^{\la}_2(k^{\varepsilon}_{\la}, v_{f_{\la}}, 0) - || \gamlamast || \sqrt{T} -  \lrr{-||\gamlamast|| \sqrt{T}} \right) \biggr)\\
        & \stackrel{\Phi(-x)=1 - \Phi(x)}{=} e^{-rT}G_T \left( \Phi\left( d^{\la}_2(k^{\varepsilon}_{\la}, v_{f_{\la}}, 0)\right) - \Phi\left( d^{\la}_2(G_T, v_{f_{\la}}, 0)\right) \right),
    \end{flalign*}
    where we used in the fourth equality $X \stackrel{d}{=} N(0, 1) \stackrel{d}{=} \frac{\gamlamast'W(T)}{|| \gamlamast ||\sqrt{T}}$.
    
    Finally, we obtain the explicit form of the left-hand side of the budget constraint:
    \begin{flalign*}
        E_1 + E_2 + E_3 &= v_{f_{\la}}\cdot \left( 1 - \Phi(d_1^{\la}(k^{\varepsilon}_{\la}, v_{f_{\la}}, 0)) + \Phi(d_1^{\la}(G_T, v_{f_{\la}}, 0)) \right) \\
        & \quad + e^{-rT}G_T \cdot \left( \Phi\left( d^{\la}_2(k^{\varepsilon}_{\la}, v_{f_{\la}}, 0)\right) - \Phi\left( d^{\la}_2(G_T, v_{f_{\la}}, 0)\right) \right).
    \end{flalign*}
    Second, we simplify the left-hand side of the VaR-constraint:
    \begin{flalign*}
        \Q & \lrr{V^{v_{f_{\la}}, \hat{\pi}_{\la}}(T)) < G_T} \\
        & \stackrel{f\,\text{def.}}{=} \Q \lrr{ V^{v_{f_{\la}}, \hat{\pi}_{\la}}(T)  + \big(G_T - V^{v_{f_{\la}}, \hat{\pi}_{\la}}(T)\big)\mathbbm{1}_{[k^{\varepsilon}_{\la},G_T]}(V^{v_{f_{\la}}, \hat{\pi}_{\la}}(T)) <G_T}\\
        & =  \Q \lrr{ V^{v_{f_{\la}}, \hat{\pi}_{\la}}(T) <k^{\varepsilon}_{\la}} 
        \stackrel{\text{Lem.}\,\ref{lem:inequalities_optimal_unconstrained_wealth}}{=} \Q \lrr{ \frac{\gamlamast' W(T)}{|| \gamlamast || \sqrt{T}} < -d_2^{\la}( k^{\varepsilon}_{\la}, v_{f_{\la}}, 0) - || \gamlamast || \sqrt{T}} \\
        & = \Phi \lrr{-d_2^{\la}(k^{\varepsilon}_{\la}, v_{f_{\la}}, 0) - || \gamlamast || \sqrt{T}} 
        \stackrel{\Phi(-x)=1 - \Phi(x)}{=} 1 - \Phi \lrr{d_2^{\la}( k^{\varepsilon}_{\la}, v_{f_{\la}}, 0) + || \gamlamast || \sqrt{T}},
    \end{flalign*}
    where we used in the second to last equality $\frac{\gamlamast'W(T)}{|| \gamlamast ||\sqrt{T}}\stackrel{d}{=} N(0, 1)$.
    
    The claim of the proposition follows.
\end{proof}
\textbf{Remark to Proposition \ref{prop:v_f_kepsilon_SNLE}:} As argued in the proof of Proposition \ref{prop: generalization no-shortsell}, the optimal portfolio for (\ref{OP_P_epsilon_K}) is also the optimal portfolio for $(P^{\la}_\varepsilon)$. Moreover, both corresponding wealth processes coincide according to \eqref{eq: Condition b} and therefore the present value of the optimal terminal payoff coincides in both $\mathcal{M}$ and $\mathcal{M}_{\la}$. This means, in Equation \eqref{eq: VaR constr. and budget} we can use either the budget equation in $\mathcal{M}$ or in $\mathcal{M}_{\la}$ to determine the parameters $(v_{f_{\la}}, k^{\varepsilon}_{\la})$. \\

    \begin{proposition}\label{prop:value_function_formula}
    The value function is given by:
    \begin{equation*}
        \begin{aligned}
            \mathbb{E}_{\Q}\bigl[ & U(\bar{V}^{v_0, \pib^*}(T))\bigr] = \frac{1}{b} \lrr{v_{f_{\la}}}^{b}e^{\Gamma_{\la}(0)(1 - b)} \lrr{1 - \Phi(d_1^{\la}(k^{\varepsilon}, v_{f_{\la}}, 0)) + \Phi(d_1^{\la}(G_T, v_{f_{\la}}, 0))} \\
            & \quad +  \frac{1}{b} \lrr{G_T}^{b} \left( \Phi\left( d^{\la}_2(k^{\varepsilon}_{\la},  v_{f_{\la}}, 0) + \Vert \gamlamast \Vert \sqrt{T}\right) - \Phi\left( d^{\la}_2(G_T,  v_{f_{\la}}, 0) + \Vert \gamlamast \Vert \sqrt{T} \right) \right).
        \end{aligned}
    \end{equation*}
\end{proposition}
\begin{proof}

\begin{flalign*}
    \mathbb{E} \bigl[ &U \lrr{\bar{V}^{v_0, \pib^*}(T)}\bigr]  \stackrel{\text{Pr.}\ref{prop:pi_relations}}{=} \EQ{U\lrr{\bar{V}^{v_0, \pi^*}(T)}} \stackrel{\text{Pr.}\ref{prop: generalization no-shortsell}}{=} \EQ{U\lrr{f_{\la}(V^{v_{f_{\la}}, \hat{\pi}_{\la}}_{\la}(T)}} \\
    & \stackrel{f \text{ def.}}{=} \EQ{U\left( V^{v_{f_{\la}}, \hat{\pi}_{\la}}_{\la}(T)  + \big(G_T - V^{v_{f_{\la}}, \hat{\pi}_{\la}}_{\la} (T)\big)\mathbbm{1}_{[k^{\varepsilon}_{\la}, G_T]}(V^{v_{f_{\la}}, \hat{\pi}_{\la}}_{\la}(T))\right)}\\
    & = \EQ{U\lrr{V^{v_{f_{\la}}, \hat{\pi}_{\la}}_{\la}(T)} \mathbbm{1}_{\lrr{0, k^{\varepsilon}_{\la}}}\lrr{V^{v_{f_{\la}}, \hat{\pi}_{\la}}_{\la}(T)} }\\
    & \qquad + \EQ{U\lrr{G_T} \mathbbm{1}_{[k^{\varepsilon}_{\la}, G_T]}\lrr{V^{v_{f_{\la}}, \hat{\pi}_{\la}}_{\la}(T)} } \\
    & \qquad +\EQ{U\lrr{V^{v_{f_{\la}}, \hat{\pi}_{\la}}_{\la}(T)} \mathbbm{1}_{\lrr{G_T, +\infty}}\lrr{V^{v_{f_{\la}}, \hat{\pi}_{\la}}_{\la}(T)} } =: E_1 + E_2 + E_3.
\end{flalign*}
Take $0 < l < k^{\varepsilon}$ and calculate:
    \begin{flalign*}
        E_1(l) &= \EQ{ \frac{1}{b} \lrr{v_{f_{\la}}  e^{-\Gamma_{\la}(0)} \left(\ZtillamastT\right)^{\frac{1}{b-1}}}^b \indF{v_{f_{\la}} e^{-\Gamma_{\la}(0)}\left(\ZtillamastT\right)^{\frac{1}{b-1}}}{(l, k^{\varepsilon}_{\la})}}\\
        & \stackrel{\eqref{eq:gamma_Z_for_auxiliary_market}}{=} \frac{v_{f_{\la}}^b}{b} e^{-b\Gamma_{\la}(0)} \EQ{\left(e^{-\lrr{r + 0.5||\gamlamast||^2}T - \gamlamast' W(T)}\right)^{\frac{b}{b - 1}}\indF{v_{f_{\la}} e^{-\Gamma_{\la}(0)}\left(\ZtillamastT\right)^{\frac{1}{b-1}}}{(l, k^{\varepsilon}_{\la})}}\\
        & \stackrel{\text{Lem.}\,\ref{lem:inequalities_optimal_unconstrained_wealth}}{=} \frac{v_{f_{\la}}^b}{b} e^{-b\Gamma_{\la}(0)} e^{\frac{b}{1 - b}\lrr{r + 0.5||\gamlamast||^2}T} \mathbb{E}_{\Q} \biggl[ e^{\frac{b}{1 - b}||\gamlamast||\sqrt{T} \frac{\gamlamast'W(T)}{|| \gamlamast ||\sqrt{T}}}  \\
        & \qquad \indF{\frac{\gamlamast' W(T)}{|| \gamlamast || \sqrt{T}}}{(-d^{\la}_2(l, v_{f_{\la}}, 0) - || \gamlamast || \sqrt{T}, -d^{\la}_2(k^{\varepsilon}_{\la}, v_{f_{\la}}, 0) - || \gamlamast || \sqrt{T}]}  \biggr]\\
        & = \frac{v_{f_{\la}}^b}{b} e^{-b\Gamma_{\la}(0)}e^{\frac{b}{1 - b}\lrr{r + 0.5||\gamlamast||^2}T}
        \mathbb{E}_{\Q} \biggl[ e^{\frac{b}{1 - b}||\gamlamast||\sqrt{T} X}  \\
        & \qquad \indF{X}{(-d^{\la}_2(l, v_{f_{\la}}, 0) - || \gamlamast || \sqrt{T}, -d^{\la}_2(k^{\varepsilon}_{\la}, v_{f_{\la}}, 0) - || \gamlamast || \sqrt{T}]}  \biggr]\\
        & \stackrel{\text{Lem.}\,\ref{lem:expectation_ztilde_interval}}{=} \frac{v_{f_{\la}}^b}{b} e^{-b\Gamma_{\la}(0)}e^{\frac{b}{1 - b}\lrr{r + 0.5||\gamlamast||^2}T}e^{\frac{1}{2}\left(\frac{b}{1 - b} ||\gamlamast|| \sqrt{T}\right)^2}\\
        &\qquad  \qquad \cdot \left( \Phi\biggl( -d^{\la}_2(k^{\varepsilon}_{\la}, v_{f_{\la}}, 0) - || \gamlamast || \sqrt{T} -  \frac{b}{1 - b} ||\gamlamast|| \sqrt{T}\right)  \\
        & \qquad \qquad  \qquad - \Phi\left( -d^{\la}_2(l, v_{f_{\la}}, 0) - || \gamlamast || \sqrt{T} -  \frac{b}{1 - b} ||\gamlamast|| \sqrt{T}\right) \biggr)\\
        & \stackrel[\eqref{eq:Gamma_d_1_d_2}]{\Phi(-x)=1 - \Phi(x)}{=} \frac{v_{f_{\la}}^b}{b} e^{(1-b)\Gamma_{\la}(0)} \cdot \biggl(1 -  \Phi\left( d^{\la}_2(k^{\varepsilon}_{\la}, v_{f_{\la}}, 0) +  \frac{1}{1 - b} ||\gamlamast|| \sqrt{T}\right) \\
        & \qquad \qquad - \left(1 -  \Phi\left( d^{\la}_2(l, v_{f_{\la}}, 0) +  \frac{1}{1 - b} ||\gamlamast|| \sqrt{T}\right)\right)\biggr)\\
        & \stackrel{\eqref{eq:Gamma_d_1_d_2}}{=} \frac{v_{f_{\la}}^b}{b} e^{(1-b)\Gamma_{\la}(0)}\cdot \left( \Phi\left( d^{\la}_1(l, v_{f_{\la}}, 0)\right) - \Phi\left( d^{\la}_1(k^{\varepsilon}_{\la}, v_{f_{\la}}, 0)\right) \right),
    \end{flalign*}
    where we used in the fourth equality $X \stackrel{d}{=} N(0, 1) \stackrel{d}{=} \frac{\gamlamast'W(T)}{|| \gamlamast ||\sqrt{T}}$. Hence, we get:
    \begin{equation*}
        \begin{aligned}
            E_1 &= \lim_{l\downarrow 0} E_1(l) = \lim_{l\downarrow 0} \lrr{\frac{v_{f_{\la}}^b}{b} e^{(1-b)\Gamma_{\la}(0)} \left( \Phi\left( d^{\la}_1(l, v_{f_{\la}}, 0)\right) - \Phi\left( d^{\la}_1(k^{\varepsilon}_{\la}, v_{f_{\la}}, 0)\right) \right)}\\
            & \stackrel{\Phi\,\text{is cts.}}{=}\frac{v_{f_{\la}}^b}{b} e^{(1-b)\Gamma_{\la}(0)} \left( \Phi\left(\lim_{l\downarrow 0} d^{\la}_1(l, v_{f_{\la}}, 0)\right) - \Phi\left( d^{\la}_1(k^{\varepsilon}_{\la}, v_{f_{\la}}, 0)\right) \right) \\
            & \stackrel[ b < 1]{\eqref{eq:Gamma_d_1_d_2}}{=}
           \frac{v_{f_{\la}}^b}{b} e^{(1-b)\Gamma_{\la}(0)} \left( 1 - \Phi\left( d^{\la}_1(k^{\varepsilon}_{\la}, v_{f_{\la}}, 0)\right) \right),
        \end{aligned}
    \end{equation*}
    where we also used in the last equality $\lim_{l\downarrow 0} \ln(l) = -\infty$ and $\lim_{u\uparrow +\infty}\Phi(u) = 1$.
    
    Replacing $0$ by $G_T$ and $k^{\varepsilon}_{\la}$ by $u$ and considering $\lim_{u\uparrow +\infty}$ we obtain:
    \begin{equation*}
        \begin{aligned}
            E_3 &= \lim_{u\uparrow +\infty} E_3(u) = \frac{v_{f_{\la}}^b}{b} e^{(1-b)\Gamma_{\la}(0)} \cdot \left( \Phi\left( d^{\la}_1(G_T, v_{f_{\la}}, 0)\right) - \Phi\left( \lim_{u\uparrow +\infty} d^{\la}_1(u, v_{f_{\la}}, 0)\right) \right)\\
            & \stackrel[ b < 1]{\eqref{eq:Gamma_d_1_d_2}}{=} \frac{v_{f_{\la}}^b}{b} e^{(1-b)\Gamma_{\la}(0)} \cdot \Phi\left( d^{\la}_1(G_T, v_{f_{\la}}, 0)\right),
        \end{aligned}
    \end{equation*}
    where we also used in the last equality $\lim_{u\uparrow +\infty} \ln(u) = +\infty$ and $\lim_{l\downarrow 0}\Phi(l) = 0$.
    
    Next we calculate $E_2$:
    \begin{flalign*}
        E_2 & = \EQ{ \frac{\lrr{G_T}^b}{b} \indF{v_{f_{\la}} e^{-\Gamma_{\la}(0)}\left(\ZtillamastT\right)^{\frac{1}{b-1}}}{[k^{\varepsilon}_{\la}, G_T]} }\\
        & \stackrel{\eqref{eq:gamma_Z_for_auxiliary_market}}{=} \frac{\lrr{G_T}^b}{b} \EQ{\indF{v_{f_{\la}} e^{-\Gamma_{\la}(0)}\left(e^{-\lrr{r + 0.5||\gamlamast||^2}T - \gamlamast' W(T)}\right)^{\frac{1}{b-1}}}{[k^{\varepsilon}_{\la}, G_T]}}\\
        & \stackrel{\text{Lem.}\,\ref{lem:inequalities_optimal_unconstrained_wealth}}{=} \frac{\lrr{G_T}^b}{b} \mathbb{E}_{\Q} \biggl[ \indF{\frac{\gamlamast' W(T)}{|| \gamlamast || \sqrt{T}}}{(-d^{\la}_2(k^{\varepsilon}_{\la}, v_{f_{\la}}, 0) - || \gamlamast || \sqrt{T}, -d^{\la}_2(G_T, v_{f_{\la}}, 0) - || \gamlamast || \sqrt{T}]}\biggr]\\
        & \stackrel{\text{Lem.}\,\ref{lem:expectation_ztilde_interval}}{=} \frac{\lrr{G_T}^b}{b} \biggl( \Phi \left(-d^{\la}_2(G_T, v_{f_{\la}}, 0) - || \gamlamast || \sqrt{T} \right) - \Phi\left( -d^{\la}_2(k^{\varepsilon}_{\la}, v_{f_{\la}}, 0) - || \gamlamast || \sqrt{T} \right) \biggr)\\
        & \stackrel{\Phi(-x)=1 - \Phi(x)}{=}\frac{\lrr{G_T}^b}{b} \left( \Phi\left( d^{\la}_2(k^{\varepsilon}_{\la}, v_{f_{\la}}, 0) + \Vert \gamlamast \Vert \sqrt{T}\right) - \Phi\left( d^{\la}_2(G_T, v_{f_{\la}}, 0) + \Vert \gamlamast \Vert \sqrt{T} \right) \right),
    \end{flalign*}
    where we used in the fourth equality $X \stackrel{d}{=} N(0, 1) \stackrel{d}{=} \frac{\gamlamast'W(T)}{|| \gamlamast ||\sqrt{T}}$.
    
    The claim of the proposition follows.
\end{proof}

\end{appendix}

\end{document}